\documentclass[preprint,prd,tightenlines,floatfix,
nofootinbib,eqsecnum,superscriptaddress]{revtex4-1}

\usepackage[T1]{fontenc}		

\usepackage{amsmath,amsfonts,amssymb,amstext,mathrsfs}
\usepackage{mathpazo}

\usepackage[dvips]{graphicx}
\usepackage{epsf,float}
\usepackage{revsymb}

\usepackage{dcolumn}
\usepackage{braket}
\usepackage{color,xcolor}
\usepackage{graphicx}
\usepackage{subfigure}
\usepackage{multirow}
\usepackage{tabularx}
\usepackage{pstricks}
\usepackage[section]{placeins}
\usepackage{booktabs}
\usepackage{array}

\usepackage{commath}

\usepackage{hyperref}

\usepackage{lineno}

\newcommand{\Pom}{\mathbb{P}}

\newcommand{\Reg}{\mathbb{R}}

\newcommand{\bdPt}{\mbox{\boldmath $dP_{t}$}}
\newcommand{\bqta}{\mbox{\boldmath $q_{t,1}$}}
\newcommand{\bqtb}{\mbox{\boldmath $q_{t,2}$}}
\newcommand{\bpta}{\mbox{\boldmath $p_{t,1}$}}
\newcommand{\bptb}{\mbox{\boldmath $p_{t,2}$}}

\newcommand{\bptat}{\mbox{\boldmath $\tilde{p}_{t,1}$}}
\newcommand{\bptbt}{\mbox{\boldmath $\tilde{p}_{t,2}$}}
\newcommand{\bkt}{\mbox{\boldmath $k_{t}$}}
\newcommand{\bktsqrt}{\mbox{\boldmath{$k_{t}^{2}$}}}

\newcommand{\bqa}{\mbox{\boldmath $q_{1}$}}

\newcommand{\bp}{\mbox{\boldmath $p$}}
\newcommand{\bk}{\mbox{\boldmath $k$}}
\newcommand{\bpa}{\mbox{\boldmath $p_{a}$}}
\newcommand{\bpb}{\mbox{\boldmath $p_{b}$}}
\newcommand{\bpaa}{\mbox{\boldmath $p_{1}$}}
\newcommand{\bpbb}{\mbox{\boldmath $p_{2}$}}

\newcommand{\bea}{\mbox{\boldmath $e_{1}$}}
\newcommand{\beb}{\mbox{\boldmath $e_{2}$}}
\newcommand{\bec}{\mbox{\boldmath $e_{3}$}}

\newcommand{\bex}{\mbox{\boldmath $e_{x}$}}
\newcommand{\bey}{\mbox{\boldmath $e_{y}$}}
\newcommand{\bez}{\mbox{\boldmath $e_{z}$}}

\newcommand{\bepsilon}{\mbox{\boldmath $\epsilon$}}

\renewcommand\slash[1]{\not \! #1}

\newcommand*\wideestimates{\mathrel{\widehat{=}}}

\newcommand{\p}{\partial}

\newcommand{\twosidep}[1]{\stackrel{\leftrightarrow}{\p}_{\! #1}}
\newcommand{\leftsidep}[1]{\stackrel{\leftarrow}{\p}_{\! #1}}
\newcommand{\rightsidep}[1]{\stackrel{\rightarrow}{\p}_{\! #1}}

\usepackage[normalem]{ulem}  

\begin{document}

\nolinenumbers

\title{\boldmath 
Central exclusive diffractive production of axial-vector $f_{1}(1285)$
and $f_{1}(1420)$ mesons in proton-proton collisions}

\vspace{0.6cm}

\author{Piotr Lebiedowicz}
 \email{Piotr.Lebiedowicz@ifj.edu.pl}
\affiliation{Institute of Nuclear Physics Polish Academy of Sciences, Radzikowskiego 152, PL-31342 Krak{\'o}w, Poland}

\author{Josef Leutgeb}
 \email{josef.leutgeb@tuwien.ac.at}
\affiliation{Institut f\"ur Theoretische Physik, Technische Universit\"at Wien,
Wiedner Hauptstrasse 8-10, A-1040 Vienna, Austria}

\author{Otto Nachtmann}
 \email{O.Nachtmann@thphys.uni-heidelberg.de}
\affiliation{Institut f\"ur Theoretische Physik, Universit\"at Heidelberg,
Philosophenweg 16, D-69120 Heidelberg, Germany}

\author{Anton Rebhan}
 \email{anton.rebhan@tuwien.ac.at}
\affiliation{Institut f\"ur Theoretische Physik, Technische Universit\"at Wien,
Wiedner Hauptstrasse 8-10, A-1040 Vienna, Austria}

\author{Antoni Szczurek
\footnote{Also at \textit{College of Natural Sciences, 
Institute of Physics, University of Rzesz{\'o}w, 
ul. Pigonia 1, PL-35310 Rzesz{\'o}w, Poland}.}}
\email{Antoni.Szczurek@ifj.edu.pl}
\affiliation{Institute of Nuclear Physics Polish Academy of Sciences, Radzikowskiego 152, PL-31342 Krak{\'o}w, Poland}

\begin{abstract}
We present a study of the central exclusive diffractive 
production of the $f_{1}(1285)$ and $f_{1}(1420)$ resonances
in proton-proton collisions.
The theoretical results are calculated within the tensor-pomeron approach.
Two pomeron-pomeron-$f_{1}$ tensorial couplings labeled 
by $(l,S) = (2,2)$ and $(4,4)$ are derived.
We adjust the model parameters (coupling constants, cutoff constant)
to the WA102 experimental data taking into account absorption effects.
Both the $(l,S) = (2,2)$ and $(4,4)$ couplings separately
allow one to describe the WA102 differential distributions.
We compare these predictions with 
those of the Sakai-Sugimoto model,
where the pomeron-pomeron-$f_{1}$ couplings are determined 
by the mixed axial-gravitational anomaly of QCD.
We derive an approximate relation between the pomeron-pomeron-$f_{1}$ 
coupling constants of this approach 
and the $(l,S) = (2,2)$ and $(4,4)$ couplings.
Then we present our predictions 
for the energies available at the RHIC and LHC.
The total cross sections and several differential distributions are presented.
Analysis of the distributions in the azimuthal angle
$\phi_{pp}$ between the transverse momenta of the outgoing protons
may be used to disentangle $f_{1}$- and $\eta$-type resonances
contributing to the same final channel.
We find for the $f_{1}(1285)$ a total cross section 
$\sim 38$~$\mu$b for $\sqrt{s} = 13$~TeV and a rapidity cut 
on the $f_{1}$ of $|{\rm y_{M}}| < 2.5$.
We predict a much larger cross section 
for production of $f_{1}(1285)$ than for production of $f_{2}(1270)$
in the $\pi^{+}\pi^{-}\pi^{+}\pi^{-}$ decay channel for the LHC energies.
This opens a possibility to study the $f_{1}(1285)$ meson 
in experiments planned at the LHC.
\end{abstract}


\maketitle

\section{Introduction}
The central exclusive production of pseudovector, 
or axial-vector, mesons with $I^{G}J^{PC} = 0^{+}1^{++}$, 
namely the $f_{1}(1285)$ and $f_{1}(1420)$, was studied
in proton-proton collisions by the WA102 Collaboration
for $\sqrt{s} = 12.7$ and $29.1$~GeV \cite{Barberis:1997ve,Barberis:1997vf,Barberis:1998by}.
The $f_{1}(1285)$ and the $f_{1}(1420)$ are well known 
but their internal structure ($q \bar{q}$, tetraquark, or molecule) remains to be established.
In~\cite{Barberis:1998by} the branching fractions of both mesons 
in all major decay modes were determined.
The $f_{1}(1280)$ was found to decay to 
$\eta \pi^{+} \pi^{-}$, $4 \pi$, $K \bar{K} \pi$, 
and $\rho^{0} \gamma$ while
the $f_{1}(1420)$ was found to decay dominantly 
to $K \bar{K} \pi$, including $K^{*}(892) \bar{K}$
+ c.c.; see \cite{Tanabashi:2018oca}.
In~\cite{Barberis:1997ve,Barberis:1999wn} 
the $\pi^{+}\pi^{-}\pi^{+}\pi^{-}$ and $\pi^{+}\pi^{-}\pi^{0}\pi^{0}$ mass spectra 
were studied and a clear peak associated 
with the $f_{1}(1285)$ meson
in the $J^{P} = 1^{+}$ $\rho \rho$ wave was observed.
Moreover, both the $f_{1}(1285)$ and $f_{1}(1420)$ mesons are suppressed 
at small glueball-filter variable ${\rm dP_{t}}$ \cite{Barberis:1998by}.
This behaviour is consistent with the signals being 
due to standard $q \bar{q}$ states \cite{Close:1997pj}.
Recent analysis of the $f_{1}(1285) \to \rho^{0} \pi^{+} \pi^{-}$ decay mode
\cite{Osipov:2018iah} favours a $q \bar{q}$ content of the $f_{1}(1285)$.
However, a glue component for the $f_{1}(1285)$ is not excluded 
\cite{Birkel:1995ct,Moreira:2017ulo}.
Though the $f_{1}(1420)$ is well established experimentally, 
its internal structure is debated in the literature;
see, e.g., \cite{Close:1997nm,Li:2000dy,Debastiani:2016xgg,
Wang:2018mjz, Liang:2020jtw}.
The study done in
\cite{Debastiani:2016xgg,Liang:2020jtw}
proposes that the $f_{1}(1420)$ 
may not be a genuine $q \bar{q}$ resonance, 
but the manifestation of the $K^{*}(892) \bar{K}$ and $\pi a_{0}(980)$ decay modes
of the $f_{1}(1285)$ resonance around 1420~MeV.
In our paper we shall treat the $f_{1}(1285)$
and the $f_{1}(1420)$ as separate objects,
we can say, as two effective resonances.
We emphasize that in this way, for most of our results,
we do not give any preference to the different views
on the precise nature of the two $f_{1}$ objects.
For some of our results we assume that the $f_{1}(1285)$
and the $f_{1}(1420)$ can be described as suitable $q \bar{q}$ states.
This assumption will then be stated explicitly at the appropriate places.
The $f_{1}(1510)$, a third $J^{P} = 1^{+}$ meson, 
is not well established; see \cite{Tanabashi:2018oca}.
The cross section as a function of center-of-mass energy 
for both the $f_{1}(1285)$ and the $f_{1}(1420)$ mesons 
was found \cite{Barberis:1998by} 
to be consistent with being produced via the double-pomeron-exchange
(i.e., $\Pom \Pom$-fusion) mechanism.

The pomeron ($\Pom$) is an essential object 
for understanding diffractive phenomena in high-energy physics. 
Within QCD the pomeron is a color singlet, predominantly gluonic, object.
The spin structure of the pomeron, in particular its coupling to hadrons, 
is, however, not yet a matter of consensus.
In the tensor-pomeron model for soft high-energy scattering
formulated in \cite{Ewerz:2013kda},
on the basis of earlier work \cite{Nachtmann:1991ua},
the pomeron exchange is effectively treated 
as the exchange of a rank-2 symmetric tensor,
as also in the holographic QCD models in 
\cite{Brower:2006ea,Domokos:2009hm,Anderson:2014jia,Ballon-Bayona:2015wra,Iatrakis:2016rvj,Anderson:2016zon}.
It is rather difficult to obtain definitive statements on
the spin structure of the pomeron from unpolarised 
elastic proton-proton scattering.
On the other hand, the results
from polarised proton-proton scattering 
by the STAR Collaboration \cite{Adamczyk:2012kn} 
provide valuable information on this question.
Three hypotheses for the spin structure of the pomeron, tensor,
vector, and scalar, were discussed in \cite{Ewerz:2016onn}
in view of the experimental results from \cite{Adamczyk:2012kn}.
Only the tensor ansatz for the pomeron was found to be compatible with the experiment.
Also some historical remarks on different views of the pomeron 
were made in \cite{Ewerz:2016onn}.
In \cite{Britzger:2019lvc} further strong evidence against 
the hypothesis of a vector character of the pomeron was given. 

In the last few years a scientific program was undertaken to analyse 
the central exclusive production (CEP) of light mesons 
in the tensor-pomeron and vector-odderon model
in several reactions: $p p \to p p M$~\cite{Lebiedowicz:2013ika},
where $M$ stands for a scalar or pseudoscalar meson, 
$p p \to p p \pi^{+}\pi^{-}$ \cite{Lebiedowicz:2014bea,Lebiedowicz:2016ioh}, 
$p p \to p n \rho^{0} \pi^{+}$ ($p p \rho^{0} \pi^{0}$) 
\cite{Lebiedowicz:2016ryp},
$p p \to p p K^{+}K^{-}$ \cite{Lebiedowicz:2018eui},
$p p \to p p (\sigma \sigma, \rho^{0} \rho^{0} \to \pi^{+}\pi^{-}\pi^{+}\pi^{-})$ 
\cite{Lebiedowicz:2016zka},
$p p \to p p p \bar{p}$ \cite{Lebiedowicz:2018sdt},
$p p \to p p (\phi \phi \to K^{+}K^{-}K^{+}K^{-})$ 
\cite{Lebiedowicz:2019jru},
and $p p \to p p (\phi \to K^{+}K^{-}, \mu^{+}\mu^{-})$ 
\cite{Lebiedowicz:2019boz}. 
Azimuthal angle correlations between the outgoing protons 
can verify the $\Pom \Pom M$ couplings for scalar 
$f_{0}(980)$, $f_{0}(1370)$, $f_{0}(1500)$, $f_{0}(1710)$ 
and pseudoscalar $\eta$, $\eta'(958)$ mesons 
\cite{Lebiedowicz:2013ika,Lebiedowicz:2018eui}. 
The couplings, being of nonperturbative nature, are difficult 
to obtain from first principles of QCD. 
The corresponding coupling constants were fitted 
to differential distributions of the WA102 Collaboration
\cite{Barberis:1998ax,Barberis:1999cq,Barberis:1999zh} 
and to the results of \cite{Kirk:2000ws}.
As was shown in \cite{Lebiedowicz:2013ika},
the tensorial $\Pom \Pom f_{0}$, $\Pom \Pom \eta$, and $\Pom \Pom \eta'$ vertices
correspond to the sum of two lowest orbital angular momentum - spin couplings, 
except for the $f_{0}(1370)$ meson.
In the tensor-meson case there are seven possible 
$\Pom \Pom f_{2}(1270)$ couplings in principle;
see the list of possible $\Pom \Pom f_{2}$ couplings in Appendix~A of \cite{Lebiedowicz:2016ioh}.
In \cite{Lebiedowicz:2019por} a study of CEP of the $f_{2}(1270)$ meson was presented.
The $f_{2}(1270)$ is expected to be abundantly produced 
in the $p p \to p p \pi^+ \pi^-$ reaction,
and it was discussed in \cite{Lebiedowicz:2019por}
how to extract the $\Pom \Pom f_{2}(1270)$ coupling
from RHIC and LHC experimental results.
We refer the reader to 
\cite{Adamczyk:2014ofa,Aaltonen:2015uva,Khachatryan:2017xsi,
Sirunyan:2020cmr,Adam:2020sap}
for the latest measurements of central $\pi^+ \pi^-$ production 
in high-energy proton-(anti)proton collisions.
In \cite{Adam:2020sap} a study of CEP of $\pi^+ \pi^-$, $K^{+}K^{-}$, and $p \bar{p}$ pairs in $pp$ collisions 
at a center-of-mass energy of $\sqrt{s} = 200$~GeV 
by the STAR Collaboration at RHIC was reported.
For the first (preliminary) STAR experimental results 
measured at $\sqrt{s} = 510$~GeV see Ref.~\cite{ICHEP_STAR}.
There are ongoing studies of CEP 
of the $\pi^+ \pi^-\pi^+ \pi^-$ channel.

In this article we consider diffractive production of 
axial-vector $f_{1}$-type mesons 
in the $pp \to pp f_{1}$ reaction within the tensor-pomeron approach.
As concrete examples we shall consider 
CEP of the $f_{1}(1285)$ and the $f_{1}(1420)$
via the pomeron-pomeron-fusion mechanism.
We shall give a detailed discussion of various ways
to write the $\Pom \Pom f_{1}$ couplings.
In the calculations we include the absorptive corrections 
and show their role in describing the data measured 
by the WA102 Collaboration \cite{Barberis:1998by}.
We will try to analyse whether our study could shed light 
on the nonperturbative $\Pom \Pom f_{1}$ couplings.
In the future the corresponding $\Pom \Pom f_{1}$ couplings
could be adjusted by comparison with precise experimental data 
from both RHIC and the LHC.

We also consider the $\Pom \Pom f_{1}$ couplings 
that follow from holographic models of QCD,
in particular the Sakai-Sugimoto model based on type IIA superstring theory \cite{Witten:1998zw}.
In the low energy regime this model is a gravitational dual to large-$N_c$ QCD, where glueballs are described by fluctuations of a confining geometry \cite{Brower:2000rp,Brunner:2015oqa,Brunner:2015yha,Brunner:2018wbv,Leutgeb:2019lqu}, 
and the pomeron can be represented by reggeization of the tensor glueball \cite{Domokos:2009hm}.
Quark degrees of freedom are introduced as probe branes in this background and their gauge field fluctuations are dual to mesons \cite{Sakai:2004cn,Sakai:2005yt}. In \cite{Anderson:2014jia} the $\Pom \Pom \eta_{0}$ couplings were derived from the bulk Chern-Simons term, which is uniquely fixed by requiring consistency of supergravity and the gravitational anomaly. Because of its universal form, the structure of the resulting couplings should be the same in all holographic models, although the strength of the couplings may vary.\footnote{The same bulk Chern-Simons action also accounts for the anomalous coupling of pseudoscalar and axial-vector mesons to photons and was used in recent studies \cite{Leutgeb:2019zpq,Leutgeb:2019gbz,Cappiello:2019hwh} for calculating
hadronic light-by-light scattering contributions to the anomalous magnetic moment of the muon in holographic QCD.} 
In a similar calculation as was done in \cite{Anderson:2014jia}, we derive the $\Pom \Pom f_{1}$ couplings relevant for this study.

The four-pion channel, 
discussed in the past by the WA91 \cite{Antinori:1995wz}
and WA102 \cite{Barberis:1997ve,Barberis:1999wn} Collaborations,
seems to be a good candidate for an $f_{1}(1285)$ study 
in high-energy $pp$ collisions.
The intermediate states that should be considered are
the $J^{P} = 1^{+}$ states $\rho^{0} \rho^{0}$ 
and $\rho^{0} (\pi^{+} \pi^{-})_{\rm P \; wave}$.
The central $\pi^{+}\pi^{-}\pi^{+}\pi^{-}$ system
in proton-proton collisions
was measured also by the ABCDHW Collaboration 
at $\sqrt{s} = 63$~GeV at the CERN Intersecting Storage Rings (ISR); 
see Ref.~\cite{Breakstone:1993ku}.
A spin-parity decomposition of the $4 \pi$, $\rho \pi \pi$,
and $\rho \rho$ states as a function of $M_{4 \pi}$
was performed with the assumption that the dominant contributions 
arise from $J^{P} = 0^{+}$ and $2^{+}$ states.
Five contributions to the four-pion spectrum were identified: 
a $4 \pi$ phase-space term with total angular momentum $J = 0$,
two $\rho \pi \pi$ terms with $J = 0$ and $J = 2$,
and two $\rho \rho$ terms ($J = 0, 2$).
Thus, an enhancement observed in the region $M_{4 \pi} \sim 1300$~MeV 
for the $J^{P} = 2^{+}$ $\rho \rho$ and $\rho \pi \pi$ terms
was assigned to the $f_{2}(1270)$ meson
and for the $J^{P} = 0^{+}$ $\rho \pi \pi$ term
to the $f_{0}(1370)$ meson [called $f_{0}(1400)$ in \cite{Breakstone:1993ku}].
However, the $J^{P} = 1^{+}$ and $J^{P} = 0^{-}$ terms, 
possible in this process (e.g., via $\Pom \Pom$ fusion),
were not considered in the spin-parity analysis.
This may invalidate the final conclusions of \cite{Breakstone:1993ku}
where the enhancement in the four-pion invariant mass region
around 1300~MeV is attributed solely to the $f_{2}(1270)$
and the $f_{0}(1400)$ with $J^{P} = 2^{+}$ and $0^{+}$,
respectively.
There is also a clear experimental contradiction to these conclusions
from \cite{Breakstone:1993ku}, 
because the $f_{1}(1285)$ meson was seen in CEP 
in the four-pion channel; 
see~\cite{Barberis:1997ve,Barberis:1998by,Barberis:1999wn}.

At high energies the $\Pom \Pom$ fusion process 
is expected to be dominant.
For the relatively low center-of-mass energies 
of the WA102 and ISR experiments the secondary exchanges
may play an important role;
see, e.g., \cite{Lebiedowicz:2013ika, Lebiedowicz:2019boz}.
That is, at low energies we should discuss $f_{1}$ production 
from $\omega_{\Reg}$-$\omega_{\Reg}$, $\rho_{\Reg}$-$\rho_{\Reg}$,
$\phi_{\Reg}$-$\phi_{\Reg}$,
$a_{2 \Reg}$-$a_{2 \Reg}$, $f_{2 \Reg}$-$f_{2 \Reg}$,
$f_{2 \Reg}$-$\Pom$, $\Pom$-$f_{2 \Reg}$ exchanges, 
in addition to the $\Pom$-$\Pom$ exchange;
see Appendix~\ref{sec:appendixD} for more detailed discussion.
Clearly, this would introduce many practically unknown parameters
in the calculations.
In this article, therefore, we shall restrict our discussions
to the $\Pom \Pom$-fusion term and
we shall try to understand 
the $pp \to pp f_{1}(1285)$ and $pp \to pp f_{1}(1420)$ reactions
by comparing our results with the WA102 experimental 
data from \cite{Barberis:1998by}.
Having fixed the parameters of the model in this way 
we will give predictions for the RHIC and LHC experiments.
Because of the possible influence of nonleading exchanges
at low energies, these predictions for cross sections 
at high energies should be viewed as an upper limit
and we try to account for this by emphasising that 
our predictions may be scaled down by a certain factor.

Some effort to measure central exclusive
four pion production at the energy $\sqrt{s} = 13$~TeV 
has been initiated by the ATLAS Collaboration; 
see, e.g., \cite{Sikora_poster,Bols:2288372}.
In Fig.~55 of \cite{Bols:2288372} a ``preliminary'' 
mass spectrum of the $\pi^{+}\pi^{-}\pi^{+}\pi^{-}$ system 
was shown.
Resonancelike structures around 1300~MeV and 1450~MeV
were seen there.
As shown in Fig.~56 of \cite{Bols:2288372}, there is a large contribution
to $4 \pi$ CEP via the intermediate $\rho \rho$ channel.
In general, a few low-mass resonances 
with different $J^{P}$ may contribute to this process, such as, 
the $1^{+}$ resonance $f_{1}(1285)$, the $2^{+}$~$f_{2}(1270)$, 
the $0^{+}$~$f_{0}(1370)$, the $0^{+}$~$f_{0}(1500)$, 
and the $0^{-}$~$\eta(1405)$.
Note that in \cite{Barberis:1999wn} the $f_{0}(1370)$ 
is found to decay dominantly to $\rho \rho$ while the $f_{0}(1500)$ 
is found to decay to $\rho \rho$ and $\sigma \sigma$.
To perform a full analysis we shall consider 
also the four-pion-continuum contributions 
discussed in Refs.~\cite{Lebiedowicz:2016zka, Kycia:2017iij}.

In Ref.~\cite{Osipov:2018iah} the decay process
$f_{1}(1285) \to \rho^{0} \pi^{+} \pi^{-}$ was analysed
in the framework of the Nambu--Jona-Lasinio model.
The effective $f_{1} \rho^{0} \rho^{0}$ vertex,
in the case when one of the vector particles is off-mass shell,
was obtained from an anomalous 
(triangle quark $f_{1} \rho^{0} \gamma$ anomaly)
$f_{1} \rho^{0} \gamma$ vertex \cite{Osipov:2017ray}.
It was found in \cite{Osipov:2018iah} that the two $\rho^{0}$-meson channel
$f_{1} \to \rho^{0} \rho^{0} \to \rho^{0} \pi^{+} \pi^{-}$
gives a smaller contribution 
than the axial-vector $a_{1}^{\pm}(1260)$-meson plus pion channel 
$f_{1} \to \pi^{\pm} a_{1}^{\mp} \to \pi^{\pm} \pi^{\mp} \rho^{0}$.
There is a large interference between the above triangle-anomaly
contributions
and the direct decay which is described by the quark box diagram.
It would be useful to measure experimentally the rate of both
the $\rho^{0} \rho^{0}$ and $\rho^{0} \pi^{+} \pi^{-}$ decay modes
in order to further clarify the situation.

An interesting proposal was discussed recently 
in \cite{Achasov:2018mzt,Achasov:2019vcs}:
to study the anomalous isospin breaking decay
$f_{1}(1285) \to \pi^{+}\pi^{-}\pi^{0}$ in CEP of the $f_{1}$.

Our paper is organised as follows. 
In Sec.~\ref{sec:pp_ppf1} we discuss the formalism behind 
the axial-vector meson production process in the tensor-pomeron approach.
Section~\ref{sec:results} contains the comparison of our
results for the $pp \to pp f_{1}(1285)$ and $pp \to pp f_{1}(1420)$ reactions
with the WA102 experimental data \cite{Barberis:1998by}.
We discuss the related theoretical uncertainties.
Then we turn to high energies and show numerical results
for total and differential cross sections calculated 
for the RHIC and LHC experiments.
We compare the cross sections for the processes $pp \to pp f_{1}(1285)$
and $pp \to pp f_{2}(1270)$ with both
$f_{1}$ and $f_{2}$ decaying 
to the $\pi^{+}\pi^{-}\pi^{+}\pi^{-}$ final state.
The main results of our study are summarised in Sec.~\ref{sec:conclusions}.
The details on the coupling of an $f_{1}$ meson to two pomerons
are given in Appendices~\ref{sec:appendixA} and \ref{sec:appendixB}.
In Appendix~\ref{sec:appendixC} we consider
the $f_1$ mixing angle and possible relations between
the $\Pom \Pom f_{1}(1285)$ and $\Pom \Pom f_{1}(1420)$ coupling constants. 
In Appendix~\ref{sec:appendixD} we discuss subleading reggeon exchanges.
In Appendix~\ref{sec:appendixE} we discuss general properties 
of the $\phi_{pp}$ azimuthal angular distributions
for CEP of $f_{1}$- and $\eta$-type mesons
which can be used to disentangle their contributions
as an addition to good mass measurements and partial wave analyses.

\section{Formalism}
\label{sec:pp_ppf1}

We study central exclusive production of $f_{1}$
in proton-proton collisions 
\begin{eqnarray}
p(p_{a},\lambda_{a}) + p(p_{b},\lambda_{b}) \to
p(p_{1},\lambda_{1}) + f_{1}(k ,\lambda) + p(p_{2},\lambda_{2}) \,,
\label{2to3_reaction}
\end{eqnarray}
where $p_{a,b}$, $p_{1,2}$ and $\lambda_{a,b}$, 
$\lambda_{1,2} = \pm \frac{1}{2}$
denote the four-momenta and helicities of the protons, 
and $k$ and $\lambda = 0, \pm 1$ 
denote the four-momentum and helicity of the $f_{1}$ meson, respectively.
Here $f_{1}$ stands for one of the pseudovector mesons with $J^{PC} = 1^{++}$,
i.e. $f_{1}(1285)$ or $f_{1}(1420)$.

In this section we shall take into account only the main process, 
the $\Pom \Pom$-fusion mechanism, 
shown at the Born level by the diagram in Fig.~\ref{fig:diagram}.
We neglect here the reggeon (e.g., $f_{2 \Reg}$) exchanges which
we discuss briefly in Appendix~\ref{sec:appendixD}.
\begin{figure}[!ht]
\includegraphics[width=6.cm]{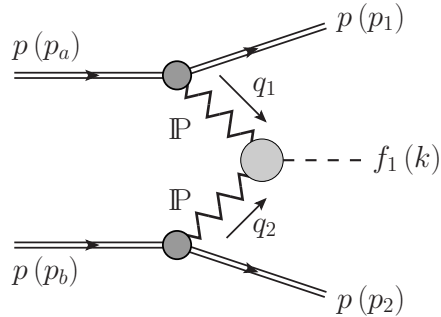}
\caption{\label{fig:diagram}
\small
The Born-level diagram for the $\Pom \Pom$-fusion mechanism 
for central exclusive diffractive production 
of an $f_{1}$-type meson in proton-proton collisions.}
\end{figure}

The kinematic variables for the reaction (\ref{2to3_reaction}) are
\begin{eqnarray}
&&q_1 = p_{a} - p_{1}, \quad q_2 = p_{b} - p_{2}, \quad k = q_{1} + q_{2}, \nonumber \\
&&t_1 = q_{1}^{2}, \quad t_2 = q_{2}^{2}, \quad m_{f_{1}}^{2} = k^{2}, \nonumber \\
&&s = (p_{a} + p_{b})^{2} = (p_{1} + p_{2} + k)^{2}, \nonumber \\
&&    s_{1} = (p_{a} + q_{2})^{2} = (p_{1} + k)^{2}, \nonumber \\
&&    s_{2} = (p_{b} + q_{1})^{2} = (p_{2} + k)^{2}\,.
\label{2to3_kinematic}
\end{eqnarray}
For the kinematics see e.g. Appendix~D of \cite{Lebiedowicz:2013ika}.

The amplitude for the reaction (\ref{2to3_reaction})
can be written as
\begin{eqnarray}
{\cal M}_{\lambda_{a} \lambda_{b} \to \lambda_{1} \lambda_{2} \lambda}
= (\epsilon_{\mu}(\lambda))^{*}\,
  {\cal M}^{\mu}_{\lambda_{a} \lambda_{b} \to \lambda_{1} \lambda_{2} f_{1}}\,,
\label{amplitude_2to3_1}
\end{eqnarray}
where $\epsilon_{\mu}(\lambda)$ is the polarisation vector of the $f_{1}$ meson.

The Born-level $\Pom\Pom$-fusion amplitude 
for exclusive production of an axial-vector meson $f_{1}$
can be written as
\begin{eqnarray}
{\cal M}^{(\Pom \Pom \to f_{1})}_{\mu, \,\lambda_{a} \lambda_{b} \to \lambda_{1} \lambda_{2} f_{1}}
&=& (-i)\,
\bar{u}(p_{1}, \lambda_{1}) 
i\Gamma^{(\Pom pp)}_{\mu_{1} \nu_{1}}(p_{1},p_{a}) 
u(p_{a}, \lambda_{a}) \nonumber \\
&& \times 
i\Delta^{(\Pom)\, \mu_{1} \nu_{1}, \alpha_{1} \beta_{1}}(s_{1},t_{1}) \,
i\Gamma^{(\Pom \Pom f_{1})}_{\alpha_{1} \beta_{1}, \alpha_{2} \beta_{2}, \mu}(q_{1},q_{2}) \,
i\Delta^{(\Pom)\, \alpha_{2} \beta_{2}, \mu_{2} \nu_{2}}(s_{2},t_{2}) \nonumber \\
&& \times 
\bar{u}(p_{2}, \lambda_{2}) 
i\Gamma^{(\Pom pp)}_{\mu_{2} \nu_{2}}(p_{2},p_{b}) 
u(p_{b}, \lambda_{b}) \,.
\label{amplitude_f1_pompom}
\end{eqnarray}
Here $\Delta^{(\Pom)}$ and $\Gamma^{(\Pom pp)}$ 
denote the effective propagator and proton vertex function, respectively, 
for the tensor-pomeron exchange.
The corresponding expressions, given in Sec.~3 of \cite{Ewerz:2013kda}, 
are
\begin{eqnarray}
&&i \Delta^{(\Pom)}_{\mu \nu, \kappa \lambda}(s,t) = 
\frac{1}{4s} \left( g_{\mu \kappa} g_{\nu \lambda} 
                  + g_{\mu \lambda} g_{\nu \kappa}
                  - \frac{1}{2} g_{\mu \nu} g_{\kappa \lambda} \right)
(-i s \alpha'_{\Pom})^{\alpha_{\Pom}(t)-1}\,,
\label{add1}\\
&&i\Gamma_{\mu \nu}^{(\Pom pp)}(p',p)
=-i 3 \beta_{\Pom NN} F_{1}(t)
\left\lbrace 
\frac{1}{2} 
\left[ \gamma_{\mu}(p'+p)_{\nu} 
     + \gamma_{\nu}(p'+p)_{\mu} \right]
- \frac{1}{4} g_{\mu \nu} (\slash{p}' + \slash{p})
\right\rbrace, \qquad
\label{add2}
\end{eqnarray}
where $t = (p'-p)^{2}$ and $\beta_{\Pom NN} = 1.87$~GeV$^{-1}$.
For simplicity we use for the pomeron-proton coupling 
the electromagnetic Dirac form factor $F_{1}(t)$ of the proton;
see also Chapter 3.2 of \cite{Donnachie:2002en}.
The pomeron trajectory $\alpha_{\Pom}(t)$
is assumed to be of standard linear form 
(see, e.g., \cite{Donnachie:1992ny,Donnachie:2002en}),
\begin{eqnarray}
&&\alpha_{\Pom}(t) = \alpha_{\Pom}(0)+\alpha'_{\Pom}\,t\,,\label{pomtrajectory}\\ 
&&\alpha_{\Pom}(0) = 1.0808\,, \;\;
  \alpha'_{\Pom} = 0.25 \; \mathrm{GeV}^{-2}\,.
\label{pomtrajectoryparms}
\end{eqnarray}

The new and unknown main ingredient of the amplitude (\ref{amplitude_f1_pompom}) 
is the pomeron-pomeron-$f_{1}$ vertex $\Gamma^{(\Pom \Pom f_{1})}$
which we want to study in the present article.
In \cite{Lebiedowicz:2013ika,Lebiedowicz:2016ioh,Lebiedowicz:2018eui,Lebiedowicz:2016zka,
Lebiedowicz:2018sdt,Lebiedowicz:2019jru,Lebiedowicz:2019por} the following strategy 
for constructing pomeron-pomeron-meson ($\Pom \Pom M$) couplings was followed.
First, one looked at the possible couplings of two fictitious ``real'' pomerons
to the meson $M$. This was easily done using elementary angular-momentum algebra;
see Appendix~A of \cite{Lebiedowicz:2013ika}.
Then $\Pom \Pom M$ couplings were written down corresponding to
the allowed values of orbital angular momentum $l$ and total 
$\Pom \Pom$ spin $S$ for a given meson $M$ in question.
Finally these couplings were also used for the CEP reaction $pp \to p M p$.
We follow this strategy also for CEP of an $f_{1}$ meson.
Thus, we investigate first the fictitious reaction
\begin{eqnarray}
\Pom(t,\epsilon^{(1)}) + \Pom(t,\epsilon^{(2)}) \to f_{1}(k,\epsilon)\,,
\label{PomPom_to_f1}
\end{eqnarray}
where $\Pom$ are ``real pomerons'' of mass squared $t > 0$
and with polarisation tensors $\epsilon^{(1)}$ and $\epsilon^{(2)}$.

From the analysis of this type of reactions presented 
in Appendix~A of \cite{Lebiedowicz:2013ika}
we find that for the $f_{1}$ with $J^{P} = 1^{+}$ there are two independent amplitudes
for the reaction (\ref{PomPom_to_f1}), labelled by 
$(l,S) = (2,2)$ and $(4,4)$.
Convenient covariant couplings leading to these amplitudes are easily constructed;
see (\ref{A6}) and (\ref{A20}) in Appendix~\ref{sec:appendixA}.
But these constructions are not unique.
In the Sakai-Sugimoto model \cite{Sakai:2004cn,Sakai:2005yt}
the coupling of an $I^{G} = 0^{+}$,
$J^{P} = 1^{+}$ axial-vector meson to two tensor glueballs is determined by
the gravitational Chern-Simons (CS) action describing axial-gravitational anomalies;
see (59) of \cite{Anderson:2014jia}.
Identifying the tensor glueballs with the fictitious ``real pomerons'' of (\ref{PomPom_to_f1})
we have derived corresponding bare coupling Lagrangians $\Pom \Pom f_{1}$
in (\ref{A27}) and (\ref{A28}) of Appendix~\ref{sec:appendixB}.

For the fictitious on-shell process (\ref{PomPom_to_f1})
the sum of the Lagrangians of (\ref{A6}) and (\ref{A20}) is strictly
equivalent to the sum of (\ref{A27}) and (\ref{A28}).
The relation of the respective coupling constants is given in (\ref{A75}).
But for the realistic case where the pomerons have invariant masses $t_{1,2} < 0$
and in general $t_{1} \neq t_{2}$ this equivalence no longer holds.
But we can expect that for small values $|t_{1}|, |t_{2}| \lesssim 0.5$~GeV$^{2}$
the off-shell effects should not be drastic.
And this, indeed, is confirmed by the explicit study presented in Appendix~\ref{sec:appendixB}.

In the following we shall present the formulas using the couplings
(\ref{A6}) and (\ref{A20}) of Appendix~\ref{sec:appendixA}.
The formulas using the couplings (\ref{A27}) and (\ref{A28})
of Appendix~\ref{sec:appendixB} are completely analogues.
Results will be shown for both types of couplings.

From the coupling Lagrangians of Appendix~\ref{sec:appendixA}
we obtain the following $\Pom \Pom f_{1}$ vertex:
\begin{eqnarray}
i\Gamma_{\kappa \lambda,\rho \sigma,\alpha}^{(\Pom \Pom f_{1})}(q_{1},q_{2}) =
\left( i\Gamma_{\kappa \lambda,\rho \sigma,\alpha}'^{(\Pom \Pom f_{1})}(q_{1},q_{2})\mid_{\rm{bare}}
      +i\Gamma_{\kappa \lambda,\rho \sigma,\alpha}''^{(\Pom \Pom f_{1})}(q_{1}, q_{2})\mid_{\rm{bare}} \right)
\tilde{F}_{\Pom \Pom f_{1}}(q_{1}^{2},q_{2}^{2},k^{2}) \,.\qquad
\label{vertex_pompomf1}
\end{eqnarray}
The $\Gamma'$ and $\Gamma''$ vertices in (\ref{vertex_pompomf1})
correspond to $(l,S) = (2,2)$ and $(4,4)$, respectively,
as derived from the corresponding coupling Lagrangians
(\ref{A6}) and (\ref{A20}) in Appendix~\ref{sec:appendixA}.
The expressions for these $\Pom \Pom f_{1}$ vertices~\footnote{Here 
the label ``bare'' is used for a vertex as derived 
from a corresponding coupling Lagrangian without a form-factor function.}
are as follows:
\newline
\hspace*{0.65cm}\includegraphics[width=125pt]{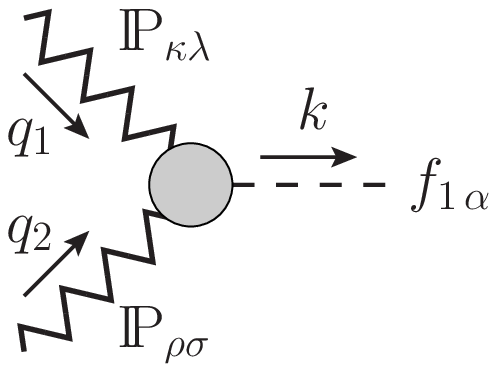}
\begin{eqnarray}
&&i\Gamma_{\kappa \lambda,\rho \sigma,\alpha}'^{(\Pom \Pom f_{1})}(q_{1},q_{2})\mid_{\rm{bare}} =
- \frac{g'_{\Pom \Pom f_{1}}}{8\, M_{0}^{2}}
(q_{1}-q_{2})^{\mu}
(q_{1}-q_{2})^{\nu}
k^{\beta}\,
\Gamma_{\kappa \lambda, \rho \sigma, \mu \nu, \alpha \beta}^{(8)}\,,
\label{vertex_pompomf1_A}\\
&&i\Gamma_{\kappa \lambda,\rho \sigma,\alpha}''^{(\Pom \Pom f_{1})}(q_{1},q_{2})\mid_{\rm{bare}} =
\frac{g''_{\Pom \Pom f_{1}}}{4 \,M_{0}^{4}}
(q_{1}-q_{2})^{\mu_{1}}
(q_{1}-q_{2})^{\mu_{2}}
(q_{1}-q_{2})^{\mu_{3}}
(q_{1}-q_{2})^{\mu_{4}}
k^{\beta} \nonumber \\
&&\qquad \qquad \times \Bigl[ ( g_{\kappa \mu_{1}}g_{\lambda \mu_{2}}
       - \frac{1}{4} g_{\kappa \lambda}g_{\mu_{1} \mu_{2}} )
       ( g_{\rho \mu_{3}} \varepsilon_{\sigma \mu_{4} \alpha \beta}
       +g_{\sigma \mu_{3}} \varepsilon_{\rho \mu_{4} \alpha \beta} )
+ (\kappa, \lambda) \leftrightarrow (\rho, \sigma) \Bigr].\qquad 
\label{vertex_pompomf1_B}
\end{eqnarray}
In (\ref{vertex_pompomf1_A}) and (\ref{vertex_pompomf1_B})
$M_{0} \equiv 1$~GeV,
$k = q_{1} + q_{2}$, $\Gamma^{(8)}$ is defined in (\ref{A2}),
and $g'_{\Pom \Pom f_{1}}$, $g''_{\Pom \Pom f_{1}}$
are dimensionless coupling constants.
The values of these coupling constants are not known 
and are not easy to obtain from first principles
of QCD, as they are of nonperturbative origin.
At the present stage the coupling constants
$g_{\Pom \Pom f_{1}}'$ and $g_{\Pom \Pom f_{1}}''$ 
should be fitted to experimental data.

For realistic applications we should multiply the ``bare'' vertices
(\ref{vertex_pompomf1_A}) and (\ref{vertex_pompomf1_B})
by a form factor $\tilde{F}^{(\Pom \Pom f_{1})}$
which we take in the factorised ansatz
as~\footnote{We are taking in (\ref{vertex_pompomf1}) the same form factor 
for each vertex $\Gamma'$ and $\Gamma''$.
In principle, we could take different form factors for each
of the vertices.}
\begin{eqnarray}
\tilde{F}^{(\Pom \Pom f_{1})}(q_{1}^{2},q_{2}^{2},k^{2}) = 
F_{M}(q_{1}^{2}) F_{M}(q_{2}^{2}) F^{(\Pom \Pom f_{1})}(k^{2})\,.
\label{Fpompommeson_FM}
\end{eqnarray}
For the on-shell meson we have $F^{(\Pom \Pom f_{1})}(m_{f_{1}}^{2}) = 1$.
In (\ref{Fpompommeson_FM}) we use
\begin{eqnarray}
F_{M}(t)=\frac{1}{1-t/\Lambda_{0}^{2}}\,,
\label{FM_t}
\end{eqnarray}
with $\Lambda_{0}^{2} = 0.5$~GeV$^{2}$;
see (3.34) of \cite{Ewerz:2013kda}
and (3.22) in Chapter 3.2 of \cite{Donnachie:2002en}.
Alternatively, we use the exponential form given as
\begin{eqnarray}
\tilde{F}^{(\Pom \Pom f_{1})}(t_{1},t_{2},m_{f_{1}}^{2}) = 
\exp\left( \frac{t_{1}+t_{2}}{\Lambda_{E}^{2}}\right) \,,
\label{Fpompommeson_exp}
\end{eqnarray}
where we have set $k^{2} = m_{f_{1}}^{2}$ and the cutoff constant $\Lambda_{E}$ 
should be adjusted to experimental data.

In the high-energy and small-angle approximation,
using (D.18) in Appendix~D of \cite{Lebiedowicz:2013ika},
the $\Pom\Pom$-fusion amplitude reads
\begin{eqnarray}
{\cal M}^{(\Pom \Pom \to f_{1})}_{\mu, \,\lambda_{a} \lambda_{b} \to \lambda_{1} \lambda_{2} f_{1}} 
&=& i\,
3 \beta_{\Pom NN} \, F_{1}(t_{1}) \,
(p_{1} + p_{a})^{\alpha_{1}}
(p_{1} + p_{a})^{\beta_{1}}\, \delta_{\lambda_{1}\lambda_{a}}\nonumber \\
&& \times \frac{1}{2 s_{1}} \left( -i s_{1} \alpha'_{\Pom} \right)^{\alpha_{\Pom}(t_{1})-1}
i\Gamma^{(\Pom \Pom f_{1})}_{\alpha_{1} \beta_{1}, \alpha_{2} \beta_{2}, \mu}(q_{1},q_{2})\, 
\frac{1}{2 s_{2}} \left( -i s_{2} \alpha'_{\Pom} \right)^{\alpha_{\Pom}(t_{2})-1}
\nonumber \\
&& \times 
3 \beta_{\Pom NN} \, F_{1}(t_{2})\,
(p_{2} + p_{b})^{\alpha_{2}}
(p_{2} + p_{b})^{\beta_{2}}\,\delta_{\lambda_{2}\lambda_{b}}
\,.
\label{amplitude_f1_pompom_he_limit}
\end{eqnarray}
For the $\Pom \Pom f_{1}$ vertex function we shall use
in the following the form (\ref{vertex_pompomf1}) 
with the bare vertices either from
(\ref{vertex_pompomf1_A}) and (\ref{vertex_pompomf1_B}) 
(corresponding to the couplings discussed in Appendix~\ref{sec:appendixA})
or those from (\ref{A40}) and (\ref{A66})
from Appendix~\ref{sec:appendixB}.

Note that the vertices (\ref{vertex_pompomf1_A}) and (\ref{vertex_pompomf1_B}) 
derived from the coupling Lagrangians (\ref{A6}) and (\ref{A20})
automatically are divergence free; i.e., they satisfy
\begin{eqnarray}
i\Gamma^{(\Pom \Pom f_{1})}_{\kappa \lambda,\rho \sigma,\alpha}(q_{1},q_{2}) \, (q_{1} + q_{2})^{\alpha} = 0\,.
\label{vertex_pompomf1_aux1}
\end{eqnarray}
For the vertices derived from (\ref{A27}) and (\ref{A28}) 
this does not hold.
Thus, in calculations of cross sections with the vertices
(\ref{A40}) and (\ref{A66}) one has to use 
for the $f_{1}$ spin sum
\begin{eqnarray}
-g_{\mu \nu} + \frac{k_{\mu}k_{\nu}}{k^{2}}\,,
\label{vertex_pompomf1_aux2}
\end{eqnarray}
since the $k_{\mu}k_{\nu}$ term will give a nonzero contribution.
With the vertices from (\ref{vertex_pompomf1_A}) and (\ref{vertex_pompomf1_B}) the $k_{\mu}k_{\nu}$ term does not contribute.

To give the full amplitude
for the reaction (\ref{2to3_reaction})
we should also include absorption effects
to the Born amplitude:
\begin{eqnarray}
{\cal {M}}_{pp \to pp f_{1}} =
{\cal {M}}_{pp \to pp f_{1}}^{\rm Born} + 
{\cal {M}}_{pp \to pp f_{1}}^{pp-\rm{rescattering}}\,.
\label{amp_full}
\end{eqnarray}

In our analysis we include the absorptive corrections
within the one-channel-eikonal approach.\footnote{We refer 
the reader to 
\cite{Schafer:2007mm,Cisek:2011vt,Cisek:2014ala,Lebiedowicz:2013vya}
for reviews of three-body processes and details concerning 
the absorptive corrections in the eikonal approximation
which takes into account the contribution of elastic $pp$ rescattering.
In Refs.~\cite{Lebiedowicz:2014bea,Lebiedowicz:2015eka}
the one-channel-eikonal approach was applied to four-body processes.}
For investigations of an eikonal model see, e.g., \cite{Gotsman:1998mm}.
The main result of \cite{Gotsman:1998mm} is that the absorption
effects become more important at higher energies;
that is, the survival probability of large rapidity gaps
decreases with increasing energy.
A two-channel eikonal model was discussed 
in \cite{Khoze:2000wk,Khoze:2002nf,Khoze:2013dha}.
A more sophisticated three-channel model was discussed in \cite{Gotsman:1999xq}.

The amplitude including the ``soft'' $pp$-rescattering corrections
which we use in the present paper can be written as
\begin{eqnarray}
{\cal M}_{pp \to pp f_{1}}^{pp-\rm{rescattering}}(s,\bpta,\bptb)=
\frac{i}{8 \pi^{2} s} \int d^{2}\bkt \,
{\cal M}_{pp \to pp}(s,-\bktsqrt)
{\cal M}_{pp\to pp f_{1}}^{\rm Born}
(s,\bptat,\bptbt)\,. \qquad \;\;
\label{abs_correction}
\end{eqnarray}
Here, in the overall center-of-mass (c.m.) system, $\bpta$ and $\bptb$
are the transverse components of the momenta of the outgoing protons
and $\bkt$ is the transverse momentum carried around the pomeron loop.
${\cal M}_{pp\to pp f_{1}}^{\rm Born}$
is the Born amplitude given by (\ref{amplitude_2to3_1}) and (\ref{amplitude_f1_pompom_he_limit})
with $\bptat = \bpta - \bkt$ and \mbox{$\bptbt = \bptb + \bkt$}.
${\cal M}_{pp \to pp}$
is the elastic $pp$ scattering amplitude given by (6.28)
in \cite{Ewerz:2013kda}
for large $s$ and with the momentum transfer $t=-\bktsqrt$.
In practice we work with the amplitudes in the high-energy approximation,
i.e. assuming $s$-channel helicity conservation
as it is realized in our model.

\section{Results}
\label{sec:results}
In this section we wish to present first results 
for the $pp \to pp f_{1}(1285)$ and $pp \to pp f_{1}(1420)$ reactions.
We will first discuss the $p p \to p p f_{1}$ reactions
at the relatively low c.m. energy $\sqrt{s} = 29.1$~GeV 
and compare our model results with the WA102 experimental data from \cite{Barberis:1998by}.
We shall try to fix the parameters of our model including at first 
only the $\Pom \Pom$-fusion mechanism.
Then we shall make predictions for the experiments at the 
RHIC and LHC.
The secondary reggeon exchanges should give small contributions at high energies and in the midrapidity region.
However, they may influence the absolute normalization 
of the cross section at low energies.
Therefore, our predictions for the RHIC and LHC experiments,
obtained in this way, should be regarded rather 
as an upper limit for the $pp \to pp f_{1}$ reactions,
but, as discussed in Appendix \ref{sec:appendixD}, 
we expect that they should overestimate the cross sections
by not more than a factor of 4.

\subsection{Comparison with the WA102 data}
\label{sec:comparison_WA102}

According to \cite{Barberis:1998by}
the WA102 experimental cross sections are 
as quoted in Table~\ref{tab:table1}.\footnote{Note that the cross sections 
for $f_{1}(1285)$ and $f_{1}(1420)$ mesons
quoted in Table~1 of \cite{Kirk:2000ws} 
correspond to $\sqrt{s} = 12.7$~GeV 
and not $\sqrt{s} = 29.1$~GeV as mentioned there.}
\begin{table}[]
\centering
\caption{Experimental results for total cross sections of $f_{1}$ mesons 
in $pp$ collisions measured by the WA102 Collaboration \cite{Barberis:1998by}.}
\label{tab:table1}
\begin{tabular}{|c|c|c|l|}
\hline
Meson & $\sqrt{s}$~(GeV) & Cuts & $\sigma_{\rm{exp.}}$~(nb)  \\ 
\hline
$f_{1}(1285)$ & 12.7 & $|x_{F,M}| \leqslant 0.2$ & $6857 \pm 1306$   \\ 
 & 29.1 & $|x_{F,M}| \leqslant 0.2$ & $6919 \pm 886$   \\ 
\hline
$f_{1}(1420)$ & 12.7 & $|x_{F,M}| \leqslant 0.2$ & $1080 \pm 385$   \\
 & 29.1 & $|x_{F,M}| \leqslant 0.2$ & $1584 \pm 145$   \\  
\hline
\end{tabular}
\end{table}
In~\cite{Barberis:1998by}~also the distributions in $|t|$ and $\phi_{pp}$
for the $f_{1}(1285)$ and $f_{1}(1420)$ meson production 
at $\sqrt{s} = 29.1$~GeV were presented.
Here, $t$ is the four-momentum transfer squared from one of the proton vertices
[we have $t = t_{1}$ or $t_{2}$; cf. (\ref{2to3_kinematic})], 
and $\phi_{pp}$ is the azimuthal angle between 
the transverse momentum vectors $\bpta$ and $\bptb$ of the outgoing protons
(see Fig.~\ref{fig:phi_pp} in Appendix~\ref{sec:appendixE}).

\begin{figure}[!ht]
\includegraphics[width=0.49\textwidth]{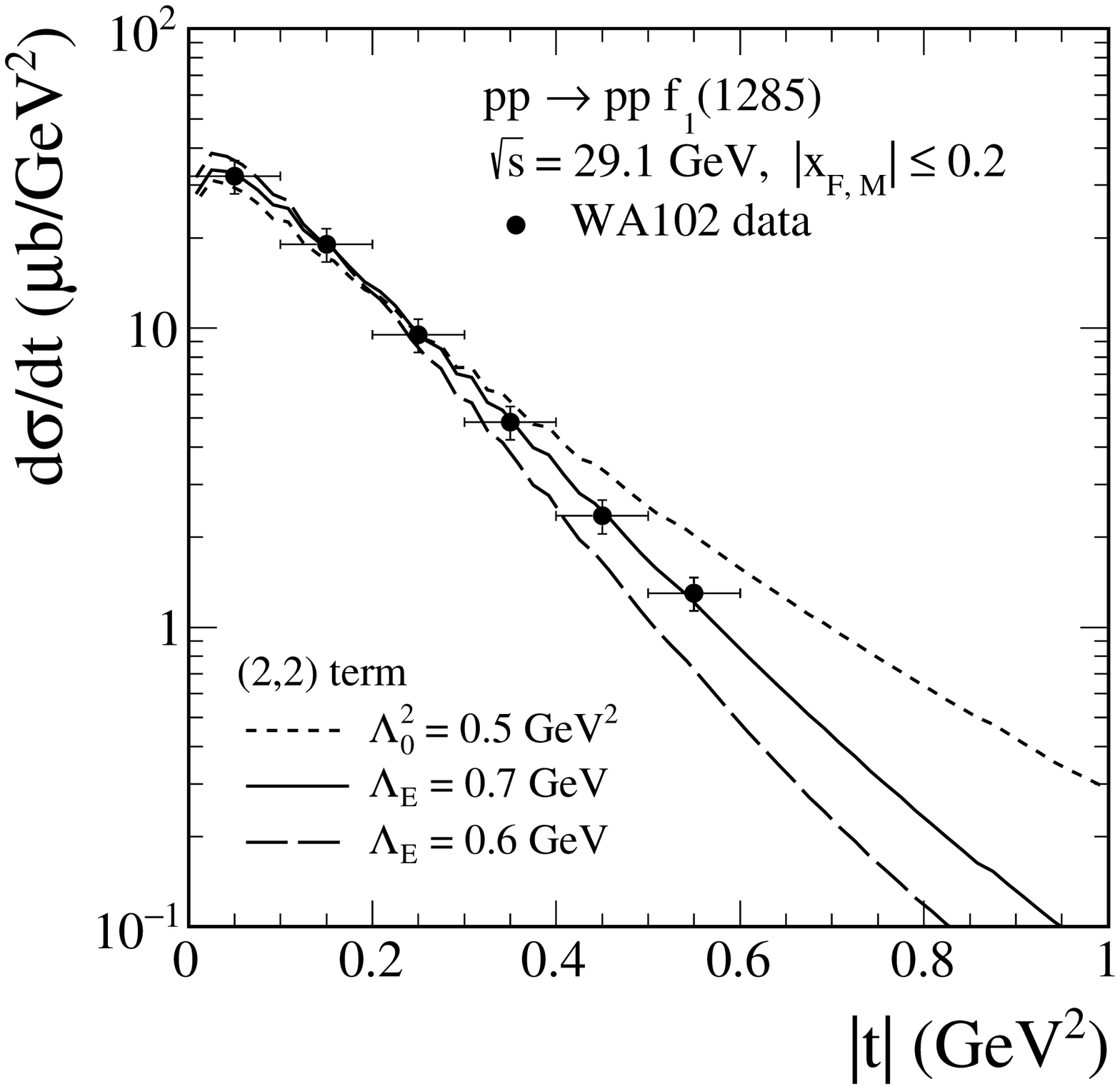}
\includegraphics[width=0.49\textwidth]{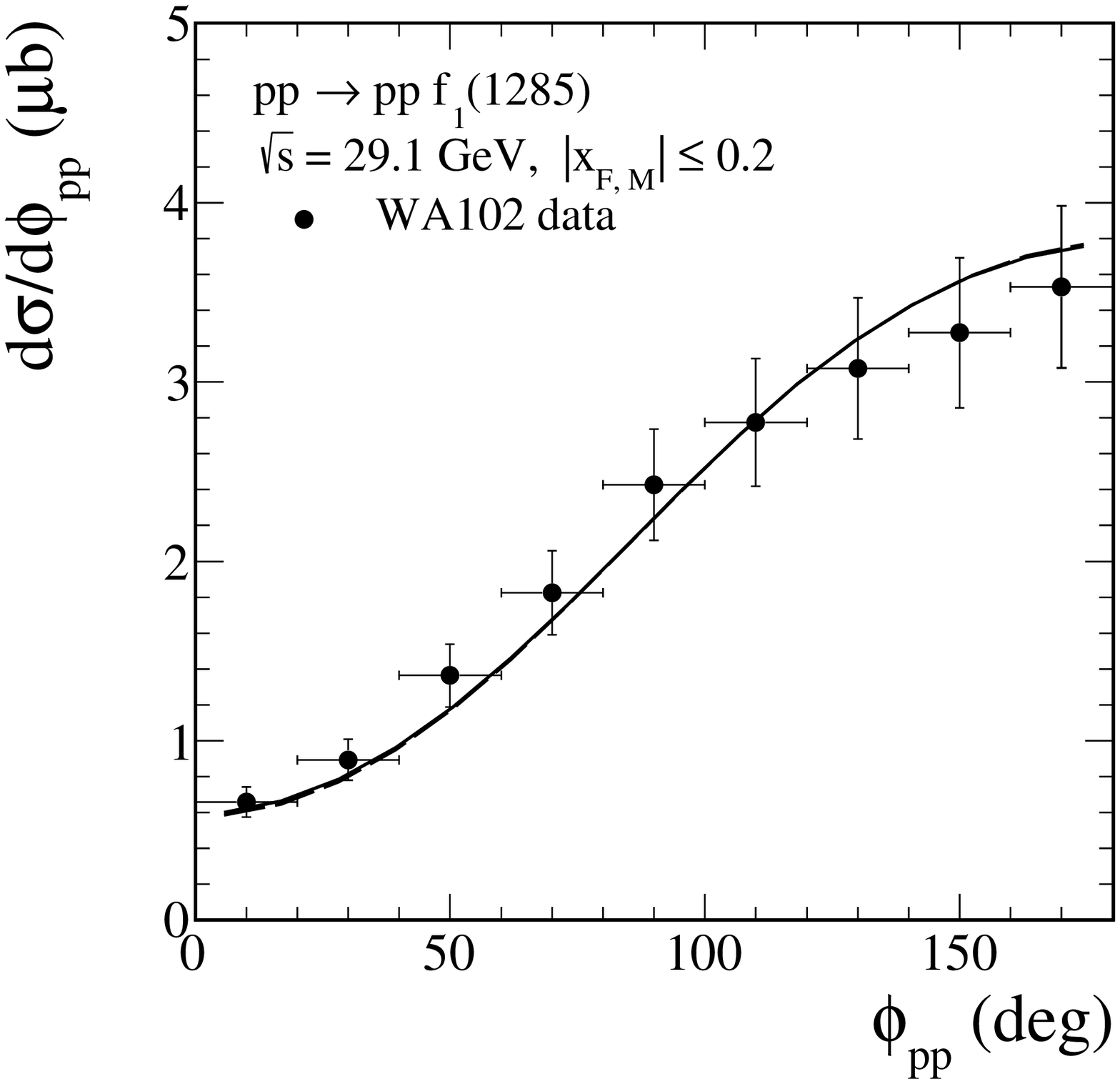}
\includegraphics[width=0.49\textwidth]{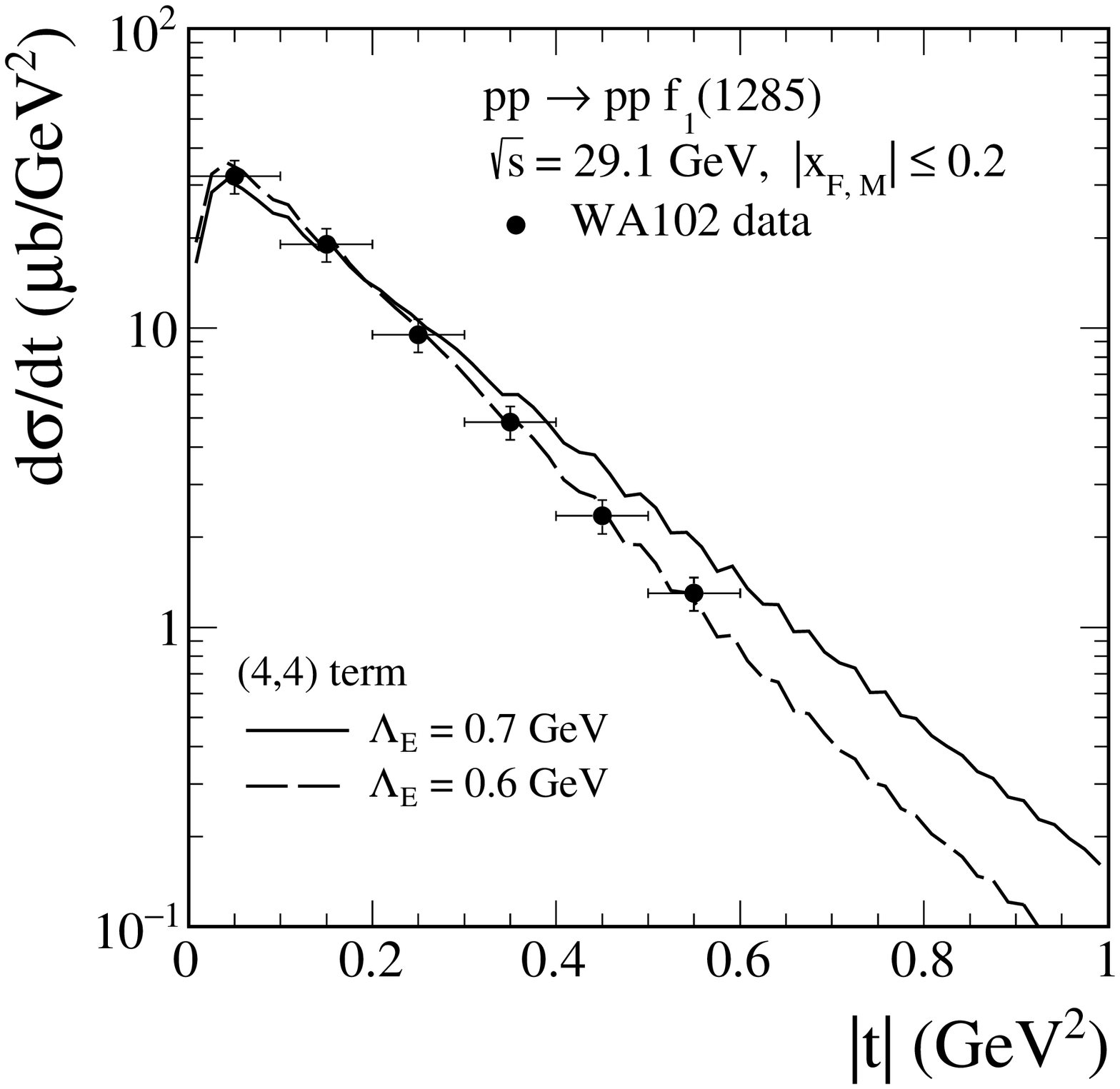}
\includegraphics[width=0.49\textwidth]{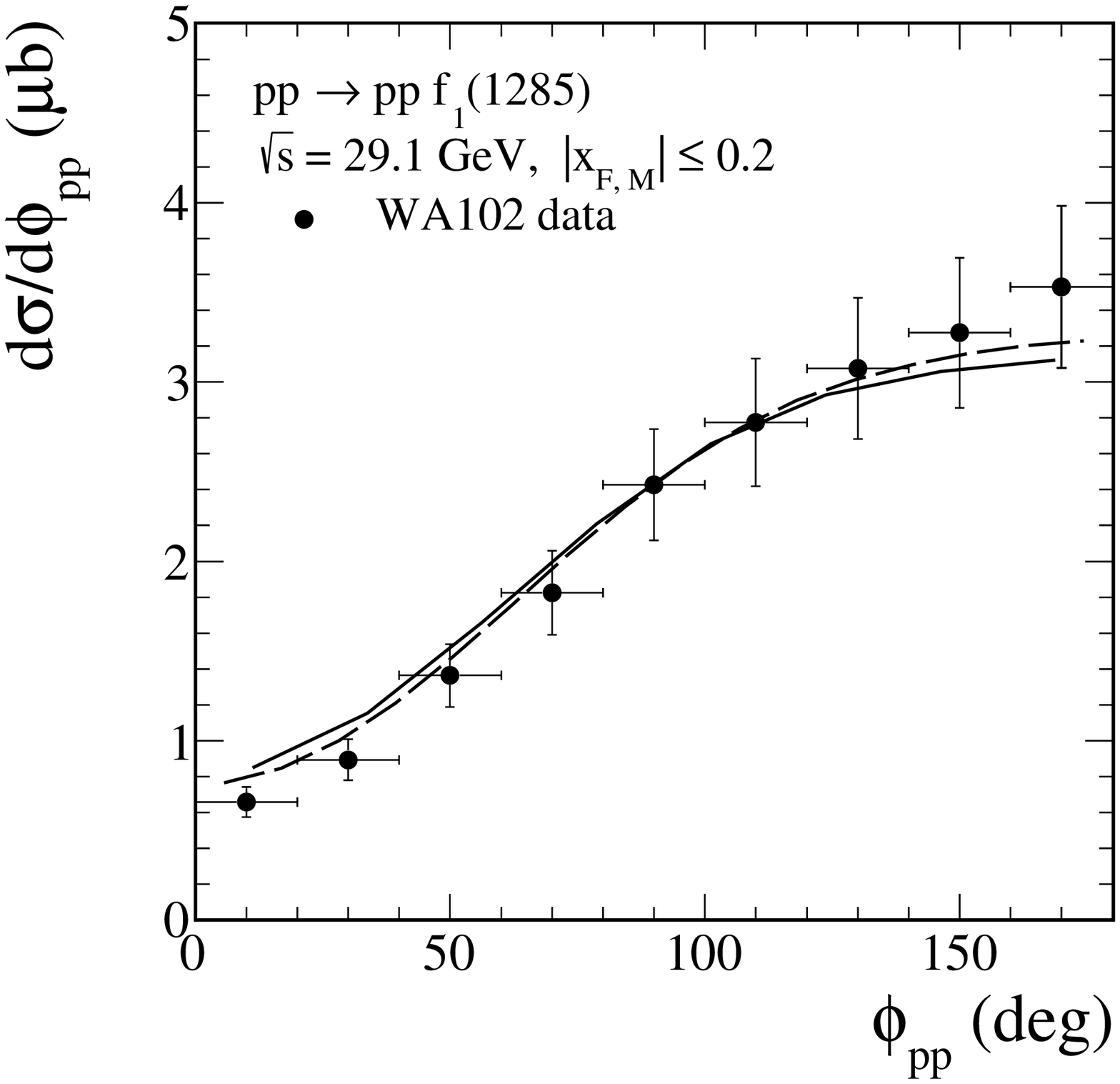}
\caption{\label{fig:1}
\small
The $|t|$ (left panels) and $\phi_{pp}$ (right panels) distributions 
for the $pp \to pp f_{1}(1285)$ reaction 
at $\sqrt{s} = 29.1$~GeV and $|x_{F,M}| \leqslant 0.2$.
The results have been normalised to the mean value
of the total cross section (\ref{WA102_f1_1285}) 
from \cite{Barberis:1998by}.
The error bars on the data correspond to 
the error on $\sigma_{\rm exp.}$ in (\ref{WA102_f1_1285}).
The separate individual contributions 
for the $(l,S) = (2,2)$
[see Eq.~(\ref{vertex_pompomf1_A})] (upper panels) and
$(l,S) = (4,4)$ [see Eq.~(\ref{vertex_pompomf1_B})] 
(lower panels) are presented.
We show results obtained 
with the exponential form factor (\ref{Fpompommeson_exp}) 
for $\Lambda_{E} = 0.7$~GeV (solid lines)
and for $\Lambda_{E} = 0.6$~GeV (long-dashed lines).
The dotted line in the top left panel is obtained
using (\ref{Fpompommeson_FM}) with (\ref{FM_t}).
The absorption effects are included in the calculations.
The oscillations in the left bottom panel are of numerical origin.}
\end{figure}

\begin{figure}[!ht]
\includegraphics[width=0.49\textwidth]{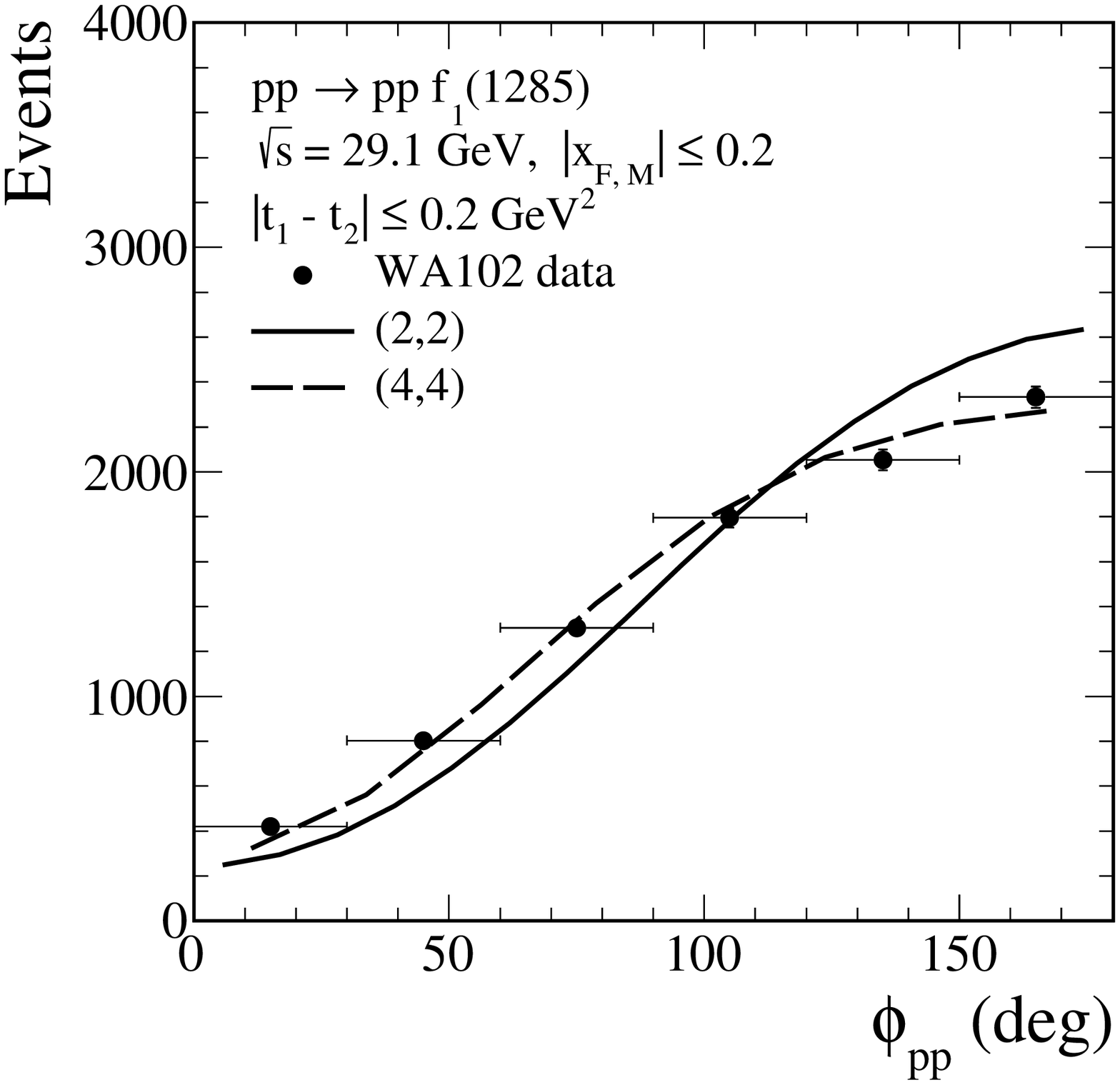}
\includegraphics[width=0.49\textwidth]{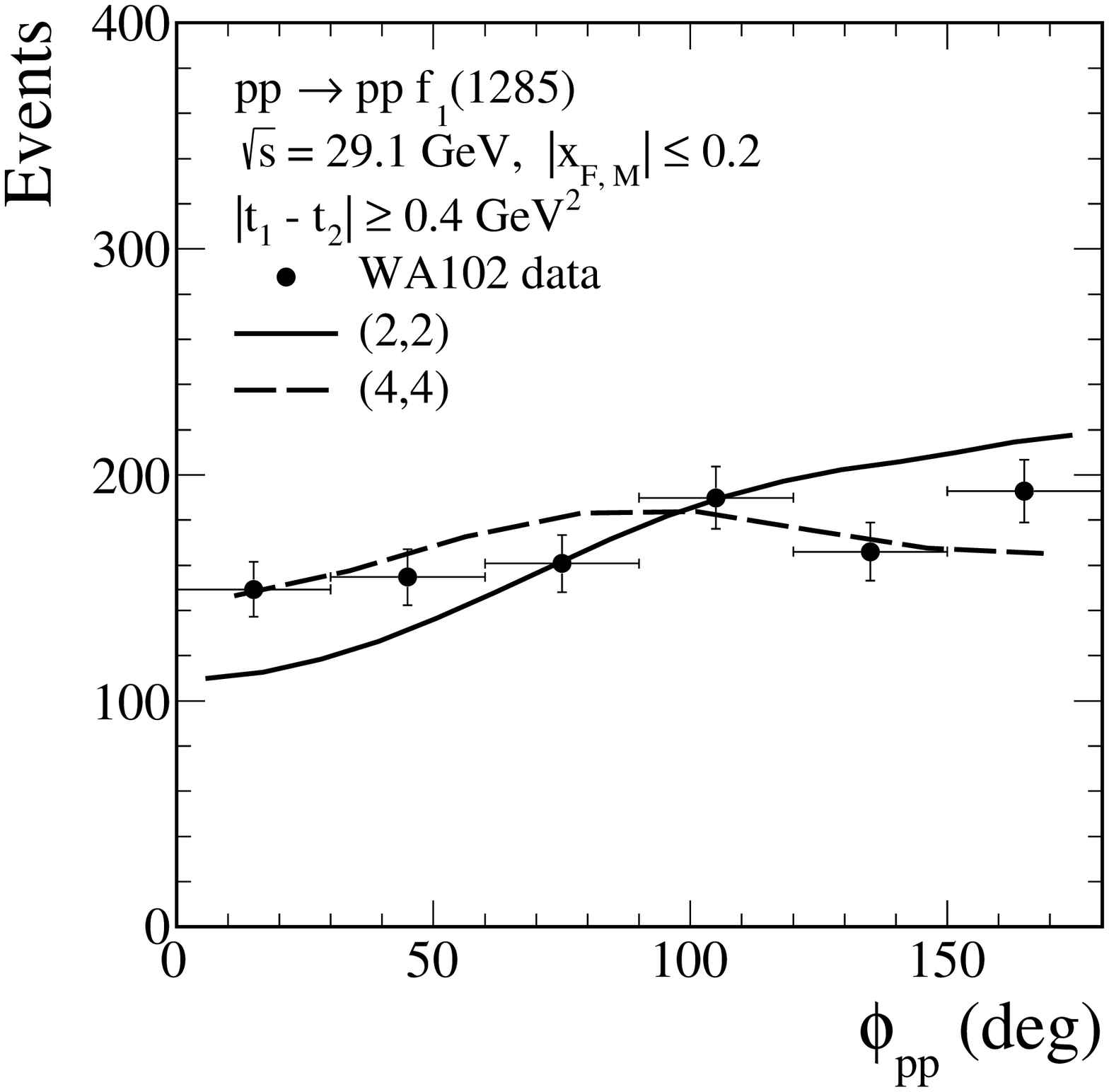}
\caption{\label{fig:2}
\small
The $\phi_{pp}$ distributions for $f_{1}(1285)$ meson production
at $\sqrt{s} = 29.1$~GeV, $|x_{F,M}| \leqslant 0.2$, and
for $|t_{1} - t_{2}| \leqslant 0.2$~GeV$^{2}$ (left panel)
and $|t_{1} - t_{2}| \geqslant 0.4$~GeV$^{2}$ (right panel).
The WA102 experimental data points are from Fig.~3 of \cite{Kirk:1999df}.
The theoretical results
have been normalised to the mean value of the number of events.
In the calculation we use here (\ref{Fpompommeson_exp}) 
with $\Lambda_{E} = 0.7$~GeV.
The absorption effects are included here.}
\end{figure}

Below we present three independent ways 
to fix the $\Pom \Pom f_{1}$ coupling parameters
in the $pp \to pp f_{1}(1285)$ reaction.
First we assume that only one of the couplings
$g'_{\Pom \Pom f_{1}}$ or $g''_{\Pom \Pom f_{1}}$ 
[$(l,S) = (2,2)$ term (\ref{vertex_pompomf1_A})
or $(l,S) = (4,4)$ term (\ref{vertex_pompomf1_B})] contributes,
and we make evaluations and comparisons with the WA102 experimental data;
see Figs.~\ref{fig:1}, \ref{fig:2} and Table~\ref{tab:ratio_dPt}.
Later we consider the combination of two terms,
the $\varkappa'$ and $\varkappa''$ couplings
calculated with the vertices (\ref{A40}) and (\ref{A66});
see Fig.~\ref{fig:2aux}.
We will also show to which values of
$g'_{\Pom \Pom f_{1}}$ and $g''_{\Pom \Pom f_{1}}$
the $(\varkappa',\varkappa'')$ values correspond.
Then we follow the analogous procedure to fix 
the $\Pom \Pom f_{1}(1420)$ couplings;
see Figs.~\ref{fig:1420_ff}, \ref{fig:1420_ff_R} and Table~\ref{tab:ratio_dPt}.

In Fig.~\ref{fig:1} we show the results for the $f_{1}(1285)$ 
meson production for $\sqrt{s} = 29.1$~GeV and 
for the Feynman variable of the meson 
$|x_{F,M}| \leqslant 0.2$.\footnote{The Feynman-$x$ variable 
is defined as $x_{F,M} = 2 p_{z,M}/\sqrt{s}$ 
with $p_{z,M}$ the longitudinal momentum 
of the outgoing meson in the center-of-mass frame.}
The WA102 data points from \cite{Barberis:1998by} 
and our model results
have been normalised to the mean value of the total cross section 
\begin{eqnarray}
\sigma_{\rm exp.} = (6919 \pm 886)\;\mathrm{nb}\,;
\label{WA102_f1_1285}
\end{eqnarray}
see Table~\ref{tab:table1}.
The experimental error of the total cross section 
is about 12.8\,\% (\ref{WA102_f1_1285}) and is dominated by systematic effects.
Correspondingly the error bars quoted in Fig.~\ref{fig:1}
are assumed to be 12.8\,\% of the cross section for each bin.

We show the results for different $\Pom \Pom f_{1}$ couplings 
discussed in the present paper.
The theoretical calculations in the top panels of Fig.~\ref{fig:1} 
correspond to the $(l,S) = (2,2)$ term (\ref{vertex_pompomf1_A})
while those in the bottom panels to the $(4,4)$ term (\ref{vertex_pompomf1_B}).
We can see from the left panels of Fig.~\ref{fig:1} 
that the $t$ dependence of $f_{1}$ production 
is very sensitive to the form factor 
$\tilde{F}^{(\Pom \Pom f_{1})}$ in the pomeron-pomeron-meson vertex.
The results with the exponential form (\ref{Fpompommeson_exp}) 
and $\Lambda_{E} = 0.7$~GeV
describe the $t$ dependence better than (\ref{Fpompommeson_FM}) with (\ref{FM_t}).
The calculations with (\ref{Fpompommeson_exp}) give a sizeable decrease of 
the cross section at large $|t|$.
Therefore, in the following we show the results calculated with (\ref{Fpompommeson_exp}).
At $t = 0$ (here $t = t_{1}$ or $t_{2}$) all contributions vanish.
Both the $(l,S) = (2,2)$ and $(4,4)$ couplings
considered separately
allow one to describe the WA102 differential distributions.

To get the mean value of the total cross section (\ref{WA102_f1_1285}) we find the following:
$g'_{\Pom \Pom f_{1}} = 4.89$ in (\ref{vertex_pompomf1_A}) for $\Lambda_{E} = 0.7$~GeV,  
$g'_{\Pom \Pom f_{1}} = 6.00$ for $\Lambda_{E} = 0.6$~GeV, 
$g''_{\Pom \Pom f_{1}} = 10.31$ in (\ref{vertex_pompomf1_B}) for $\Lambda_{E} = 0.7$~GeV,
$g''_{\Pom \Pom f_{1}} = 12.90$ for $\Lambda_{E} = 0.6$~GeV,
$\varkappa' = 8.58$ in (\ref{A40})
for $\Lambda_{E} = 0.7$~GeV,
and $\varkappa' = 7.40$ for $\Lambda_{E} = 0.8$~GeV.
Here we assumed the value of coupling constants to be positive as we employ them separately.

In \cite{Kirk:1999df} an interesting behaviour 
of the $\phi_{pp}$ distribution for $f_{1}(1285)$ meson production 
for two different values of $|t_{1} - t_{2}|$ was presented.
In Fig.~\ref{fig:2} we show the $\phi_{pp}$ distribution of events
from \cite{Kirk:1999df}
for $|t_{1} - t_{2}| \leqslant 0.2$~GeV$^{2}$ (left panel) and 
$|t_{1} - t_{2}| \geqslant 0.4$~GeV$^{2}$ (right panel).
Our model results have been normalised to the mean value of 
the number of events.
The results for $\Lambda_{E} = 0.7$~GeV 
in (\ref{Fpompommeson_exp}) are shown.
We have checked that for $\Lambda_{E} = 0.6$~GeV
the shape of the $\phi_{pp}$ distributions is almost the same.
An almost ``flat'' distribution at large values of $|t_{1} - t_{2}|$ can be observed.
It seems that the $(l,S) = (4,4)$ term best reproduces 
the shape of the WA102 data.
As we will show below in Fig.~\ref{fig:2abs},
the absorption effects play a significant role there.

Note that in \cite{Kirk:1999df} 
also the number of events for the $f_{1}(1285)$ meson
for the two kinematical conditions 
(a) $|t_{1} - t_{2}| \leqslant 0.2$~GeV$^{2}$ and
(b) $|t_{1} - t_{2}| \geqslant 0.4$~GeV$^{2}$ was given.
The experimental ratio is $R_{\rm exp.} = N_{a}/N_{b} \simeq 8.6$,
where $N_{a}$ and $N_{b}$ are the number of events 
from Figs.~3(a) and 3(b) of \cite{Kirk:1999df}, respectively.
Then, we define the ratio
\begin{eqnarray}
R = \frac{\sigma(|t_{1} - t_{2}| \leqslant 0.2\;{\rm GeV}^{2})}
         {\sigma(|t_{1} - t_{2}| \geqslant 0.4\;{\rm GeV}^{2})}\,.
\label{ratio_phi}
\end{eqnarray}
From our model using $\Lambda_{E} = 0.7$~GeV in (\ref{Fpompommeson_exp})
we get for the $(2,2)$ term (\ref{vertex_pompomf1_A}) 
the ratio $R = 8.6$,
while for the $(4,4)$ term (\ref{vertex_pompomf1_B}) we get $R = 5.6$.
If we use $\Lambda_{E} = 0.6$~GeV 
we get $R = 15.9$ 
and $R = 10.3$, respectively.
Therefore, for the $(2,2)$ term, $\Lambda_{E} = 0.7$~GeV is a good choice,
while for the $(4,4)$ term we should use a bit smaller value.
For the $\varkappa'$ term given by (\ref{A40}) 
and $\Lambda_{E} = 0.7$~GeV we get $R = 13.2$
while for $\Lambda_{E} = 0.8$~GeV we get $R = 8.8$.
For the $(\varkappa',\varkappa'')$ terms,
respectively for $\varkappa''/ \varkappa' = -(6.25, 3.76, 2.44, 1.0) \;\mathrm{GeV}^{-2}$ and $\Lambda_{E} = 0.7$~GeV
we get $R = (7.6, 10.5, 11.9, 13.2)$.

In Fig.~\ref{fig:2abs} we show the results for the $\phi_{pp}$ distributions
for different cuts on $|t_{1} - t_{2}|$ without and with the absorption
effects included in the calculations.
The results for the two $(l,S)$ couplings are shown.
The absorption effects lead to a large reduction of the cross section.
We obtain the ratio of full and Born cross sections,
the survival factor, as $\langle S^{2}\rangle = 0.5$--$0.7$.
Note that $\langle S^{2}\rangle$ depends on the kinematics.
We can see a large damping of the cross section 
in the region of $\phi_{pp} \sim \pi$,
especially for $|t_{1} - t_{2}| \geqslant 0.4$~GeV$^{2}$.
We notice that our results for the $(4,4)$ term 
have similar shapes as those presented in \cite{Petrov:2004hh} 
[see Figs.~3(c) and 3(d)]
where the authors also included the absorption corrections.
\begin{figure}[!ht]
\includegraphics[width=0.49\textwidth]{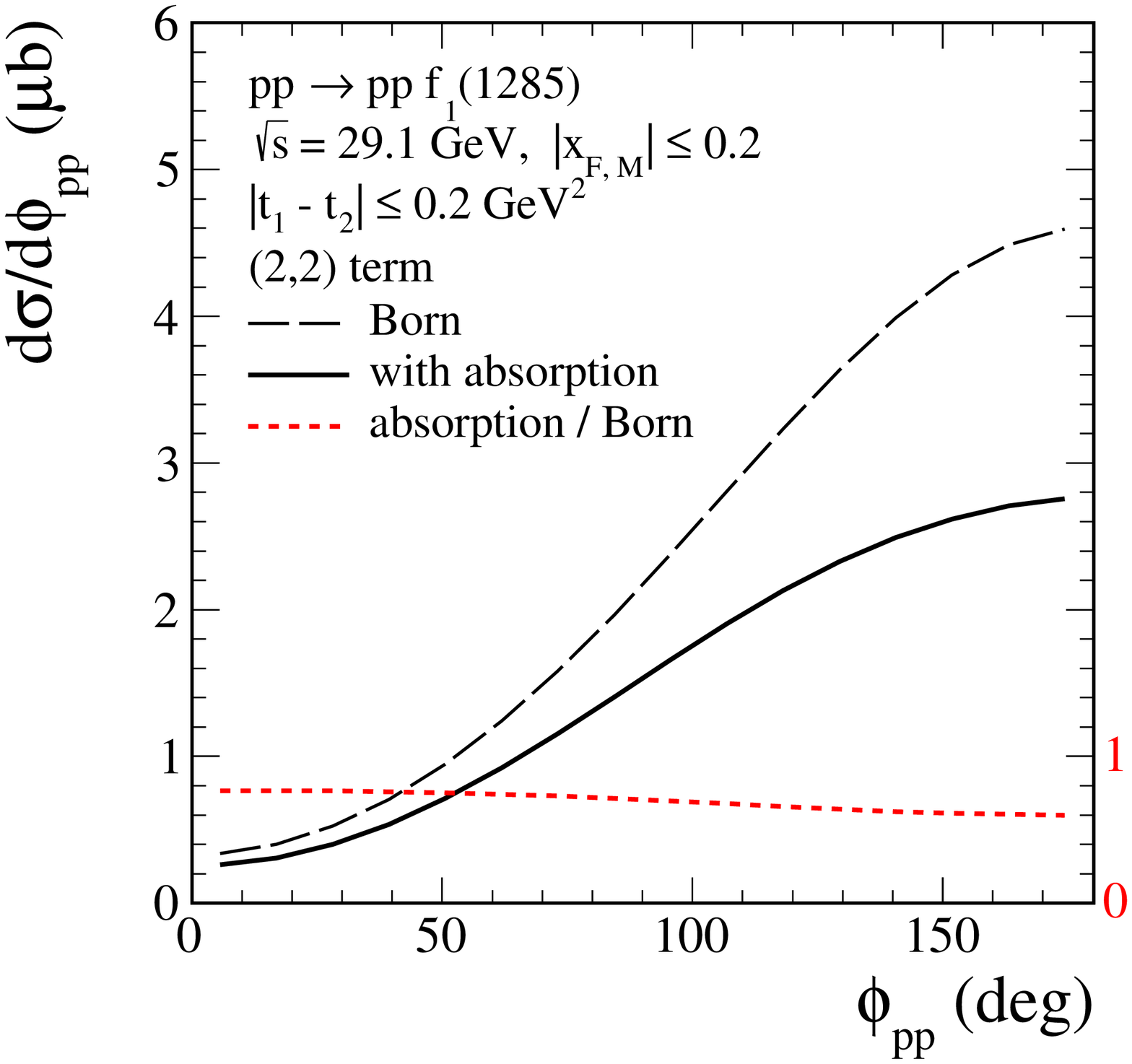}
\includegraphics[width=0.49\textwidth]{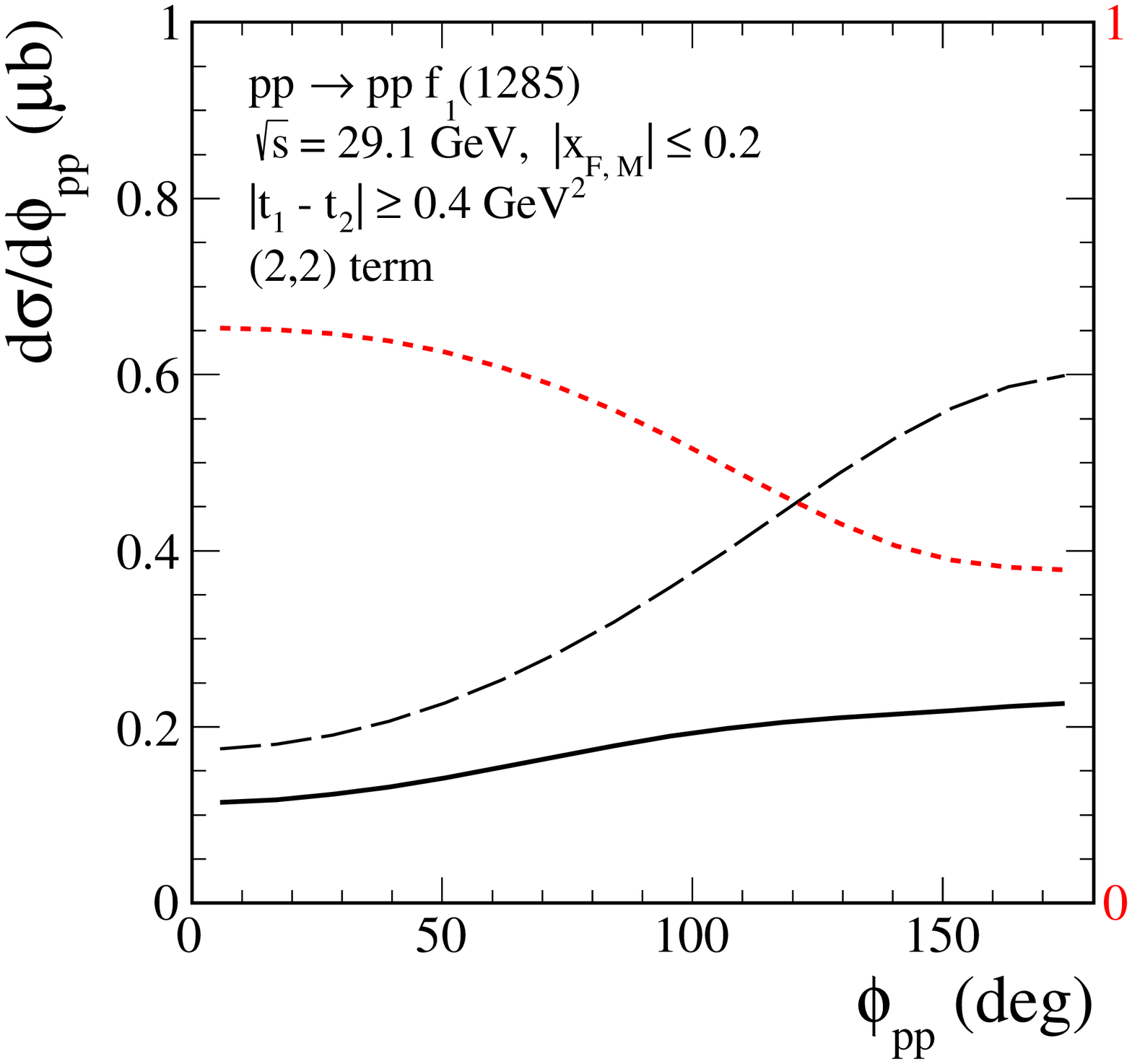}
\includegraphics[width=0.49\textwidth]{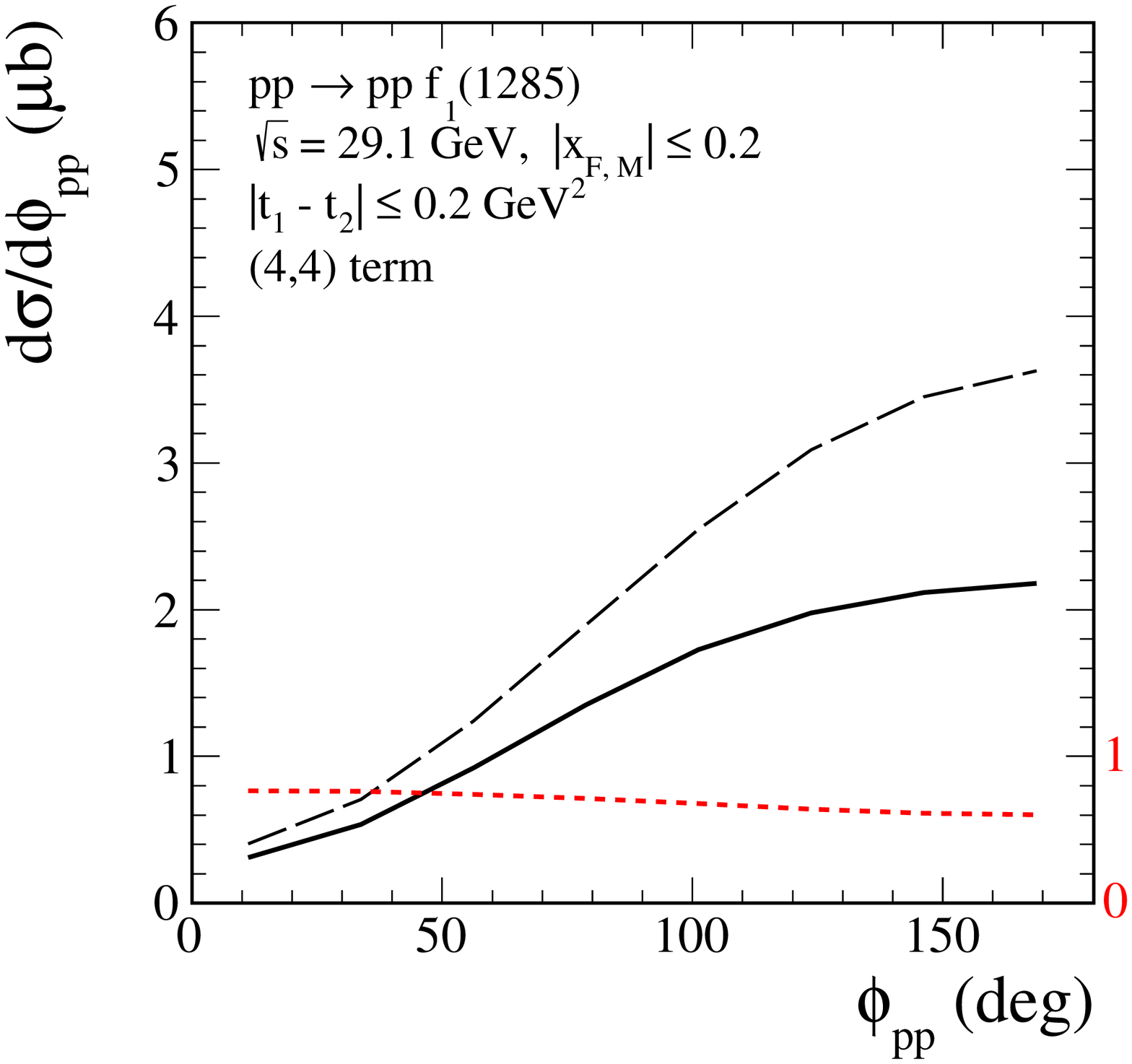}
\includegraphics[width=0.49\textwidth]{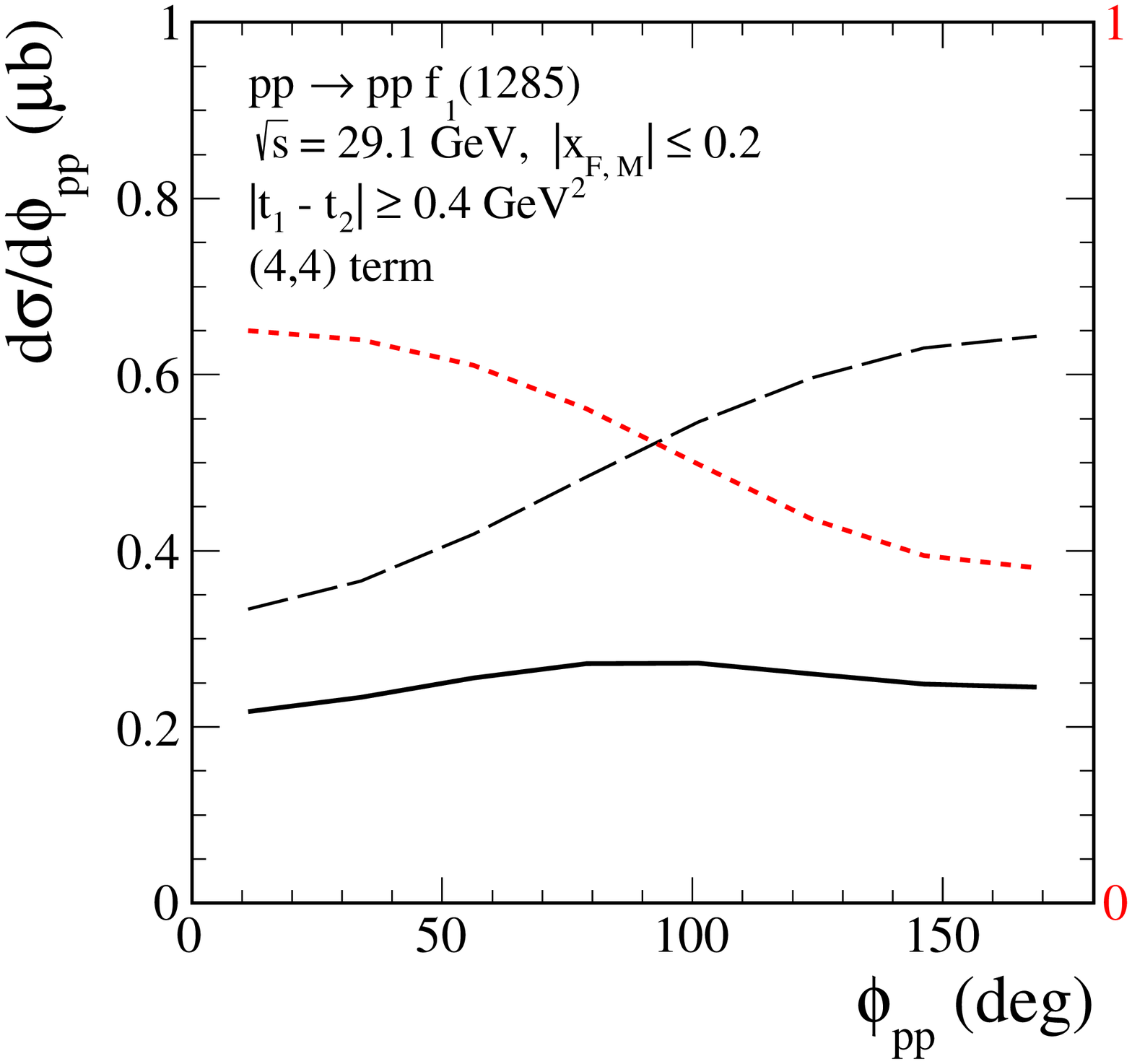}
\caption{\label{fig:2abs}
\small
The $\phi_{pp}$ distributions for $f_{1}(1285)$ meson production
at $\sqrt{s} = 29.1$~GeV, $|x_{F,M}| \leqslant 0.2$, and
for $|t_{1} - t_{2}| \leqslant 0.2$~GeV$^{2}$ (left panels) and 
for $|t_{1} - t_{2}| \geqslant 0.4$~GeV$^{2}$ (right panels).
In the calculation here we use (\ref{Fpompommeson_exp}) 
with $\Lambda_{E} = 0.7$~GeV.
The top panels show the results for the $(l,S) = (2,2)$ term 
and $g'_{\Pom \Pom f_{1}} = 4.89$ [see Eq.~(\ref{vertex_pompomf1_A})]
and the bottom panels show the $(l,S) = (4,4)$ term 
and $g''_{\Pom \Pom f_{1}} = 10.31$ [see Eq.~(\ref{vertex_pompomf1_B})].
The long-dashed black lines represent the Born results
and the solid black lines correspond to the results with the absorption effects included.
The dotted red lines represent the ratio of full and Born cross sections on the scale indicated by the red numbers on the right-hand side 
of the panels.}
\end{figure}

In \cite{Barberis:1998by} also the ${\rm dP_{t}}$ dependence
for both the $f_{1}(1285)$ and the $f_{1}(1420)$ mesons was presented.
Here, ${\rm dP_{t}}$ (the so-called ``glueball-filter variable'' 
\cite{Close:1997pj,Barberis:1996iq}) is defined as 
\begin{eqnarray}
\bdPt = \bqta - \bqtb = \bptb - \bpta \,, 
\quad {\rm dP_{t}} = |\bdPt|\,.
\label{dPt_variable}
\end{eqnarray}
The experimental values for the cross sections
in three ${\rm dP_{t}}$ intervals and for
the ratio of $f_{1}$ production at small 
${\rm dP_{t}}$ to large ${\rm dP_{t}}$ are given there.
In Table~\ref{tab:ratio_dPt} we show the WA102 data 
and our corresponding results for the different $\Pom \Pom f_{1}$ couplings.
The small values of the experimental ratios
for the $f_{1}(1285)$ and the $f_{1}(1420)$ as listed 
in the last column may signal that these two mesons are
predominantly $q \bar{q}$ states \cite{Close:1997pj}.
From the comparison of the first four rows we see again that
the exponential form of the $t$ dependences 
in the $(l,S) = (2,2)$ $\Pom \Pom f_{1}(1285)$ vertex is preferred.
For the $(4,4)$ term an optimal value of the $\Lambda_{E}$ parameter
is in the range of (0.6--0.7)~GeV.
There are also shown the results obtained 
for the couplings (\ref{A40}) and (\ref{A66})
and for the ratio of coupling constants from (\ref{kapparatiorange});
see (\ref{kapparatio_with_MKK}) of Appendix~\ref{sec:appendixB}.
For comparison, the results for
$\varkappa''/\varkappa'=-1.0$~GeV$^{-2}$ are also presented.
We use here the form factor (\ref{Fpompommeson_exp}) 
with $\Lambda_{E} = 0.7$~GeV.
\begin{table}
\caption{Results of $f_{1}$-meson production as a function of ${\rm dP_{t}}$
(\ref{dPt_variable}), in three ${\rm dP_{t}}$ intervals,
expressed as a percentage of the total contribution 
at the WA102 collision energy $\sqrt{s}=29.1$~GeV 
and for $|x_{F,M}| \leqslant 0.2$.
In the last column the ratios of 
$\sigma ({\rm dP_{t}} \leqslant \,0.2~\mathrm{GeV})/
 \sigma ({\rm dP_{t}} \geqslant \,0.5~\mathrm{GeV})$ are given.
The experimental numbers are from \cite{Barberis:1998by}. 
The theoretical numbers correspond to the separate individual coupling
terms $(l,S) = (2,2)$ and $(4,4)$
[see (\ref{vertex_pompomf1_A}) and (\ref{vertex_pompomf1_B}), respectively]
for different $\Lambda_{E}$ parameters
in the relevant type of the $\Pom \Pom f_{1}$ form factor.
The $\varkappa'$ and $\varkappa''$ results
were calculated from (\ref{A40}) and (\ref{A66}), respectively.
We show the results for the coupling range given by Eq.~(\ref{kapparatiorange})
and the result for 
$\varkappa''/\varkappa'=-1.0$~GeV$^{-2}$
from our fit to the WA102 data.
The absorption effects have been included in our analysis
within the one-channel-eikonal approach.}
\label{tab:ratio_dPt}
\begin{tabular}{|c|l|c|c|c|c|}
\hline
Meson& 
& 
${\rm dP_{t}} \leqslant 0.2$~GeV & 
$0.2 \leqslant {\rm dP_{t}} \leqslant 0.5$~GeV & 
${\rm dP_{t}} \geqslant 0.5$~GeV & Ratio\\
\hline
$f_{1}(1285)$ &Experiment \cite{Barberis:1998by}& $3 \pm 1$ & $35 \pm 2$ & $61 \pm 4$ & $0.05 \pm 0.02$\\
 &$(2,2)$, $\Lambda_{0}^{2} = 0.5$~GeV$^{2}$ & 1.5 & 30.3 & 68.1 & 0.02 \\
 &$(2,2)$, $\Lambda_{E} = 0.6$~GeV & 2.6 & 43.9 & 53.5 & 0.05 \\
 &$(2,2)$, $\Lambda_{E} = 0.7$~GeV & 2.0 & 37.1 & 60.9 & 0.03 \\
 &$(4,4)$, $\Lambda_{E} = 0.6$~GeV & 2.5 & 43.7 & 53.7 & 0.05 \\
 &$(4,4)$, $\Lambda_{E} = 0.7$~GeV & 1.9 & 36.8 & 61.3 & 0.03 \\
 &$\varkappa'$, $\Lambda_{E} = 0.7$~GeV & 2.0 & 37.5 & 60.5 & 0.03 \\
 &$\varkappa'$, $\Lambda_{E} = 0.8$~GeV & 1.7 & 32.5 & 65.8 & 0.03 \\
\hline
 &$(\varkappa',\varkappa'')$, $\Lambda_{E} = 0.7$~GeV:
 &  &  &  &  \\
 &$\varkappa''/\varkappa'=-6.25$~GeV$^{-2}$
 & 3.7 & 55.9 & 40.4 & 0.09 \\
 &$\varkappa''/\varkappa'=-3.76$~GeV$^{-2}$
 & 3.2 & 54.1 & 42.7 & 0.08\\
 &$\varkappa''/\varkappa'=-2.44$~GeV$^{-2}$
 & 2.8 & 50.1 & 47.1 & 0.06 \\
 &$\varkappa''/\varkappa'=-1.0$~GeV$^{-2}$
 & 2.4 & 41.8 & 55.8 & 0.04 \\
\hline
$f_{1}(1420)$&Experiment \cite{Barberis:1998by}& $2 \pm 2$ & $38 \pm 2$ & $60 \pm 4$ & $0.03 \pm 0.03$\\
 &$(2,2)$, $\Lambda_{0}^{2} = 0.5$~GeV$^{2}$ & 1.6 & 30.7 & 67.7 & 0.02 \\
 &$(2,2)$, $\Lambda_{E} = 0.6$~GeV & 2.7 & 44.3 & 53.0 & 0.05 \\
 &$(2,2)$, $\Lambda_{E} = 0.7$~GeV & 2.0 & 37.5 & 60.5 & 0.03 \\
 &$(2,2)$, $\Lambda_{E} = 0.8$~GeV & 1.6 & 32.7 & 65.7 & 0.02 \\
 &$(4,4)$, $\Lambda_{E} = 0.6$~GeV & 2.6 & 44.0 & 53.4 & 0.05 \\
 &$(4,4)$, $\Lambda_{E} = 0.7$~GeV & 2.0 & 37.1 & 60.9 & 0.03 \\
 &$\varkappa'$, $\Lambda_{E} = 0.7$~GeV & 2.0 & 37.8 & 60.2 & 0.03 \\
 &$\varkappa'$, $\Lambda_{E} = 0.8$~GeV & 1.7 & 33.0 & 65.3 & 0.03 \\
\hline
 &$(\varkappa',\varkappa'')$, $\Lambda_{E} = 0.7$~GeV:
 &  &  &  &  \\
 &$\varkappa''/\varkappa'=-6.25$~GeV$^{-2}$
 & 3.7 & 56.2 & 40.1 & 0.09 \\
 &$\varkappa''/\varkappa'=-3.76$~GeV$^{-2}$
 & 3.3 & 54.2 & 42.5 & 0.08 \\
 &$\varkappa''/\varkappa'=-2.44$~GeV$^{-2}$
 & 2.9 & 50.3 & 47.8 & 0.06 \\
 &$\varkappa''/\varkappa'=-1.0$~GeV$^{-2}$
 & 2.4 & 44.4 & 53.2 & 0.04 \\
\hline
\end{tabular}
\end{table}

Up to now, in Figs.~\ref{fig:1}, \ref{fig:2} and \ref{fig:2abs},
we have shown the contributions 
of the individual $(l,S)$ terms (couplings),
calculated with the vertices (\ref{vertex_pompomf1_A}) and (\ref{vertex_pompomf1_B}),
separately.

In Fig.~\ref{fig:2aux} we examine the combination of two 
$\Pom \Pom f_{1}$ couplings
$\varkappa'$ and $\varkappa''$
calculated with the vertices (\ref{A40}) and (\ref{A66}),
respectively.
We can see that the best fit is for the ratio
$\varkappa''/\varkappa' \simeq -1.0$~GeV$^{-2}$ 
(see the red dotted lines on the top panels),
which roughly agrees with the preliminary analysis performed in
\cite{Hechenberger:thesis} (cf. Eq.~(2.68) in \cite{Hechenberger:thesis}).

As discussed in Appendix~\ref{sec:appendixB},
the prediction for $\varkappa''/\varkappa'$
obtained in the Sakai-Sugimoto model is
\begin{equation}
\varkappa''/ \varkappa' = -(6.25 \cdots 2.44) \;\mathrm{GeV}^{-2}\label{kapparatiorange}
\end{equation}
for $M_\mathrm{KK}=(949 \cdots 1532)\;\mathrm{MeV}$.
This agrees with the above fit 
($\varkappa''/\varkappa' = -1.0$~GeV$^{-2}$) 
as far as the sign of this ratio is
concerned, but not in its magnitude.
Other than a simple inadequacy of the Sakai-Sugimoto model, this
could indicate that the Sakai-Sugimoto model needs a more complicated
form of reggeization of the tensor glueball propagator
as indeed discussed in \cite{Anderson:2014jia}
in the context of CEP of $\eta$ and $\eta'$ mesons.
It could also be an indication of the importance of secondary
reggeon exchanges.

Fitting the mean value of the total cross section 
(\ref{WA102_f1_1285}) we find
\begin{equation}
 (\varkappa', \varkappa'') = \begin{cases}
  (-8.88, 8.88\;{\rm GeV}^{-2}) & \mbox{for}\; \varkappa''/ \varkappa' = -1.0~\mathrm{GeV}^{-2} \,,\\
  (-9.14, 22.30\;{\rm GeV}^{-2}) & \phantom{\mbox{for}\;} \varkappa''/ \varkappa' = -2.44~\mathrm{GeV}^{-2} \,,\\
  (-9.22, 34.67\;{\rm GeV}^{-2}) & \phantom{\mbox{for}\;} \varkappa''/ \varkappa' = -3.76~\mathrm{GeV}^{-2} \,,\\
  (-8.81, 55.06\;{\rm GeV}^{-2}) & \phantom{\mbox{for}\;} \varkappa''/ \varkappa' = -6.25~\mathrm{GeV}^{-2} \,.
 \end{cases}
 \label{kappa_couplings_1285}
\end{equation}
Taking into account the experimental errors (\ref{WA102_f1_1285})
assumed to be 12.8\,\% of the cross section for each bin
(see the bottom panels of Fig.~\ref{fig:2aux}),
we get an error of our result for
$\varkappa''/\varkappa' = -1.0$~GeV$^{-2}$
of about 6\,\%. Thus the 1 standard deviation (s.d.) interval is here
\begin{equation}
(\varkappa', \varkappa'') = 
(-8.35, \;8.35\;{\rm GeV}^{-2})
\cdots 
(-9.41, \;9.41\;{\rm GeV}^{-2}) 
\quad \mbox{for}\; \varkappa''/ \varkappa' = -1.0~\mathrm{GeV}^{-2}\,.
\label{kappa_couplings_1285_error}
\end{equation}

In the bottom right panel of Fig.~\ref{fig:2aux}
we show results for the total
$\phi_{pp}$ distribution for the individual 
$\varkappa'$ and $\varkappa''$ coupling terms
and for their coherent sum.
Here we take $(\varkappa', \varkappa'') = (-8.88, \;8.88\;{\rm GeV}^{-2})$.
The interference effect of the $\varkappa'$ and $\varkappa''$ 
terms is clearly seen there.
As we see from (\ref{A78}) the $\varkappa''$ term
corresponds (approximately)
to a superposition of the $(l,S) = (2,2)$ and $(4,4)$ terms
with opposite signs.
We expect then destructive interference of the two $(l,S)$ terms,
and indeed, the $\varkappa''$ contribution
shows such a behaviour; i.e., there is a complete cancellation of the 
two $(l,S)$ terms for $\phi_{pp} \simeq 90^{\circ}$.
Hence, the option $\varkappa' = 0$, $\varkappa'' \neq 0$ is clearly ruled out
by the data for the $\phi_{pp}$ distribution.
In fact, this option is also incompatible
with the result (\ref{kapparatio_with_MKK}) obtained in the Sakai-Sugimoto model,
since it would correspond to the limit
$M_\mathrm{KK} \to 0$ where the holographic model ceases 
to have large-$N_c$ QCD as its infrared limit.
\begin{figure}[!ht]
\includegraphics[width=0.49\textwidth]{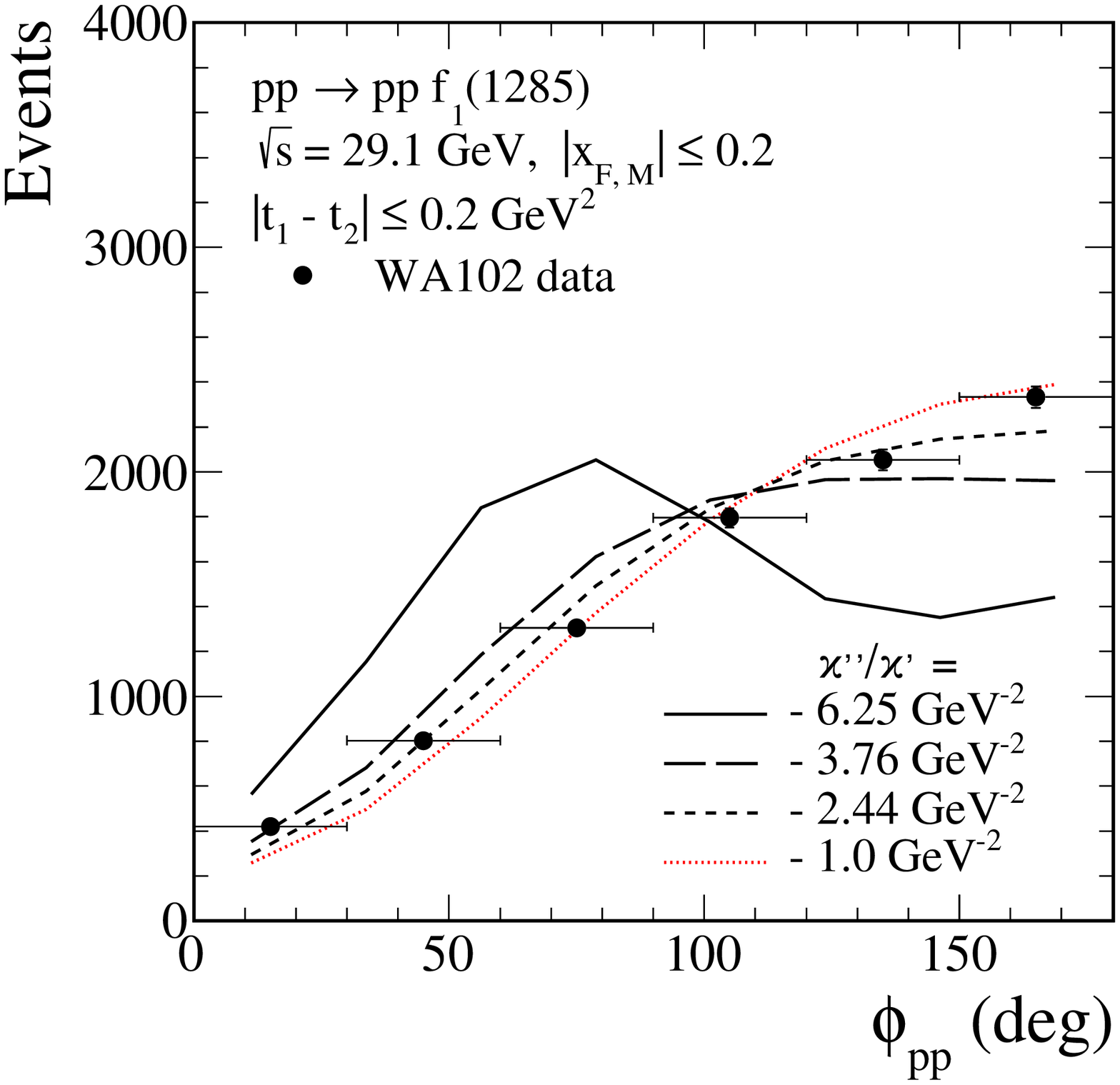}
\includegraphics[width=0.49\textwidth]{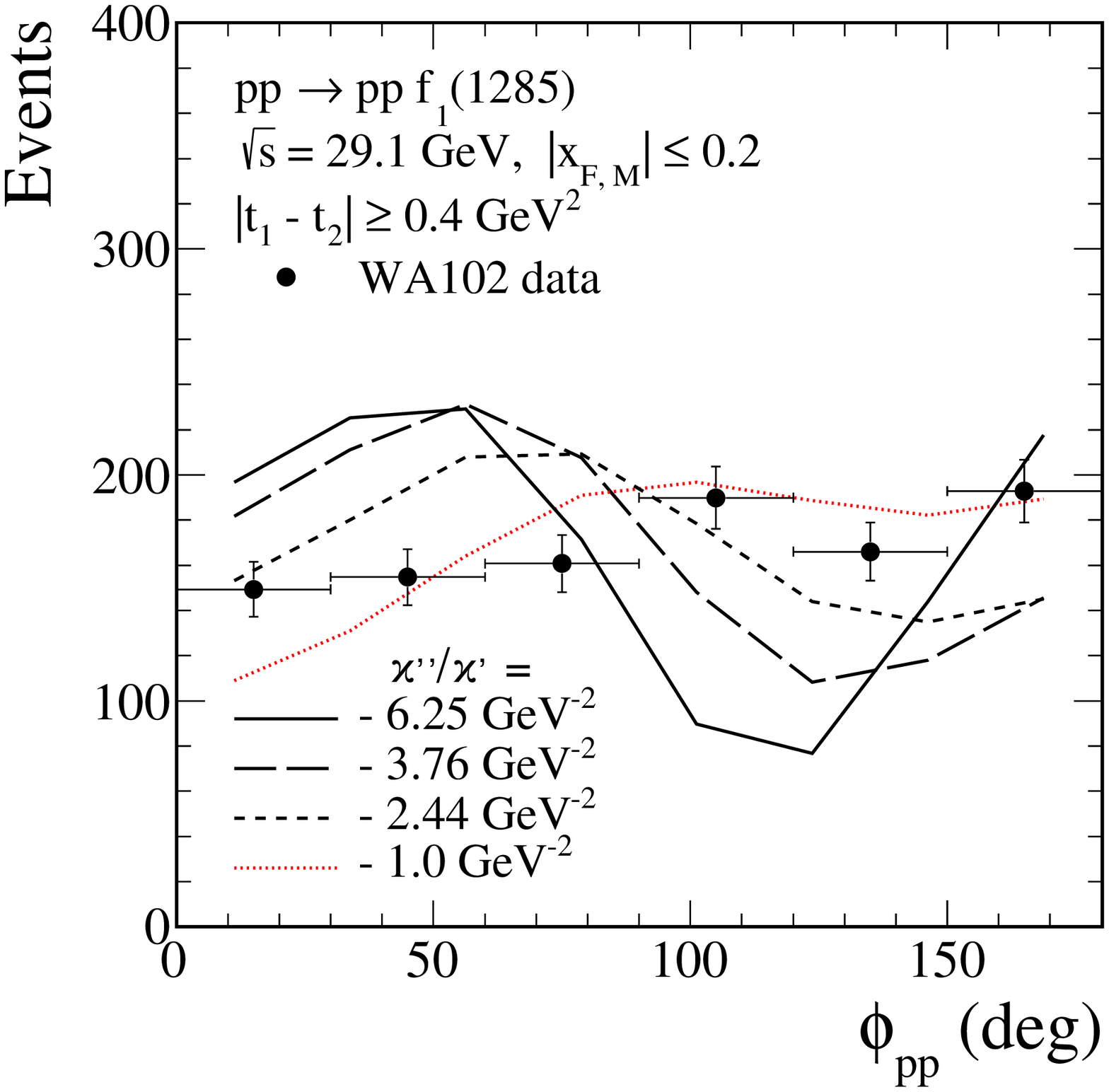}
\includegraphics[width=0.49\textwidth]{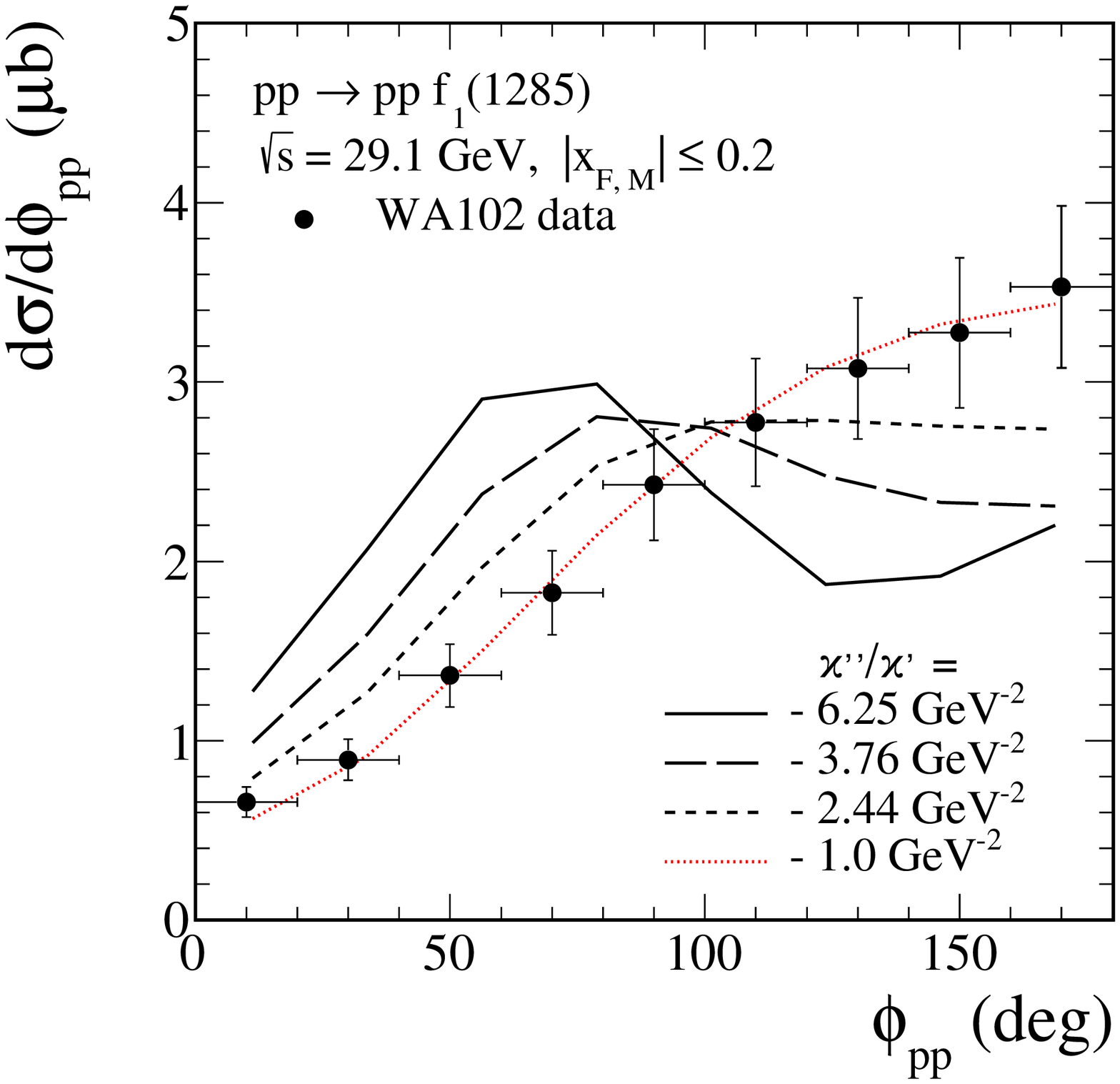}
\includegraphics[width=0.49\textwidth]{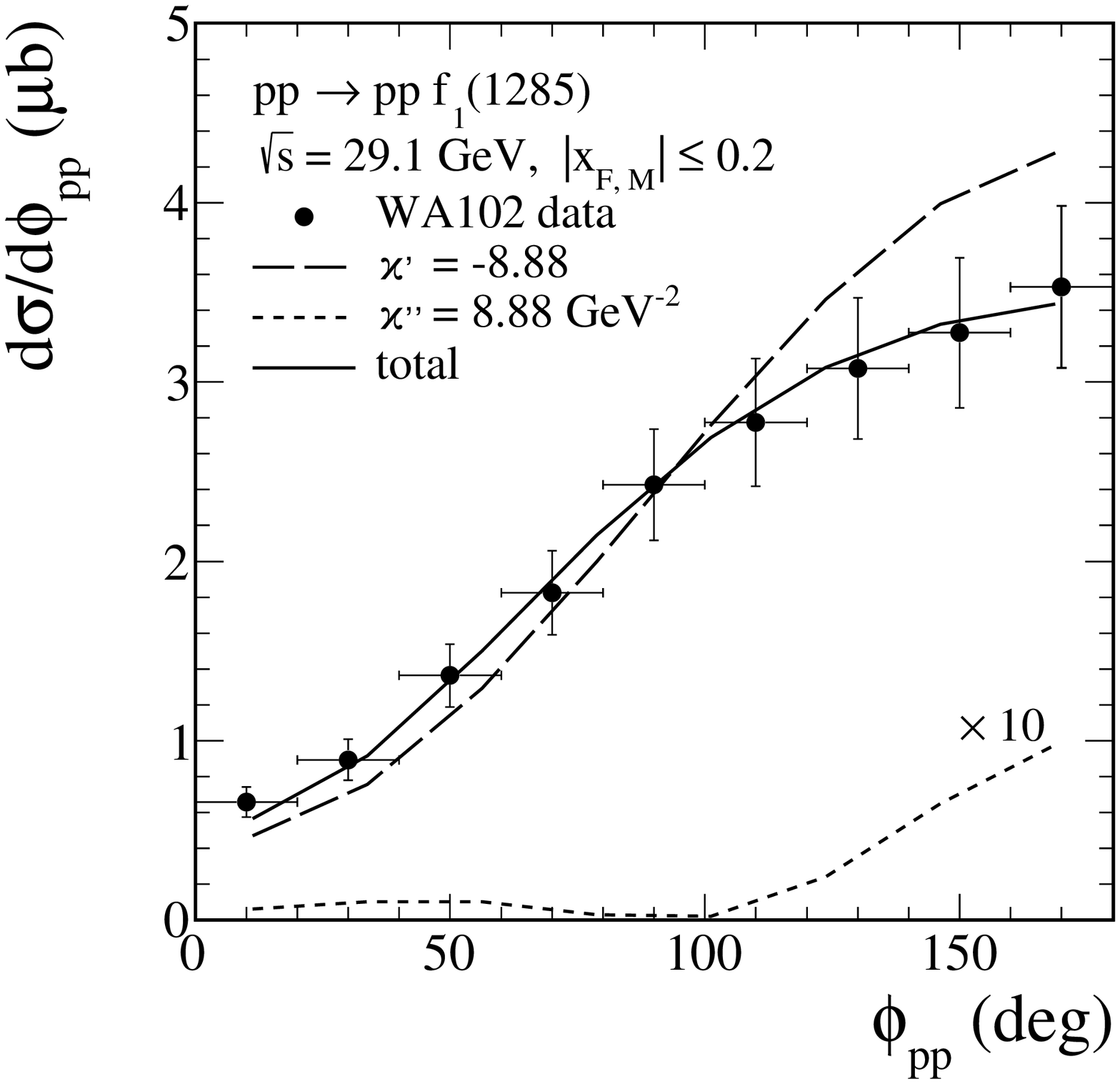}
\caption{\label{fig:2aux}
\small
The $\phi_{pp}$ distributions for $f_{1}(1285)$ meson production
at $\sqrt{s} = 29.1$~GeV.
Results for the $(\varkappa',\varkappa'')$ term calculated with
the vertices (\ref{A40}) and (\ref{A66}) are shown.
We use here the form factor (\ref{Fpompommeson_exp}) 
with $\Lambda_{E} = 0.7$~GeV.
In the top panels the theoretical results 
have been normalised to the mean value of the number of events
from \cite{Kirk:1999df}.
In the bottom panels we compare
the theoretical curves with the WA102 data from \cite{Barberis:1998by}.
Here the results have been normalised to the mean value
of the total cross section (\ref{WA102_f1_1285})
and the error bars on the data have been calculated as in Fig.~\ref{fig:1}.
In the bottom right panel we show the
results 
for $(\varkappa', \varkappa'') = (-8.88, \;8.88\;{\rm GeV}^{-2})$
for the individual 
$\varkappa'$ and $\varkappa''$ coupling terms
and for their coherent sum.
The $\varkappa''$ contribution has been
enhanced by a factor of 10 for better visibility.
The absorption effects are included in the calculations.}
\end{figure}

Summarizing our findings for $f_{1}(1285)$ CEP, we 
have obtained a reasonable description of the WA102 data
with either a pure $(l,S) = (2,2)$ or a pure $(l,S) = (4,4)$ coupling,
as well as with the $\varkappa',\varkappa''$ couplings with parameters,
\begin{eqnarray}
(l,S) = (2,2)\;\mathrm{term}:&& \;\,
g'_{\Pom \Pom f_{1}} = 4.89\,, \;
g''_{\Pom \Pom f_{1}} = 0\,, \;
\Lambda_{E} = 0.7\;\mathrm{GeV}\,;
\label{parameters_1285_a}\\
(l,S) = (4,4)\;\mathrm{term}:&& \;\,
g'_{\Pom \Pom f_{1}} = 0\,, \;
g''_{\Pom \Pom f_{1}} = 10.31\,, \;
\Lambda_{E} = 0.7\;\mathrm{GeV}\,;
\label{parameters_1285_b}\\
\varkappa' \;\mathrm{term\; only}:&& \;\,
|\varkappa'| = 8.58\,, \;
\varkappa'' = 0\,, \;
\Lambda_{E} = 0.7\;\mathrm{GeV}\,;
\label{parameters_1285_c}\\
(\varkappa' , \varkappa'') \;\mathrm{term}:&& \;\,
\varkappa' = -8.88\,, \;
\varkappa'' = 8.88\;\mathrm{GeV}^{-2}\,, \;
\Lambda_{E} = 0.7\;\mathrm{GeV}\,.
\label{parameters_1285_d}
\end{eqnarray}
As discussed in (\ref{kappa_couplings_1285_error}) the purely
statistical errors on the coupling parameters
(\ref{parameters_1285_a})--(\ref{parameters_1285_d}) are
estimated to be around 6\,\%.

It is also interesting to compare the results
(\ref{parameters_1285_a}) and (\ref{parameters_1285_c})
with the approximate relation (\ref{A78})
for $\varkappa'' = 0$ and $k^{2} = m_{f_{1}}^{2}$.
We note that we see no way to fix the overall sign of the $f_{1}$
couplings from experiment.
The states $\ket{f_{1}}$ and $-\ket{f_{1}}$ are clearly
equivalent from quantum mechanics.
Of course, relative signs of couplings have physical significance,
for instance, the relative sign of $g'_{\Pom \Pom f_{1}}$
and $g''_{\Pom \Pom f_{1}}$.
Keeping this in mind we compare the absolute values of the
left-hand side (l.h.s.) and right-hand side (r.h.s.) of (\ref{A78}).
With $m_{f_{1}} = (1281.9 \pm 0.5)$~MeV \cite{Tanabashi:2018oca}
we get
\begin{eqnarray}
\abs{\frac{g'_{\Pom \Pom f_{1}}}{\varkappa'}} = 0.57\,, \quad 
\frac{M_{0}^{2}}{m_{f_{1}}^{2}} = 0.61\,.
\label{parameters_1285_comp}
\end{eqnarray}
This shows that the approximate relation (\ref{A78}) is
here satisfied to an accuracy of around 10\,\%.

Using (\ref{A78}) we can also see to which values of
$g'_{\Pom \Pom f_{1}}$ and $g''_{\Pom \Pom f_{1}}$
the $(\varkappa',\varkappa'')$ values of (\ref{parameters_1285_d})
roughly correspond.
With (\ref{parameters_1285_d}) and setting 
$t_{1} = t_{2} = -0.1$~GeV$^{2}$ in (\ref{A78}) we get
\begin{eqnarray}
g'_{\Pom \Pom f_{1}} = 0.42\,, \quad 
g''_{\Pom \Pom f_{1}} = 10.81\,.
\label{comparison_couplings_1285}
\end{eqnarray}
Thus, $(\varkappa',\varkappa'')$ from (\ref{parameters_1285_d})
corresponds practically to a pure $(l,S) = (4,4)$ term
and the values for $g''_{\Pom \Pom f_{1}}$
from (\ref{parameters_1285_b}) and (\ref{comparison_couplings_1285})
agree to within 5\,\% accuracy.

Now we present a comparison of our theoretical results 
also for the $f_{1}(1420)$ meson 
with relevant data from the WA102 experiment \cite{Barberis:1998by}.
In Fig.~\ref{fig:1420_ff} we show the $|t|$ (left panels) 
and $\phi_{pp}$ (right panels) distributions
for $\sqrt{s} = 29.1$~GeV and $|x_{F,M}| \leqslant 0.2$.
The WA102 data points from \cite{Barberis:1998by} and our model results
have been normalised to the mean value of the total cross section 
\begin{eqnarray}
\sigma_{\rm exp.} = (1584 \pm 145)\;\mathrm{nb}\,;
\label{WA102_f1_1420}
\end{eqnarray}
see Table~\ref{tab:table1}.
The experimental error bars
are assumed to be 9.2\,\% corresponding to the error 
of $\sigma_{\rm exp.}$ in (\ref{WA102_f1_1420}).

From Fig.~\ref{fig:1420_ff} we can see 
that the $(l,S) = (2,2)$ term
is sufficient to describe the WA102 data.
We have checked that the shape of $\phi_{pp}$ distributions
almost does not depend on the choice of the cutoff parameter 
$\Lambda_{E}$, in particular for the $(l,S) = (2,2)$ term.
Taking into account the results listed 
in Table~\ref{tab:ratio_dPt} we conclude that 
$\Lambda_{E} = 0.7$~GeV is an optimal choice.
To get the mean value of the total cross section 
(\ref{WA102_f1_1420}) we find 
(assuming positive values of the coupling constants):
$g'_{\Pom \Pom f_{1}(1420)} = 2.06$ in (\ref{vertex_pompomf1_A}) 
for $\Lambda_{E} = 0.8$~GeV,  
2.39 for $\Lambda_{E} = 0.7$~GeV,
2.94 for $\Lambda_{E} = 0.6$~GeV,
$g''_{\Pom \Pom f_{1}(1420)} = 4.20$ in (\ref{vertex_pompomf1_B}) 
for $\Lambda_{E} = 0.7$~GeV,
5.24 for $\Lambda_{E} = 0.6$~GeV,
{$\varkappa' = 5.08$ in (\ref{A40}) 
for $\Lambda_{E} = 0.7$~GeV,
and 4.39 for $\Lambda_{E} = 0.8$~GeV.

In Fig.~\ref{fig:1420_ff_R} we show 
the results for $\varkappa'$ plus $\varkappa''$ terms
calculated with the vertices (\ref{A40}) and (\ref{A66})
and for different values of $\varkappa''/ \varkappa'$.
As for the $f_{1}(1285)$ CEP
a reasonable fit is obtained for 
$\varkappa''/ \varkappa' = -1$~GeV$^{-2}$.
Fitting the mean value of the total cross section 
(\ref{WA102_f1_1420}) we find for the $f_{1}(1420)$ meson
\begin{equation}
 (\varkappa', \varkappa'') = \begin{cases}
  (-5.23, 5.23\;{\rm GeV}^{-2}) & \mbox{for}\; \varkappa''/ \varkappa' = -1.0~\mathrm{GeV}^{-2} \,,\\
  (-5.40, 13.18\;{\rm GeV}^{-2}) & \phantom{\mbox{for}\;} \varkappa''/ \varkappa' = -2.44~\mathrm{GeV}^{-2} \,,\\
  (-5.44, 20.45\;{\rm GeV}^{-2}) & \phantom{\mbox{for}\;} \varkappa''/ \varkappa' = -3.76~\mathrm{GeV}^{-2} \,,\\
  (-5.19, 32.44\;{\rm GeV}^{-2}) & \phantom{\mbox{for}\;} \varkappa''/ \varkappa' = -6.25~\mathrm{GeV}^{-2} \,.\\
 \end{cases}
 \label{kappa_couplings_1420}
\end{equation}

It is interesting to see whether the couplings $\Pom \Pom f_{1}(1285)$
and $\Pom \Pom f_{1}(1420)$ are similar or very different.
Reasonable fits are obtained for the $f_{1}(1420)$ with parameters
\begin{eqnarray}
(l,S) = (2,2)\;\mathrm{term}:&& \;\,
g'_{\Pom \Pom f_{1}} = 2.39\,, \;
g''_{\Pom \Pom f_{1}} = 0\,, \;
\Lambda_{E} = 0.7\;\mathrm{GeV}\,;
\label{parameters_1420_a}\\
(l,S) = (4,4)\;\mathrm{term}:&& \;\,
g'_{\Pom \Pom f_{1}} = 0\,, \;
g''_{\Pom \Pom f_{1}} = 4.20\,, \;
\Lambda_{E} = 0.7\;\mathrm{GeV}\,;
\label{parameters_1420_b}\\
\varkappa' \;\mathrm{term\; only}:&& \;\,
|\varkappa'| = 5.08\,, \;
\varkappa'' = 0\,, \;
\Lambda_{E} = 0.7\;\mathrm{GeV}\,;
\label{parameters_1420_c}\\
(\varkappa' , \varkappa'') \;\mathrm{term}:&& \;\,
\varkappa' = -5.23\,, \;
\varkappa'' = 5.23\;\mathrm{GeV}^{-2} \,, \;
\Lambda_{E} = 0.7\;\mathrm{GeV}\,,
\label{parameters_1420_d}
\end{eqnarray}
with statistical errors on the coupling parameters around 5\,\% [cf.~(\ref{WA102_f1_1420})].

Here we get for the comparison of (\ref{parameters_1420_a}) and (\ref{parameters_1420_c}) with (\ref{A78}),
using $m_{f_{1}} = (1426.3 \pm 0.9)$~MeV 
from \cite{Tanabashi:2018oca},
\begin{eqnarray}
\abs{\frac{g_{\Pom \Pom f_{1}}^{\prime}}{\varkappa^{\prime}}} 
= 0.47\,, \quad 
\frac{M_{0}^{2}}{m_{f_{1}}^{2}} = 0.49\,.
\label{parameters_1420_comp}
\end{eqnarray}
Clearly, the agreement here is quite satisfactory.
Using in (\ref{A78}) $t_{1} = t_{2} = -0.1$~GeV$^{2}$ 
we find that (\ref{parameters_1420_d}) should roughly correspond to
\begin{eqnarray}
g'_{\Pom \Pom f_{1}} = -0.30\,, \quad 
g''_{\Pom \Pom f_{1}} = 5.14\,.
\label{comparison_couplings_1420}
\end{eqnarray}
As for the $f_{1}(1285)$ we find that for the $f_{1}(1420)$
the $(\varkappa',\varkappa'')$ term with
$\varkappa''/\varkappa' = -1$~GeV$^{-2}$
corresponds practically to a pure $(l,S) = (4,4)$ coupling.
The values of $g''_{\Pom \Pom f_{1}}$ from
(\ref{parameters_1420_b}) and (\ref{comparison_couplings_1420})
agree here to an accuracy of around 20\,\%.

We can also compare the relative strength of the coupling constants found for the
$f_{1}(1285)$ and $f_{1}(1420)$
with theoretical expectations assuming that 
these two $f_{1}$ mesons are separate $q \bar{q}$ states 
with mixing as parametrized in (\ref{mixing_f1_states}).

In Appendix~\ref{sec:appendixC} we derive the ratio of the coupling
constants for the two axial-vector mesons resulting from the assumption
that the pomeron couples only to the flavour-SU(3) singlet components, which
would be the case in the chiral limit for couplings that are exclusively determined by
the axial-gravitational anomaly (as in the Sakai-Sugimoto model).
For $f_1$-mixing angles that are often considered in the literature, namely
ideal mixing ($\phi_f=0^\circ$) and $\phi_f\gtrsim 20^\circ$, the ratio of all couplings
for $f_{1}(1420)$ over those for $f_{1}(1285)$ would then be given 
uniformly by a factor $1/\sqrt{2}=0.71$ and $\gtrsim 1.44$, respectively.

However, from (\ref{parameters_1285_a}) and (\ref{parameters_1420_a}),
(\ref{parameters_1285_b}) and (\ref{parameters_1420_b}),
(\ref{parameters_1285_d}) and (\ref{parameters_1420_d}),
we get
\begin{equation}
\frac{g'_{\Pom \Pom f_{1}(1420)}}{g'_{\Pom \Pom f_{1}(1285)}} 
= 0.49\,,\quad
\frac{g''_{\Pom \Pom f_{1}(1420)}}{g''_{\Pom \Pom f_{1}(1285)}} 
= 0.41\,,\quad
\frac{\varkappa^{\prime,\,\prime\prime}_{\Pom \Pom f_{1}(1420)}}{\varkappa^{\prime,\,\prime\prime}_{\Pom \Pom f_{1}(1285)}} 
= 0.59\,,
\label{f1_vs_f1p}
\end{equation}
respectively.

If at the WA102 energy of $\sqrt{s} = 29.1$~GeV only
$\Pom \Pom$ fusion contributes to the CEP of both $f_{1}$ mesons,
this means that pomerons 
do not couple predominantly to the flavour-SU(3) singlet components 
that are involved in the axial-gravitational anomaly. 
However, if the breaking of the SU(3) flavour symmetry 
by the strange quark mass has a large effect
for $\Pom \Pom f_1$ couplings, this presents
a problem for the chiral Sakai-Sugimoto model.
The discrepancy could, however, be partly due to 
important contributions from
subleading reggeon exchanges at WA102 energies.
Another possibility \cite{Debastiani:2016xgg,Liang:2020jtw} 
would be that the $f_1(1420)$ is not a separate resonance,
but rather the manifestation of the opening of additional decay channels 
in the tail of the $f_1(1285)$.
\begin{figure}[!ht]
\includegraphics[width=0.49\textwidth]{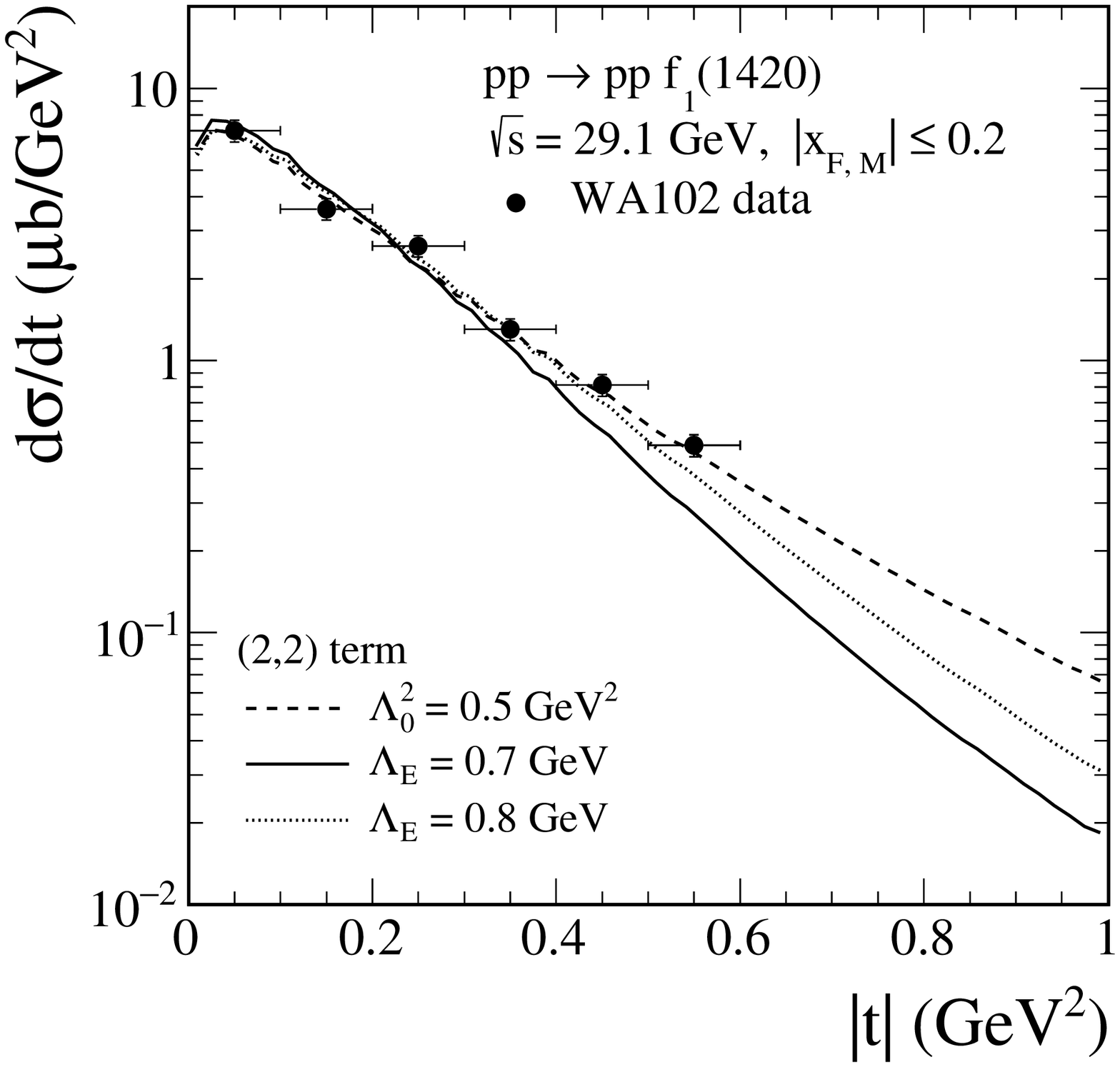}
\includegraphics[width=0.49\textwidth]{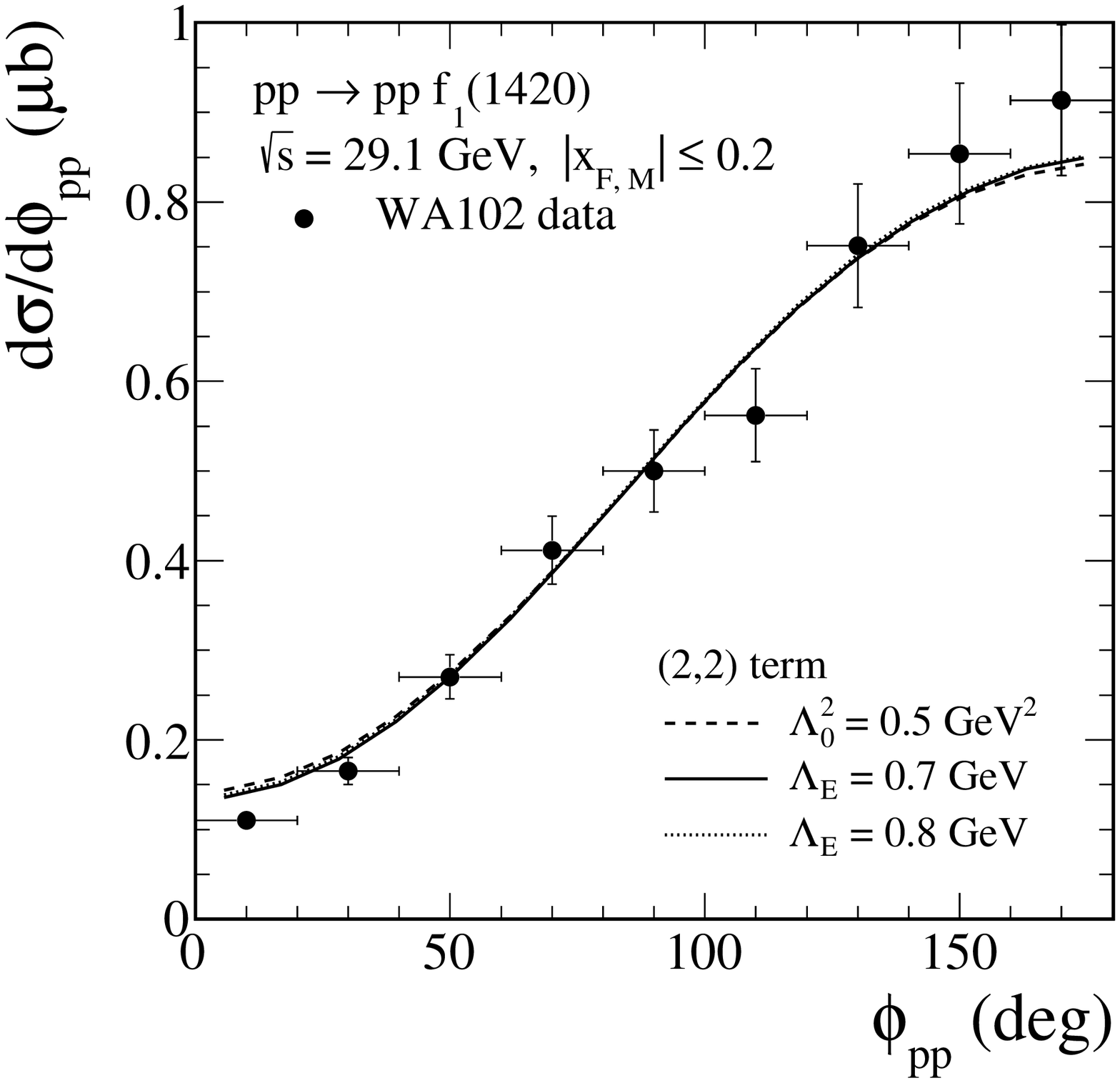}
\includegraphics[width=0.49\textwidth]{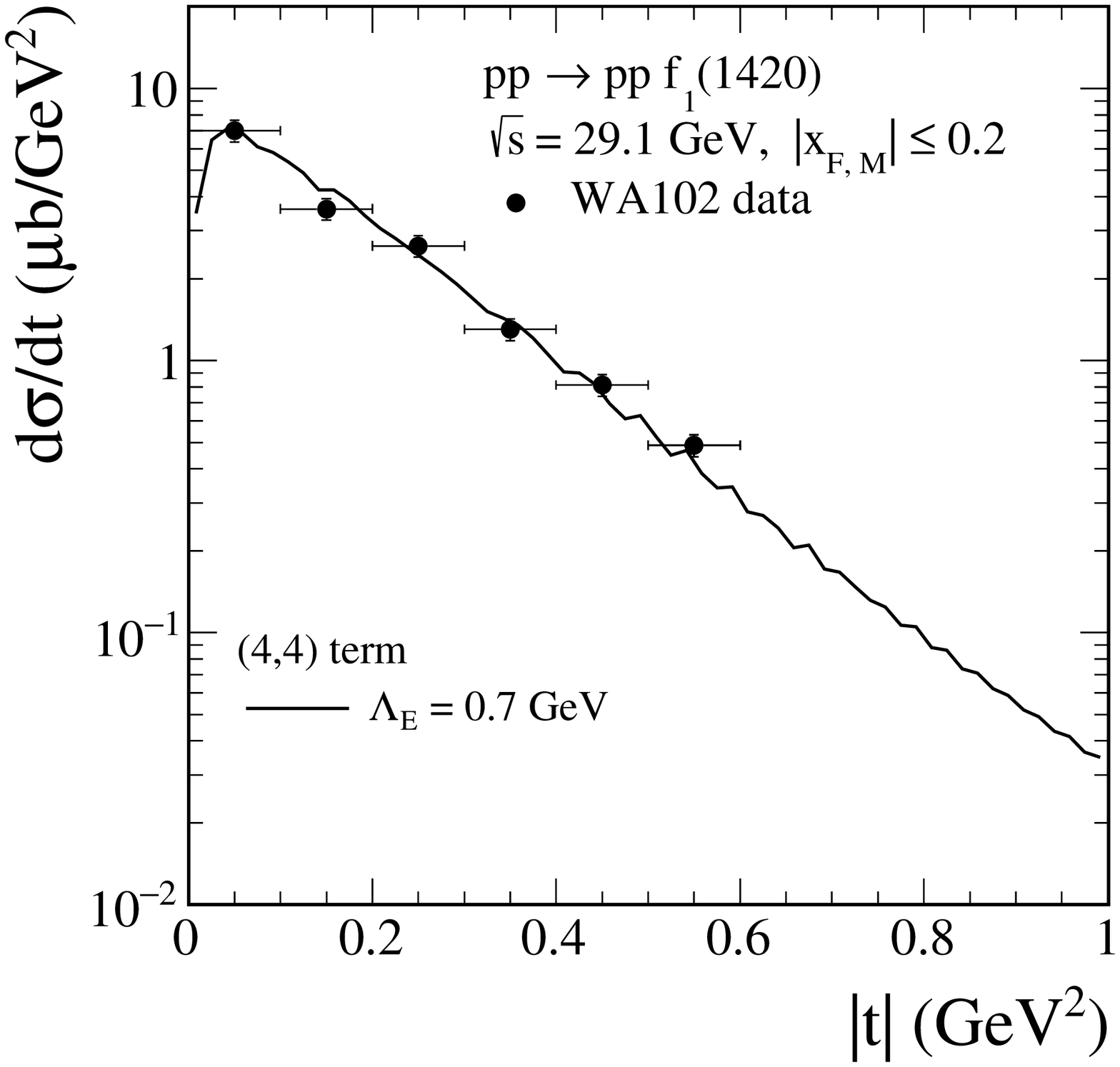}
\includegraphics[width=0.49\textwidth]{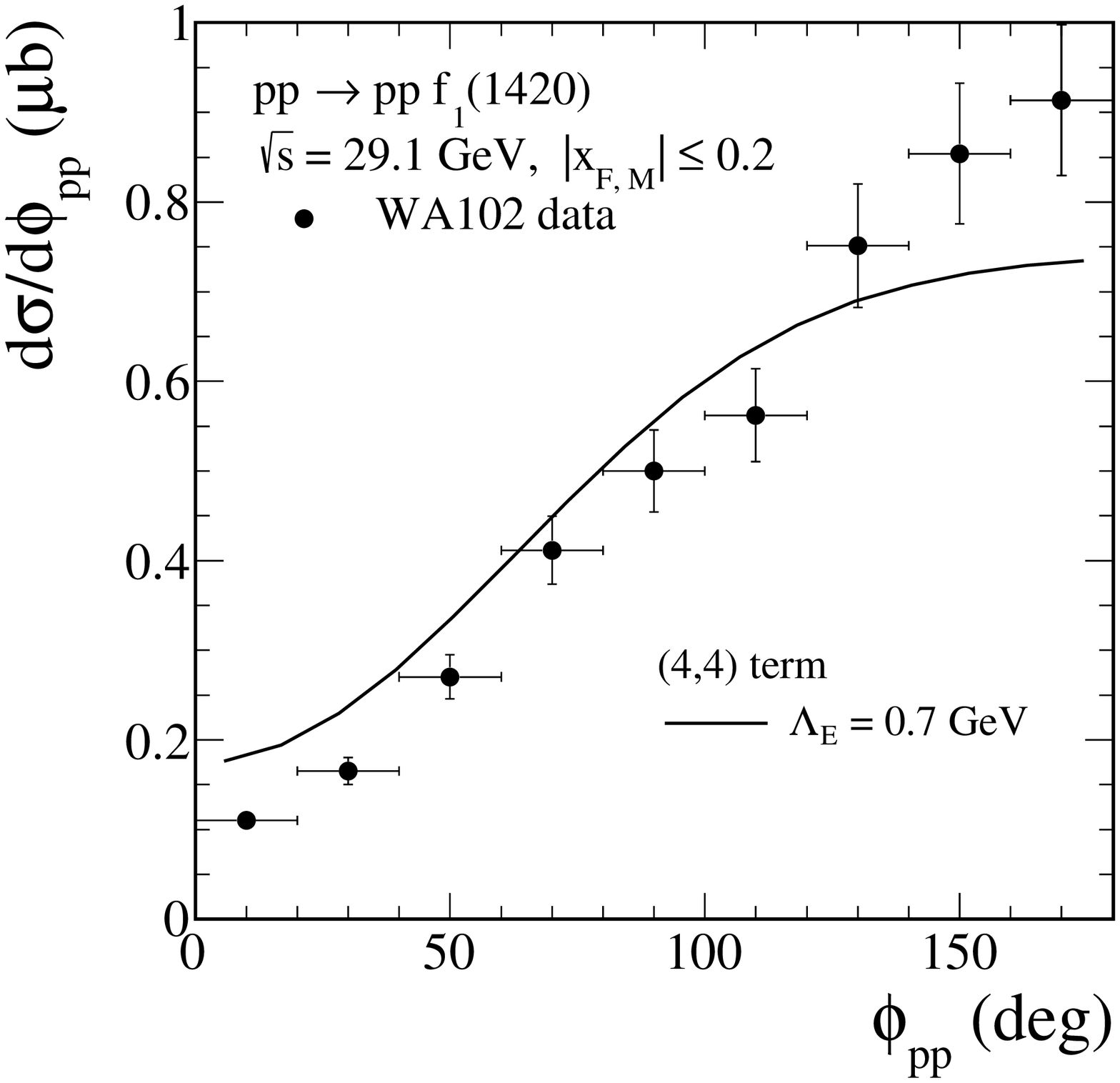}
\caption{\label{fig:1420_ff}
\small
The $|t|$ (left panels) and $\phi_{pp}$ (right panels) distributions 
for the $pp \to pp f_{1}(1420)$ reaction 
at $\sqrt{s} = 29.1$~GeV and $|x_{F,M}| \leqslant 0.2$.
The theoretical results and the WA102 data points 
from \cite{Barberis:1998by}
have been normalised to the mean value
of the total cross section (\ref{WA102_f1_1420}).
The error bars on the data correspond to the error on 
$\sigma_{\rm exp.}$ in (\ref{WA102_f1_1420}).
The separate individual coupling contributions 
for different cutoff parameters are shown.
The absorption effects are included in the calculations.
The oscillations in the left bottom panel are of numerical origin.}
\end{figure}

\begin{figure}[!ht]
\includegraphics[width=0.49\textwidth]{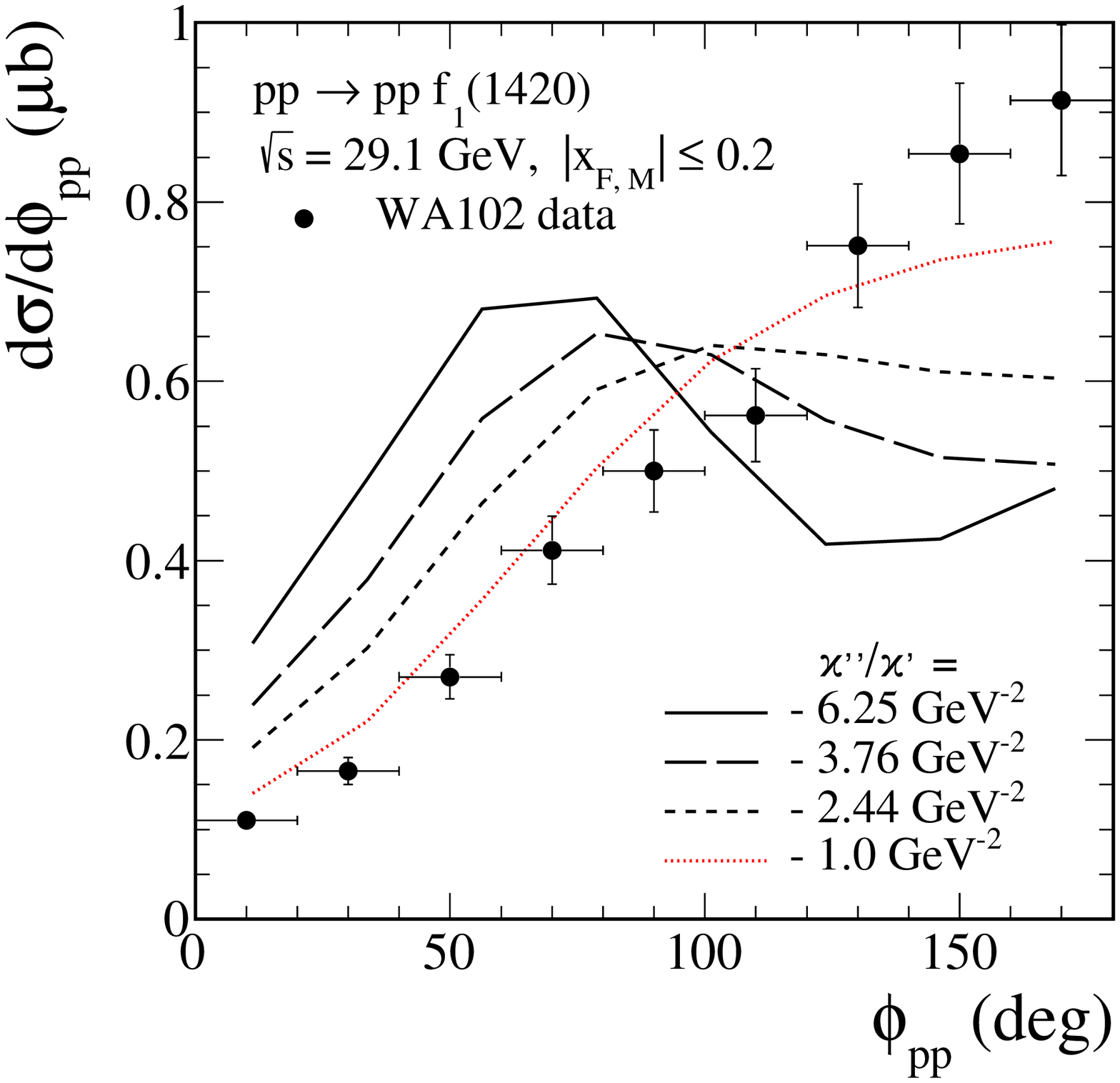}
\caption{\label{fig:1420_ff_R}
\small
The $\phi_{pp}$ distributions 
for the $pp \to pp f_{1}(1420)$ reaction 
at $\sqrt{s} = 29.1$~GeV and $|x_{F,M}| \leqslant 0.2$.
The theoretical results and the WA102 data points 
from \cite{Barberis:1998by}
have been normalised to the mean value
of the total cross section (\ref{WA102_f1_1420}).
The meaning of the lines is as in Fig.~\ref{fig:2aux}.}
\end{figure}

To summarize,
we have seen in this section that $\Pom \Pom$ fusion with suitable
$\Pom \Pom f_{1}$ couplings can give a reasonable description of
the WA102 data.
We have also seen that with the distributions explored
it is very hard to discriminate between the various possible couplings,
that is, to see which combination of coupling constants
is preferred experimentally. 
In addition we have the problem that at the relatively low c.m. energy
of $\sqrt{s} = 29.1$~GeV subleading reggeon exchanges may still be 
rather important. This topic will be dealt with in Appendix~\ref{sec:appendixD}.

In the next sections we shall show our results for RHIC and LHC energies
where subleading reggeon exchanges should be negligible,
at least, for the midrapidity region.
For these results we shall use the $\Pom \Pom f_{1}$ couplings
as determined in the present section.
But we must emphasize that our results for the RHIC and LHC
obtained in this way should be considered as upper limits
of the cross sections.
If at the WA102 energies there are important contributions
from subleading reggeon exchanges, 
the cross sections at the RHIC and LHC energies could be significantly smaller.
As we discuss in Appendix~\ref{sec:appendixD}, we estimate that the reduction could
be by a factor of up to 4 relative to the predictions given below.

\subsection{Predictions for the LHC experiments}
\label{sec:predictions_LHC}

Now we wish to show our results (predictions) for the LHC.

Here we consider only the $\Pom \Pom$ fusion 
with the coupling parameters found in Sec.~\ref{sec:comparison_WA102}
from the comparison with the WA102 data.

In Table~\ref{tab:table_LHC} we have collected cross sections 
in $\mu$b for
the reactions $pp \to pp f_{1}(1285)$ and $pp \to pp f_{1}(1420)$ 
at $\sqrt{s}=13$~TeV.
We show results for some kinematical cuts on the rapidity of the mesons,
$|{\rm y_{M}}| < 2.5$,
and also with an extra cut on momenta of leading protons
$0.17\;{\rm GeV} < |p_{y,p}| < 0.50\;{\rm GeV}$ that will be
applied when using the ALFA subdetector on both sides of the ATLAS detector.
We also show results for larger (forward) rapidities 
and without a measurement of outgoing protons
relevant for the LHCb experiment.
The calculations have been done in the Born approximation
and with the absorption corrections included. 
For the $f_{1}(1285)$ we show the individual results for the $(l,S) = (2,2)$ and $(4,4)$ terms 
with $g'_{\Pom \Pom f_{1}(1285)} = 4.89$ in (\ref{vertex_pompomf1_A})
and $g''_{\Pom \Pom f_{1}(1285)} = 10.31$ in (\ref{vertex_pompomf1_B});
see (\ref{parameters_1285_a}) and (\ref{parameters_1285_b}), respectively.
For the $(\varkappa',\varkappa'')$ terms, 
(\ref{A40}) plus (\ref{A66}),
we use (\ref{kappa_couplings_1285}).
We have taken here the form factor (\ref{Fpompommeson_exp}) 
with $\Lambda_{E} = 0.7$~GeV.
For the $f_{1}(1420)$ we show the results 
for the $(l,S) = (2,2)$ term with $g'_{\Pom \Pom f_{1}} = 2.39$,
see (\ref{parameters_1420_a}),
and the $(\varkappa',\varkappa'')$ option from (\ref{kappa_couplings_1420}).
As we see from comparing the last two columns of 
Table~\ref{tab:table_LHC} the absorption effects lead to
a sizeable reduction of the cross sections compared to the Born
results.
\begin{table}[]
\centering
\caption{The integrated cross sections in $\mu$b for CEP of $f_{1}$ mesons 
in $pp$ collisions for $\sqrt{s}=13$~TeV for some kinematical cuts 
on the rapidity ${\rm y_{M}}$ of the meson, and also when
limitations on the outgoing protons are imposed.
The results for the $(l,S) = (2,2)$ and $(4,4)$ terms 
calculated from (\ref{vertex_pompomf1_A}) and (\ref{vertex_pompomf1_B}),
respectively,
and for the $\varkappa'$ plus $\varkappa''$ terms calculated with
the vertices (\ref{A40}) plus (\ref{A66}) are shown.
The parameter values for $(\varkappa',\varkappa'')$
are taken from (\ref{kappa_couplings_1285}) for the $f_{1}(1285)$
and from (\ref{kappa_couplings_1420}) for the $f_{1}(1420)$.
We have taken here the form factor (\ref{Fpompommeson_exp}) 
with $\Lambda_{E} = 0.7$~GeV.
The results without and with absorption effects are presented.}
\label{tab:table_LHC}
\begin{tabular}{|c|l|c|c|c|c|}
\hline
Meson  & Cuts  & Contribution & Parameters & $\sigma_{{\rm Born}}$~($\mu$b) & $\sigma_{{\rm abs.}}$~($\mu$b) \\ 
\hline
$f_{1}(1285)$  & $|{\rm y_{M}}| < 1.0$  & $(2,2)$ &  Eq.~(\ref{parameters_1285_a})  & 36.11 & 14.83 \\ 
&  & $(4,4)$ & Eq.~(\ref{parameters_1285_b})      & 32.95 & 13.82 \\ 
&& $(\varkappa',\varkappa'')$ & $\varkappa''/\varkappa' = -6.25\;{\rm GeV}^{-2}$
& 27.17 & 18.63 \\ 
&& $(\varkappa',\varkappa'')$ & $\varkappa''/\varkappa' = -2.44\;{\rm GeV}^{-2}$
& 34.25 & 17.54 \\ 
&& $(\varkappa',\varkappa'')$ & $\varkappa''/\varkappa' = -1.0\;{\rm GeV}^{-2}$
& 36.27 & 16.56 \\ 
\hline
               & $|{\rm y_{M}}| < 2.5$ & $(2,2)$ & Eq.~(\ref{parameters_1285_a})       & 90.63 & 37.54 \\ 
               &                       & $(4,4)$ & Eq.~(\ref{parameters_1285_b})       & 83.97 & 34.01 \\ 
&& $(\varkappa',\varkappa'')$ & $\varkappa''/\varkappa' = -6.25\;{\rm GeV}^{-2}$
& 69.08 & 45.79 \\ 
&& $(\varkappa',\varkappa'')$ & $\varkappa''/\varkappa' = -2.44\;{\rm GeV}^{-2}$
& 86.05 & 43.44 \\ 
&& $(\varkappa',\varkappa'')$ & $\varkappa''/\varkappa' = -1.0\;{\rm GeV}^{-2}$
& 91.47 & 41.00 \\ 
\hline
               & $|{\rm y_{M}}| < 2.5$, & $(2,2)$ & Eq.~(\ref{parameters_1285_a})      & 19.37 & \;\,6.46 \\ 
& $0.17\;{\rm GeV} < |p_{y,p}| < 0.50\;{\rm GeV}$ 
                                        & $(4,4)$ & Eq.~(\ref{parameters_1285_b})      & 18.07 & \;\,6.06 \\ 
&& $(\varkappa',\varkappa'')$ & $\varkappa''/\varkappa' = -6.25\;{\rm GeV}^{-2}$
& 11.64 & \;\,7.14 \\ 
&& $(\varkappa',\varkappa'')$ & $\varkappa''/\varkappa' = -2.44\;{\rm GeV}^{-2}$
& 16.71 & \;\,7.10 \\ 
&& $(\varkappa',\varkappa'')$ & $\varkappa''/\varkappa' = -1.0\;{\rm GeV}^{-2}$
& 19.71 & \;\,7.09 \\ 
\hline
               & $2.0 < {\rm y_{M}} < 4.5$ & $(2,2)$ & Eq.~(\ref{parameters_1285_a})   & 46.63 & 18.89 \\ 
               &                           & $(4,4)$ & Eq.~(\ref{parameters_1285_b})   & 43.58 & 18.07 \\ 
&& $(\varkappa',\varkappa'')$ & $\varkappa''/\varkappa' = -6.25\;{\rm GeV}^{-2}$
& 35.32 & 23.13 \\ 
&& $(\varkappa',\varkappa'')$ & $\varkappa''/\varkappa' = -2.44\;{\rm GeV}^{-2}$
& 44.28 & 22.14 \\ 
&& $(\varkappa',\varkappa'')$ & $\varkappa''/\varkappa' = -1.0\;{\rm GeV}^{-2}$
& 46.52 & 20.50 \\ 
\hline
$f_{1}(1420)$  & $|{\rm y_{M}}| < 1.0$ & $(2,2)$ & Eq.~(\ref{parameters_1420_a})        & \;\,8.80 & \;\,3.66 \\
&& $(\varkappa',\varkappa'')$ & $\varkappa''/\varkappa' = -1.0\;{\rm GeV}^{-2}$
& \;\,8.75 & \;\,4.10 \\ 
\hline
              & $|{\rm y_{M}}| < 2.5$ & $(2,2)$ & Eq.~(\ref{parameters_1420_a})        & 22.22 & \;\,9.20 \\
&& $(\varkappa',\varkappa'')$ & $\varkappa''/\varkappa' = -1.0\;{\rm GeV}^{-2}$
& 22.16 & \;\,9.85 \\   
\hline
& $|{\rm y_{M}}| < 2.5$,
                                        & $(2,2)$ & Eq.~(\ref{parameters_1420_a})      & \;\,5.14 & \;\,1.77 \\ 
& $0.17\;{\rm GeV} < |p_{y,p}| < 0.50\;{\rm GeV}$ 
& $(\varkappa',\varkappa'')$ & $\varkappa''/\varkappa' = -1.0\;{\rm GeV}^{-2}$
& \;\,4.65 & \;\,1.67 \\
\hline
& $2.0 < {\rm y_{M}} < 4.5$ & $(2,2)$ & Eq.~(\ref{parameters_1420_a})   & 11.37 & \;\,4.68 \\
&& $(\varkappa',\varkappa'')$ & $\varkappa''/\varkappa' = -1.0\;{\rm GeV}^{-2}$
& 11.35 & \;\,4.92 \\
\hline
\end{tabular}
\end{table}

In Fig.~\ref{fig:ATLAS-ALFA} we show our predictions 
for the $pp \to pp f_{1}(1285)$ reaction
for $\sqrt{s} = 13$~TeV, $|{\rm y_{M}}| < 2.5$,
and for the cut on the leading protons of
$0.17\;{\rm GeV} < |p_{y,p}| < 0.50\;{\rm GeV}$.
Here the distribution of $p_{t, M}$ does not require, 
whereas those of $\phi_{pp}$, $|t|$, and ${\rm dP_{t}}$ do require
the detection of the leading protons.
The results calculated with the vertices 
(\ref{vertex_pompomf1_A}) [$(2,2)$ term],
(\ref{vertex_pompomf1_B}) [$(4,4)$ term], and 
(\ref{A40}) plus (\ref{A66}) 
[$\varkappa''/\varkappa' = -1$~GeV$^{2}$ and $-2.44$~GeV$^{2}$] 
give quite similar distributions.
The contribution with $\varkappa''/\varkappa' = -6.25$~GeV$^{2}$
gives a significantly different shape 
in the distributions of $\phi_{pp}$
and of the transverse \mbox{momentum of the $f_{1}(1285)$}.

In all cases the absorption effects are included.
Inclusion of absorption effects modifies the differential distributions
because their shapes depend on the kinematics of outgoing protons. 
We have checked numerically that the absorption effects 
decrease the distributions mostly at higher values 
of the variables $\phi_{pp}$ and ${\rm dP_{t}}$
and at smaller values of $p_{t,M}$ and $|t|$.
The measurement of such distributions would allow one to better
understand absorption effects.
This could be tested in future in experiments at the LHC,
when both protons are measured, such as ATLAS-ALFA and CMS-TOTEM.
The {\tt GenEx} \cite{Kycia:2014hea,Kycia:2017ota}
and {\tt GRANIITTI} \cite{Mieskolainen:2019jpv}
Monte Carlo event generators could be used in this context.

\begin{figure}[!ht]
\includegraphics[width=0.49\textwidth]{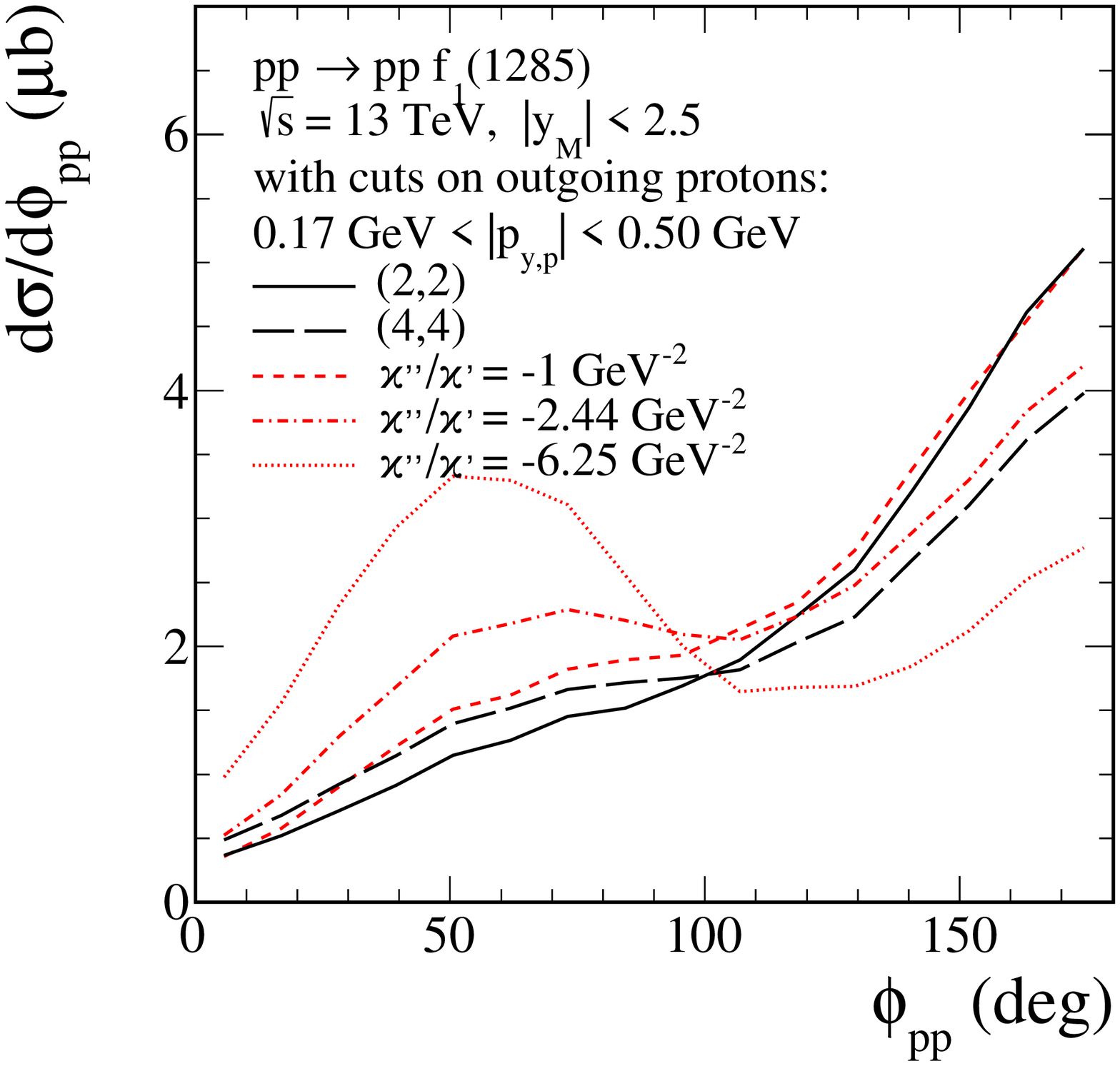}
\includegraphics[width=0.49\textwidth]{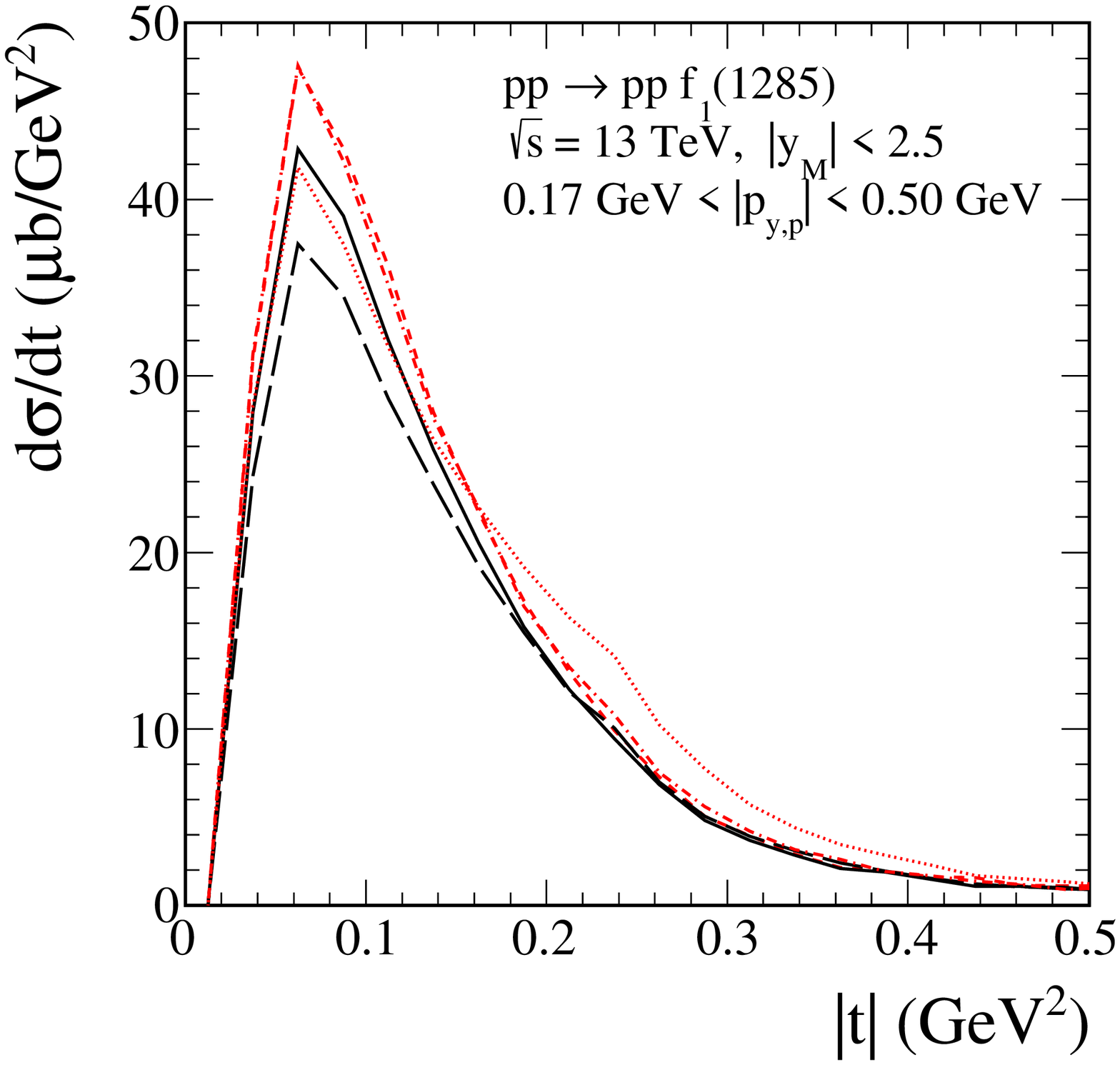}
\includegraphics[width=0.49\textwidth]{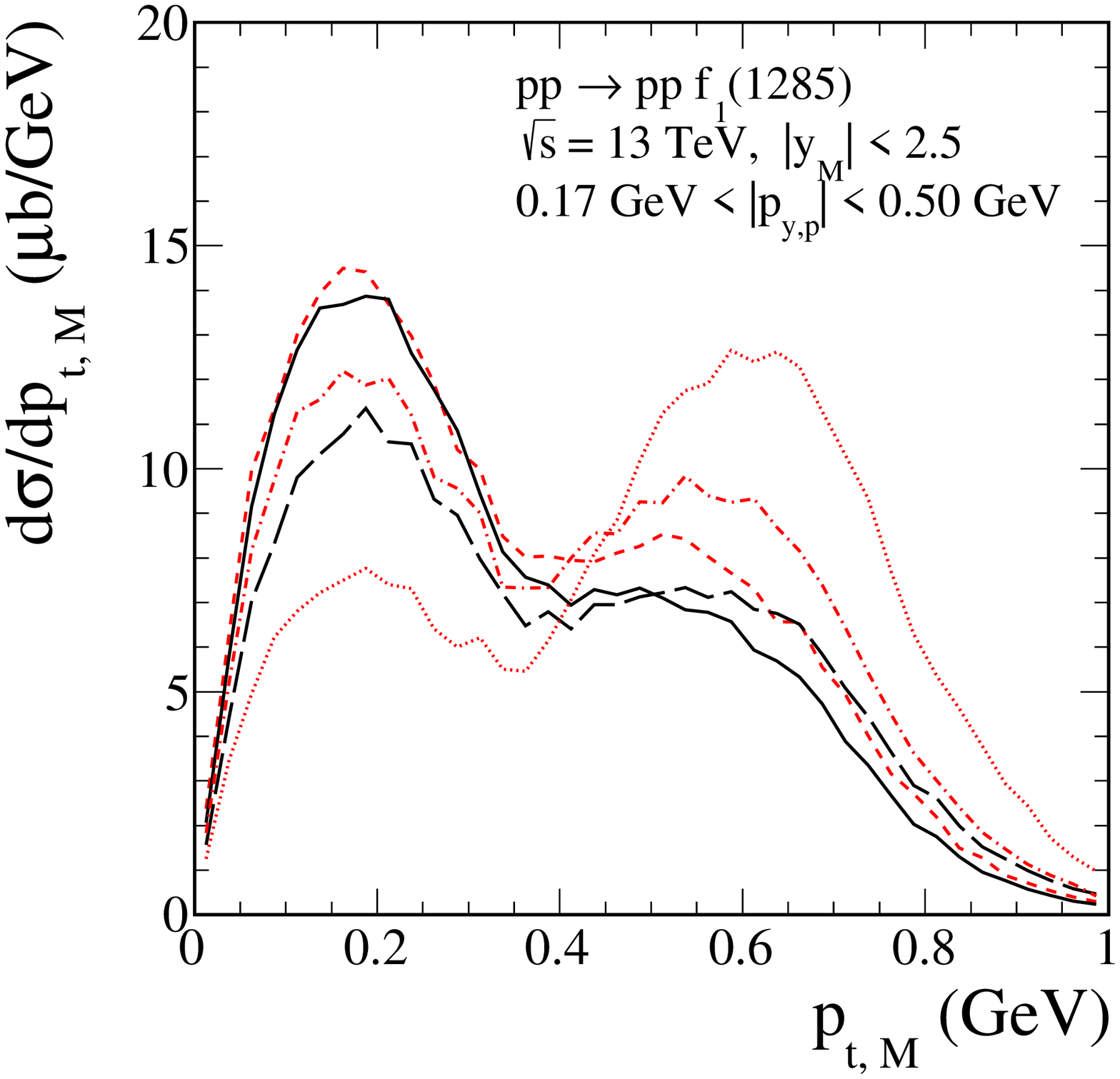}
\includegraphics[width=0.49\textwidth]{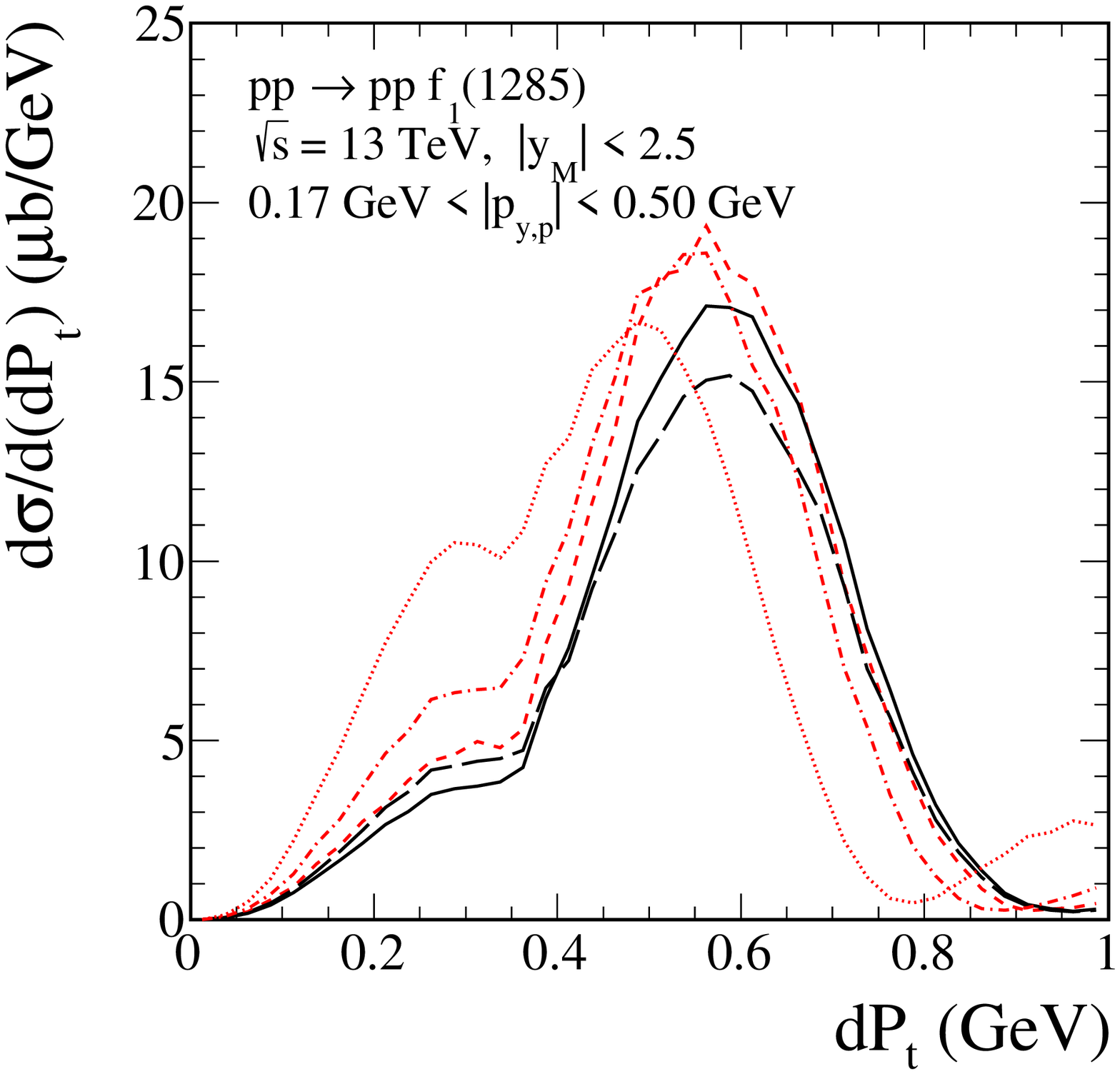}
\caption{\label{fig:ATLAS-ALFA}
\small
The differential cross sections for
the $f_{1}(1285)$ production at $\sqrt{s} = 13$~TeV and
\mbox{$|{\rm y_{M}}| < 2.5$}.
The results for $(l,S) = (2,2)$, $(4,4)$, 
and $(\varkappa',\varkappa'')$ contributions are shown.
Here we use for the $(2,2)$ and $(4,4)$ terms
(\ref{parameters_1285_a}) and (\ref{parameters_1285_b}), respectively.
For the $(\varkappa',\varkappa'')$ terms we use 
(\ref{kappa_couplings_1285}).
The absorption effects are included in all the calculations.}
\end{figure}

Now we discuss one of the most prominent decay modes of the $f_{1}(1285)$,
the decay $f_{1}(1285) \to \pi^{+}\pi^{-}\pi^{+}\pi^{-}$.
This four-pion decay channel seems well suited to measure 
the $f_{1}(1285)$ meson in CEP.
However, the $f_{1}(1285)$ is rather close 
in mass to the $f_{2}(1270)$ which also decays into four pions.
In principle, the $f_{1}(1285)$ and $f_{2}(1270)$ decays 
will interfere in the four-pion final state.
Note that this interference could be used to determine the relative sign
of the $f_{1}$ and $f_{2}$ production times decay amplitudes.
But the interference terms will drop out in the total decay rates.

In PDG \cite{Tanabashi:2018oca} the following branching fractions are listed:
\begin{eqnarray}
&&{\cal BR}(f_{1}(1285) \to \pi^{+}\pi^{-}\pi^{+}\pi^{-}) 
= (11.2^{+0.7}_{-0.6}) \,\%\,, 
\label{BR_f1_1285}\\
&&{\cal BR}(f_{2}(1270) \to \pi^{+}\pi^{-}\pi^{+}\pi^{-}) 
= (2.8 \pm 0.4) \,\%\,.
\label{BR_f1_1270}
\end{eqnarray}
Note that 
$\Gamma(f_{2}(1270)) = 186.7^{+2.2}_{-2.5}$~MeV, 
$\Gamma(f_{1}(1285)) = (22.7 \pm 1.1)$~MeV.
Thus we have $\Gamma(f_{2}(1270)) \gg \Gamma(f_{1}(1285))$.

In the following, for CEP of the $f_{1}(1285)$ meson,
we assume the $(l,S) = (2,2)$ coupling 
and $\Lambda_{E} = 0.7$~GeV; 
see (\ref{parameters_1285_a}) 
and $\sigma_{{\rm abs.}}$ in Table~\ref{tab:table_LHC}.
For CEP of the $f_{2}(1270)$ meson
the cross section is
$\sigma_{pp \to pp f_{2}(1270)} = 11.25$~$\mu$b
with the parameters from Ref.~\cite{Lebiedowicz:2019por}:
$(g_{\Pom \Pom f_{2}}^{(2)}, g_{\Pom \Pom f_{2}}^{(5)}) = (-4.0, 16.0)$, 
$\Lambda_{0}^{2} = 0.5$~GeV$^{2}$.
The absorption effects are taken into account in the calculation.
We obtain the integrated cross sections 
for $\sqrt{s} = 13$~TeV and $|{\rm y_{M}}| < 2.5$,
including the PDG branching fractions (\ref{BR_f1_1285}) and (\ref{BR_f1_1270}), as follows
\begin{eqnarray}
\sigma_{pp \to pp f_{1}(1285)} \times {\cal BR}(f_{1}(1285) \to \pi^{+}\pi^{-}\pi^{+}\pi^{-}) = 4.20\; \mu{\rm b}
\end{eqnarray}
and
\begin{eqnarray}
\sigma_{pp \to pp f_{2}(1270)} \times {\cal BR}(f_{2}(1270) \to \pi^{+}\pi^{-}\pi^{+}\pi^{-}) = 0.32\; \mu{\rm b}\;,
\end{eqnarray}
respectively.
Thus we predict a large cross section 
for the exclusive axial-vector $f_{1}(1285)$ production 
compared to the production of the tensor $f_{2}(1270)$ meson
in the $\pi^{+}\pi^{-}\pi^{+}\pi^{-}$ channel.
Even if we scale down the $f_{1}$ cross section by a factor of 4,
it will still be larger than our result for the $f_{2}$ cross section.
In addition, $\Gamma(f_{2}(1270)) \gg \Gamma(f_{1}(1285))$, so
$f_{1}(1285)$ will be seen as a sharp peak on top 
of a smaller bump corresponding to the $f_{2}(1270)$.

\subsection{Predictions for the STAR experiment at RHIC}
\label{sec:predictions_STAR}

The STAR experiments at RHIC measure CEP reactions
at $\sqrt{s} = 200$~GeV \cite{Adam:2020sap}
and at $\sqrt{s} = 510$~GeV \cite{ICHEP_STAR}.
It has the possibility to observe the outgoing protons
at least in a certain phase space region.
We shall present the predictions of our model for
the cut on the rapidity of the meson $|{\rm y_{M}}| < 0.7$
and for limitations on the outgoing protons,
for $\sqrt{s}=200$~GeV,
\begin{eqnarray}
&&(p_{x,p} + 0.3 \;{\rm GeV})^{2} + p_{y,p}^{2} < 0.25\;{\rm GeV}^{2} \,, \nonumber \\
&&0.2\;{\rm GeV} < |p_{y,p}| < 0.4\;{\rm GeV}\,, \nonumber \\
&&p_{x,p}>-0.2\;{\rm GeV}\,,
\label{STAR_cuts_on_protons_200}
\end{eqnarray}
as specified in Eq.~(6.1) of \cite{Adam:2020sap},
and for $\sqrt{s}=510$~GeV,
\begin{eqnarray}
&&(p_{x,p} + 0.6 \;{\rm GeV})^{2} + p_{y,p}^{2} < 1.25\;{\rm GeV}^{2} \,, \nonumber \\
&&0.4\;{\rm GeV} < |p_{y,p}| < 0.8\;{\rm GeV}\,, \nonumber \\
&&p_{x,p}>-0.27\;{\rm GeV}\,,
\label{STAR_cuts_on_protons_510}
\end{eqnarray}
as specified in \cite{ICHEP_STAR}.

In Table~\ref{tab:table_RHIC} we give the analog of
Table~\ref{tab:table_LHC} but for the STAR experiments.
\begin{table}[]
\centering
\caption{The integrated cross sections in nb for CEP of $f_{1}$ mesons 
in $pp$ collisions for the STAR experiments
for $|{\rm y_{M}}| < 0.7$ and when
in addition limitations on the outgoing protons are imposed;
see Eq.~(\ref{STAR_cuts_on_protons_200}) for $\sqrt{s}=200$~GeV
and Eq.~(\ref{STAR_cuts_on_protons_510}) for $\sqrt{s}=510$~GeV.
The parameter values for $(\varkappa',\varkappa'')$
are taken from (\ref{kappa_couplings_1285}) for the $f_{1}(1285)$
and from (\ref{kappa_couplings_1420}) for the $f_{1}(1420)$.
We have taken here the form factor (\ref{Fpompommeson_exp}) 
with $\Lambda_{E} = 0.7$~GeV.
The results without and with absorption effects are presented.}
\label{tab:table_RHIC}
\begin{tabular}{|c|c|l|c|c|c|c|}
\hline
$\sqrt{s}$ (GeV) & Meson  & Cuts  & Contribution & Parameters & $\sigma_{{\rm Born}}$~(nb) & $\sigma_{{\rm abs.}}$~(nb) \\ 
\hline
200 & $f_{1}(1285)$  & $|{\rm y_{M}}| < 0.7$,  & $(2,2)$ & Eq.~(\ref{parameters_1285_a})  
& 204.2 & 127.5 \\ 
&& and Eq.~(\ref{STAR_cuts_on_protons_200})
& $(4,4)$ & Eq.~(\ref{parameters_1285_b})                                              
& 163.7 & 103.1 \\ 
&&& $(\varkappa',\varkappa'')$ & $\varkappa''/\varkappa' = -6.25\;{\rm GeV}^{-2}$
& \;\,88.5 & \;\,76.1 \\ 
&&& $(\varkappa',\varkappa'')$ & $\varkappa''/\varkappa' = -2.44\;{\rm GeV}^{-2}$
& 178.8 & 122.8 \\ 
&&& $(\varkappa',\varkappa'')$ & $\varkappa''/\varkappa' = -1.0\;{\rm GeV}^{-2}$
& 210.5 & 136.4 \\ 
\hline
200 & $f_{1}(1420)$  & $|{\rm y_{M}}| < 0.7$,  & $(2,2)$ & Eq.~(\ref{parameters_1420_a})  
& \;\,50.0 & \;\,31.3 \\ 
&& and Eq.~(\ref{STAR_cuts_on_protons_200})
& $(\varkappa',\varkappa'')$ & $\varkappa''/\varkappa' = -1.0\;{\rm GeV}^{-2}$
& \;\,50.3 & \;\,31.9 \\ 
\hline
510 & $f_{1}(1285)$  & $|{\rm y_{M}}| < 0.7$,  & $(2,2)$ & Eq.~(\ref{parameters_1285_a})  
& 127.5 & \;\,27.8 \\ 
&& and Eq.~(\ref{STAR_cuts_on_protons_510})
& $(4,4)$ & Eq.~(\ref{parameters_1285_b})                                              
& 111.5 & \;\,27.0 \\ 
&&& $(\varkappa',\varkappa'')$ & $\varkappa''/\varkappa' = -6.25\;{\rm GeV}^{-2}$
& \;\,98.9 & \;\,89.4 \\ 
&&& $(\varkappa',\varkappa'')$ & $\varkappa''/\varkappa' = -2.44\;{\rm GeV}^{-2}$
& \;\,41.0 & \;\,29.6 \\ 
&&& $(\varkappa',\varkappa'')$ & $\varkappa''/\varkappa' = -1.0\;{\rm GeV}^{-2}$
& \;\,90.3 & \;\,26.3 \\ 
\hline
510 & $f_{1}(1420)$  & $|{\rm y_{M}}| < 0.7$,  & $(2,2)$ & Eq.~(\ref{parameters_1420_a})  
& \;\,30.7 & \;\,\;\,6.8 \\ 
&& and Eq.~(\ref{STAR_cuts_on_protons_510})
& $(\varkappa',\varkappa'')$ & $\varkappa''/\varkappa' = -1.0\;{\rm GeV}^{-2}$
& \;\,21.3 & \;\,\;\,6.2 \\ 
\hline
\end{tabular}
\end{table}

In Fig.~\ref{fig:STAR_200} we show as an example various predictions 
for $f_{1}(1285)$ CEP at $\sqrt{s} = 200$~GeV, 
at $|{\rm y_{M}}| < 0.7$,
and with extra cuts on the leading protons (\ref{STAR_cuts_on_protons_200}).
The experimental cuts have crucial influence on the shape of
the differential distributions. 
In particular, the result that the distributions
(nearly) vanish for certain values of the variables
$\phi_{pp}$, $p_{t,M}$ and ${\rm dP_{t}}$ 
is caused by the specific cuts (\ref{STAR_cuts_on_protons_200}).
\begin{figure}[!ht]
\includegraphics[width=0.49\textwidth]{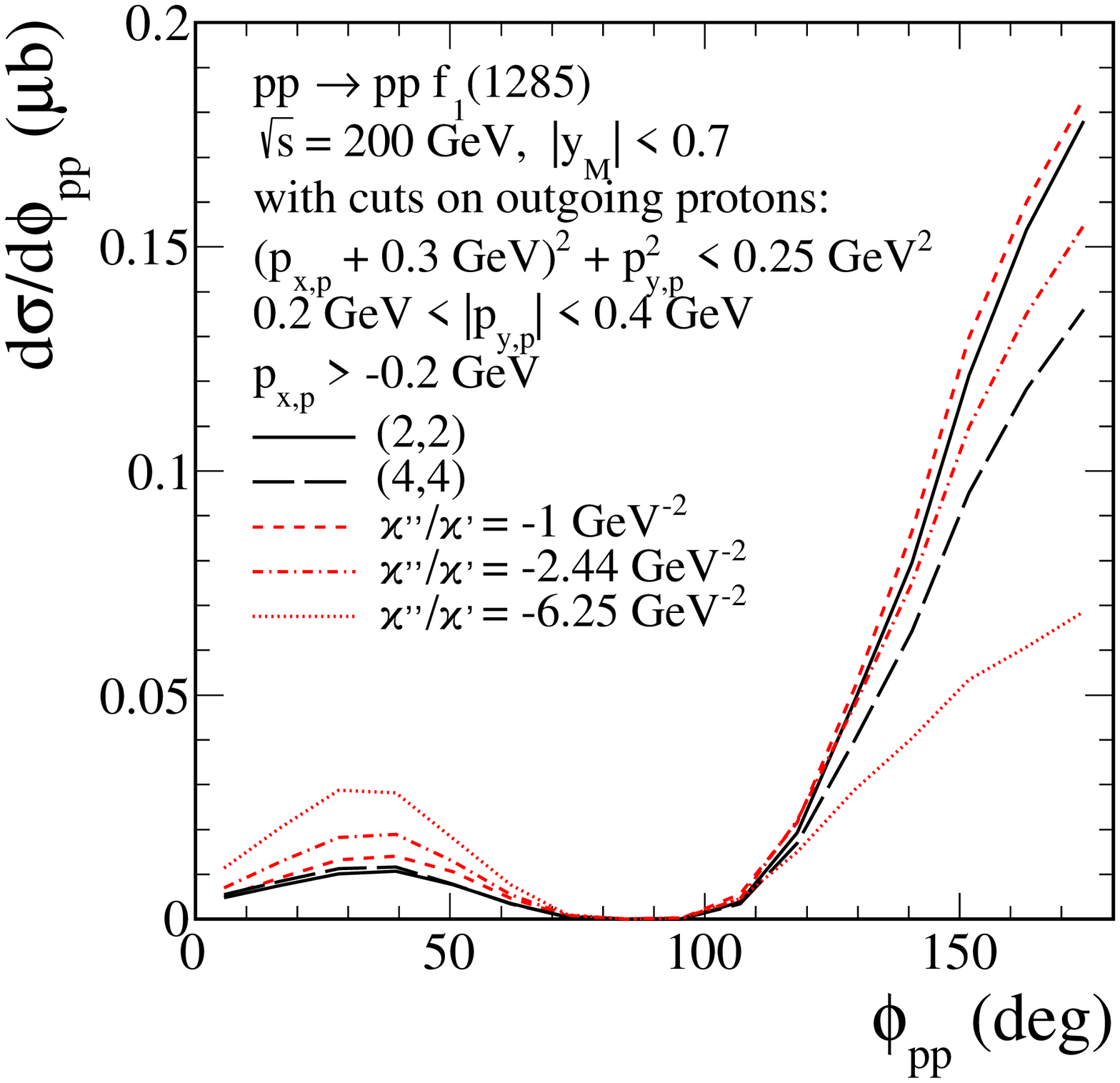}
\includegraphics[width=0.49\textwidth]{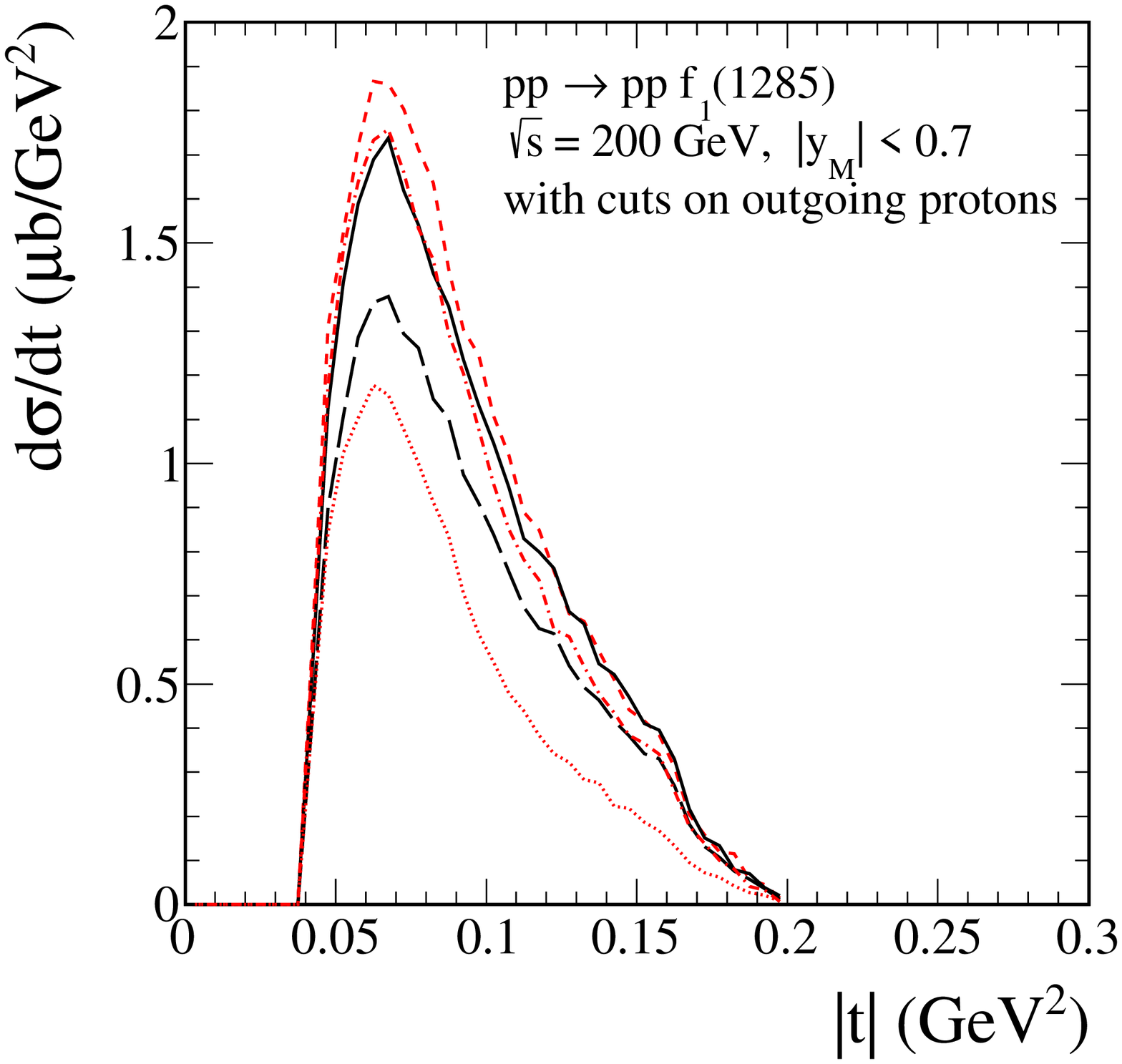}
\includegraphics[width=0.49\textwidth]{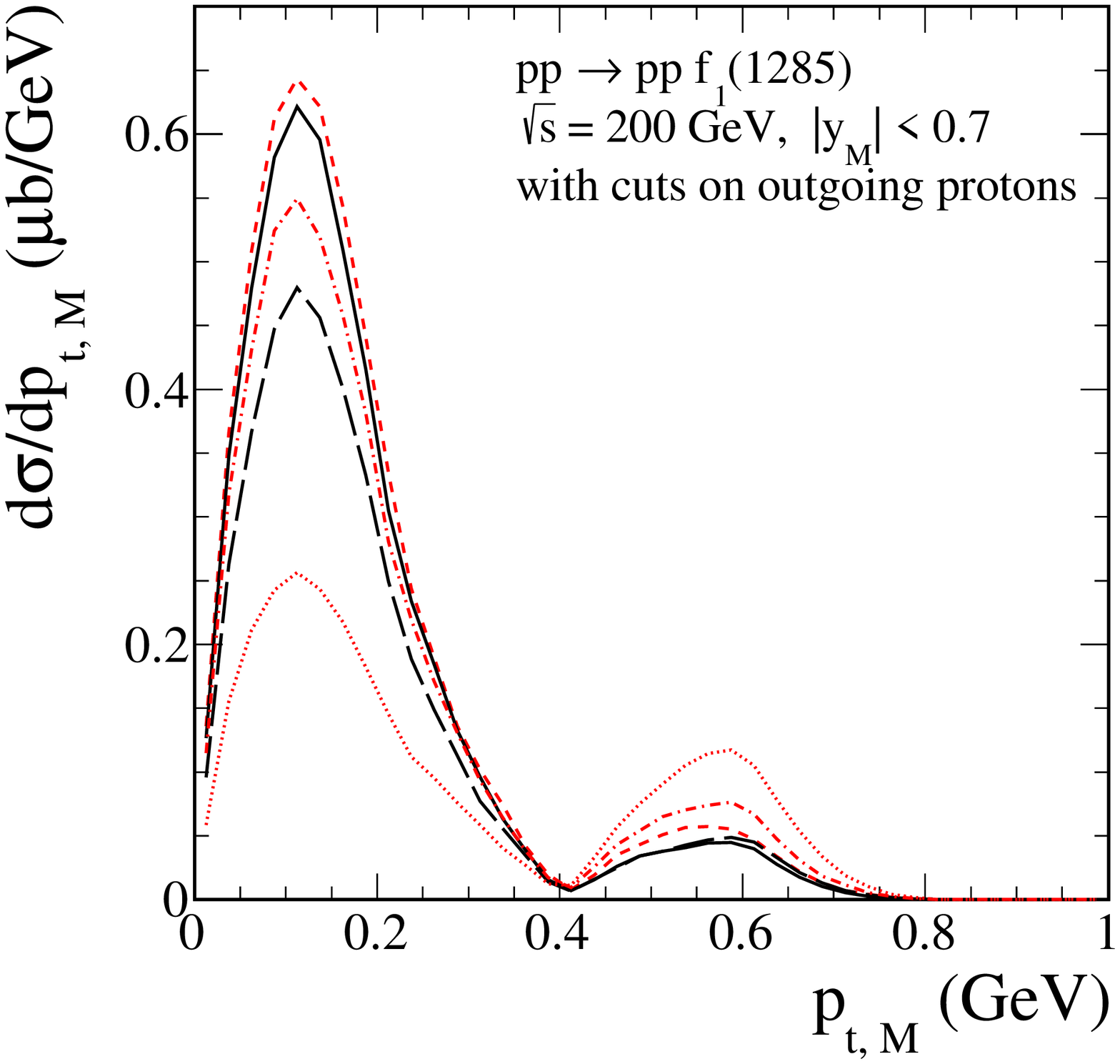}
\includegraphics[width=0.49\textwidth]{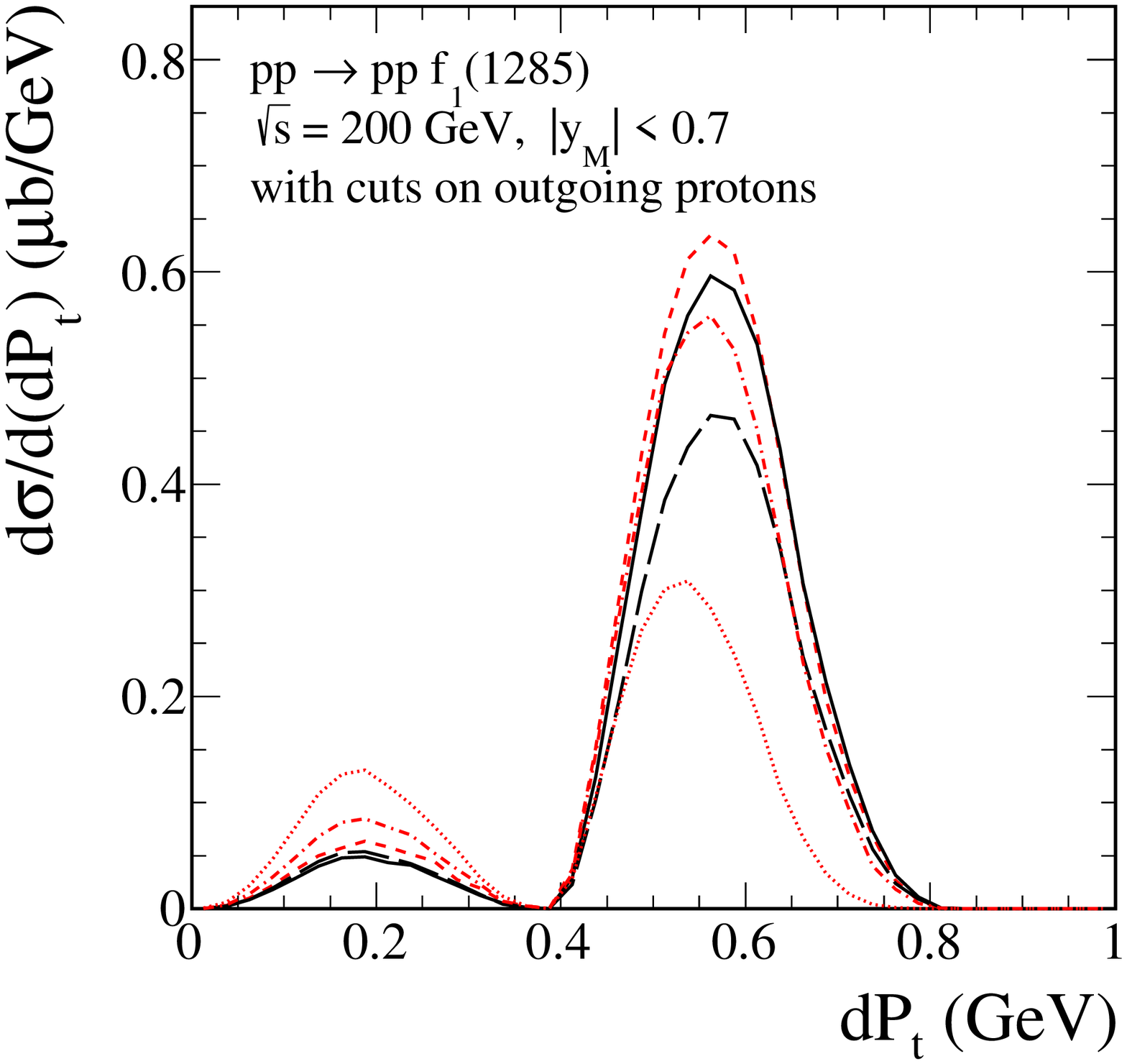}
\caption{\label{fig:STAR_200}
\small
The differential cross sections for
the $f_{1}(1285)$ production at $\sqrt{s} = 200$~GeV for
$|{\rm y_{M}}| < 0.7$ and with cuts on the leading protons
specified in (\ref{STAR_cuts_on_protons_200}).
The meaning of the lines is the same as 
in Fig.~\ref{fig:ATLAS-ALFA}.
The absorption effects are included in all the calculations.}
\end{figure}

In Fig.~\ref{fig:STAR_510} 
we show our predictions for $f_{1}(1285)$ CEP at $\sqrt{s} = 510$~GeV, 
$|{\rm y_{M}}| < 0.7$,
and with extra cuts on the leading protons (\ref{STAR_cuts_on_protons_510}).
The suppression of the differential cross sections
$d\sigma/ d\phi_{pp}$ close to $90^\circ$
is due to the specific cuts (\ref{STAR_cuts_on_protons_510})
applied to the forward scattered protons.
The general situation 
for $d\sigma/ d\phi_{pp}$ and $d\sigma/ dt$ at $\sqrt{s} = 510$~GeV 
is similar to that of $\sqrt{s} = 200$~GeV 
but there are some noticeable differences 
due to the different cuts on the outgoing protons.
A clear difference is seen for the option 
$\varkappa''/\varkappa' = -6.25\;{\rm GeV}^{-2}$.
This is due to the kinematics-dependent absorption effects.
\begin{figure}[!ht]
\includegraphics[width=0.49\textwidth]{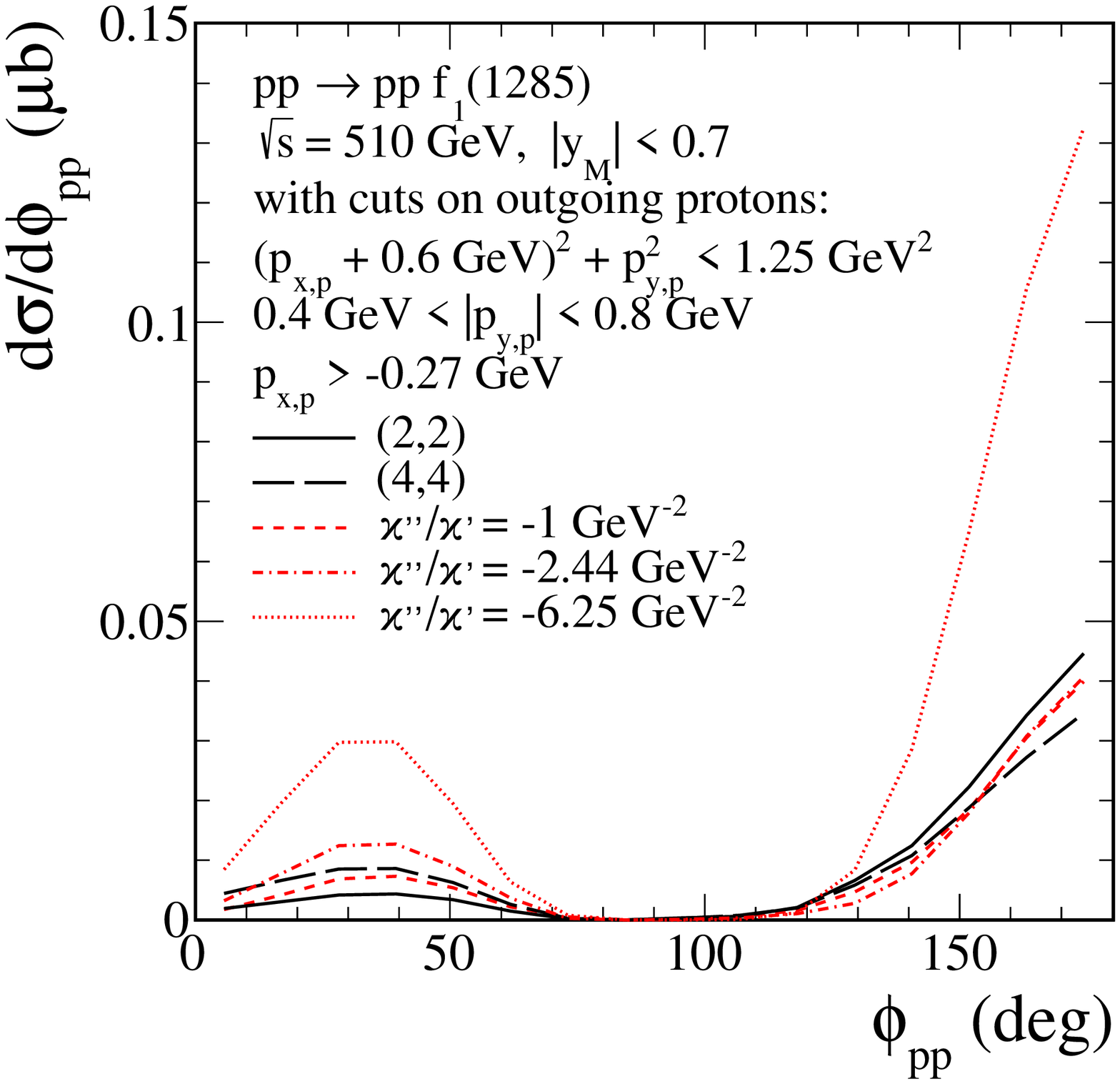}
\includegraphics[width=0.49\textwidth]{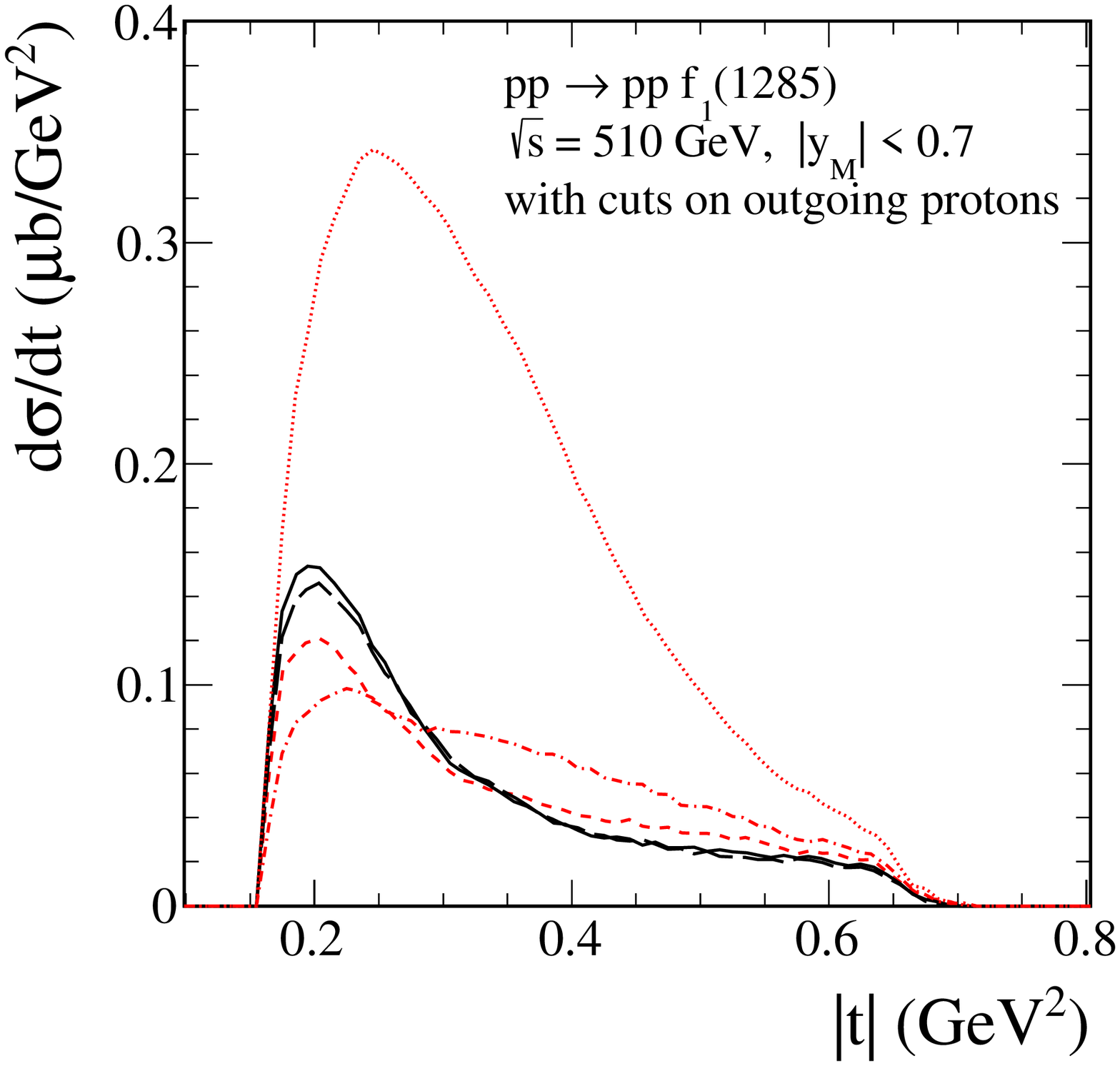}
\caption{\label{fig:STAR_510}
\small
The differential cross sections for
the $f_{1}(1285)$ production at $\sqrt{s} = 510$~GeV for
$|{\rm y_{M}}| < 0.7$ and with cuts on the leading protons
specified in (\ref{STAR_cuts_on_protons_510}).
The meaning of the lines is the same as 
in Fig.~\ref{fig:ATLAS-ALFA}.
The absorption effects are included in all the calculations.}
\end{figure}

\section{Conclusions}
\label{sec:conclusions}

In this paper, we have discussed in detail 
the exclusive central production of the pseudovector 
$f_{1}(1285)$ and $f_{1}(1420)$ mesons in proton-proton collisions.
The calculations for the $pp \to ppf_{1}(1285)$ and 
$pp \to ppf_{1}(1420)$ reactions
have been performed in the tensor-pomeron approach \cite{Ewerz:2013kda}.
In general, two $\Pom \Pom f_{1}$ couplings with different orbital
angular momentum and spin of two ``pomeron particles'' are possible,
namely $(l,S) = (2,2)$ and $(4,4)$.
We have presented explicitly amplitudes and formulas 
for the $\Pom \Pom f_{1}$ vertices
as derived from corresponding coupling Lagrangians.
Two different approaches for the $\Pom \Pom f_{1}$ coupling have been 
considered.
\begin{enumerate}
 \item[(1)] 
 In the first approach, two independent $\Pom \Pom f_{1}$ coupling
 constants, $g'_{\Pom \Pom f_{1}}$ and $g''_{\Pom \Pom f_{1}}$ 
 that correspond to the $(l,S) = (2,2)$ and $(l,S) = (4,4)$ couplings
 [see Eqs.~(\ref{vertex_pompomf1_A}) 
 and (\ref{vertex_pompomf1_B}), respectively], 
 not known \textit{a priori} 
 as they are of nonperturbative origin,
 have been fitted to existing data from the WA102 experiment. 
 A reasonable agreement with the WA102 data can be obtained 
 with either a pure $(l,S) = (2,2)$ or a pure $(l,S) = (4,4)$
 coupling.
 \item[(2)] The second approach is based on holographic QCD, namely
 the (chiral) Sakai-Sugimoto model, 
 where the pomeron-pomeron-$f_{1}$ couplings (\ref{A27}) and (\ref{A28})
 are obtained from a Chern-Simons action representing
 the mixed axial-gravitational anomaly of QCD. 
 This also involves two coupling constants,
 with a prediction for their ratio 
 in terms of the Kaluza-Klein mass scale of the model as given by 
 (\ref{kapparatiorange}).
 Comparing the $\phi_{pp}$ distribution 
 for different values of this ratio
 confirms the sign of this ratio as predicted 
 by the Sakai-Sugimoto model, but not its magnitude.
 However, freely fitting the magnitude of the couplings, 
 reasonable agreement with the WA102 data is again obtained.
\end{enumerate}

Assuming that the WA102 data are already dominated by pomeron exchanges,
we have presented various predictions for experiments at the 
RHIC and the LHC.
The total cross sections and several differential distributions
for the $pp \to pp f_{1}(1285)$ reaction have been presented.
In our opinion the $\pi^{+}\pi^{-}\pi^{+}\pi^{-}$ channel seems the best
to observe $f_1(1285)$ for both the RHIC and the LHC experiments.
We have shown that independent of the $\Pom \Pom f_{1}$ coupling decomposition
the cross section for the 
$pp \to pp (f_{1}(1285) \to \pi^{+}\pi^{-}\pi^{+}\pi^{-})$
reaction is much larger than for 
the $pp \to pp (f_{2}(1270) \to \pi^{+}\pi^{-}\pi^{+}\pi^{-})$ reaction.
As the $f_1(1285)$ has a much narrower width than the $f_2(1270)$ 
it would be seen in the mass distribution
as a narrow peak on a somewhat broader bump corresponding 
to the $f_2(1270)$.

The question can be asked if CEP of the $f_{1}(1285)$
and $f_{1}(1420)$ mesons may be confounded in experiments with CEP
of $\eta$-type mesons which are nearby in mass.
The $f_{1}(1285)$ and the $\eta(1295)$ are close in mass.
For the $f_{1}(1420)$ we have the $\eta(1405)$ and the $\eta(1475)$
as potential background candidates.~\footnote{We thank
a referee for raising this question and for pointing out 
Ref.~\cite{Achasov:2011xc} where some puzzles of $\eta(1475)$ and $f_{1}(1420)$
production and decay reactions are discussed.}

Let us first discuss the $f_{1}(1285)$ and $\eta(1295)$ issue.
These two mesons have a common decay mode ($\eta \pi \pi$) but
only the $f_{1}(1285)$ decays to $4 \pi$ 
and $K \bar{K} \pi$ \cite{Tanabashi:2018oca}.
Thus, concentrating in an experiment on these latter final states
there can be no confusion between the $f_{1}(1285)$ and the $\eta(1295)$.

For the $f_{1}(1420)$ and the nearby $f_{1}(1405)$ and $\eta(1475)$ mesons
things are more complicated.
The channel where the $f_{1}(1420)$ is to be observed is $K \bar{K} \pi$,
and this channel is also prominent for the $\eta(1405)$ and 
$\eta(1475)$ decays. Thus, here experimentalists will have to
rely on precise mass measurements and partial-wave analyses
in order to distinguish $f_{1}$- and $\eta$-type resonances.
Now we discuss that the distributions in the azimuthal angle
$\phi_{pp}$ between the transverse momenta of the outgoing protons
may also be used to disentangle $f_{1}$ and $\eta$ contributions.
In Appendix~\ref{sec:appendixE} we show that for CEP of an $\eta$-type
meson at high energies $\sqrt{s}$ the $\phi_{pp}$ distribution
must vanish for $\phi_{pp} = 0$ and $\phi_{pp} = \pi$.
For CEP of an $f_{1}$ meson there is no such restriction and,
indeed, the $\phi_{pp}$ distributions measured by the WA102 
Collaboration are nonzero
for $\phi_{pp} = 0$ and $\phi_{pp} = \pi$; 
see Figs.~\ref{fig:1}, \ref{fig:2aux}, and \ref{fig:1420_ff}.

Our predictions can be tested by the STAR
Collaboration at RHIC and by all collaborations
(ALICE, ATLAS, CMS, LHCb) working at the LHC.

In all cases considered we have included absorption effects.
We have found that the absorption effects strongly depend on kinematics,
i.e., also on experimental cuts, 
as well as on the type of the $\Pom \Pom f_1$ coupling used in the
calculation.
Different tensorial couplings discussed in the present paper lead 
to different dependences on $t_1$ and $t_2$ which are crucial for 
the size of absorption effects.
The effect of absorption was not the primary aim of this study;
therefore, the discussion of this point was kept rather short in our
present paper.

To summarize, we think that a study of CEP of the axial vector mesons
$f_{1}$ should be quite rewarding for experimentalists.
We have analysed in detail the results of the WA102 experiment
which worked at $\sqrt{s} = 29.1$~GeV,
and we have shown that we get a good description of the results
with the pomeron-pomeron fusion mechanism.
Such studies could be extended, for instance by the COMPASS experiment
\cite{Abbon:2014aex,Ketzer:2019wmd}, 
where presumably one could study the influence of
reggeon-pomeron and reggeon-reggeon fusion terms.
At high energies, at RHIC and LHC, pomeron-pomeron fusion
is expected to dominate.
We have given predictions for CEP of $f_{1}$ mesons there.
Comparing them with future experimental results
should allow a good determination of the $\Pom \Pom f_{1}$
coupling constants. These are nonperturbative QCD parameters.
Their theoretical calculation is a challenge.
The holographic methods applied to QCD already give some predictions here,
as we have shown in our paper.
We can envisage a fruitful interplay of experiment and theory
in this field in the future leading finally to a satisfactory
picture of the couplings of two pomerons 
to the axial vector $f_{1}$ mesons
studied here and, quite generally, to single mesons.

\appendix

\section{The coupling of an $f_1$-type meson to two pomerons}
\label{sec:appendixA}

Here we study the coupling of a meson $f_{1}$ with $I^{G}J^{PC} = 0^{+}1^{++}$
to two tensor pomerons.
We use the relations for the tensor pomeron 
from \cite{Ewerz:2013kda, Lebiedowicz:2013ika}.

In Appendix~A of \cite{Lebiedowicz:2013ika} the fictitious reaction of two
``real spin-2 pomerons`` annihilating to a meson was studied.
This was done in order to get an idea what type of 
pomeron-pomeron-meson ($\Pom \Pom M$) couplings we would have to expect.
Looking at Table~6 of \cite{Lebiedowicz:2013ika} we see
that for the production of a $J^{P} = 1^{+}$ meson we can have the following values
of angular momentum $l$ and total spin $S$ 
of the two tensor pomerons:
\begin{eqnarray}
(l,S) = (2,2)\,, \;(4,4)\,.
\label{A1}
\end{eqnarray}
We find only these two possibilities.

The task is now to construct $\Pom \Pom f_{1}$ coupling
Lagrangians which, applied to the above
``real spin-2 pomeron'' annihilation,
give the $(l,S) = (2,2)$ and $(4,4)$ amplitudes.
We emphasize that such constructions are not unique.
We give in our paper, in this and the following appendix,
two possibilities for such constructions and
we discuss their relations.
Here we shall rely on the experience gained
with the construction of pomeron-pomeron-meson couplings
in \cite{Ewerz:2013kda,Ewerz:2016onn,Lebiedowicz:2013ika, Lebiedowicz:2014bea,Lebiedowicz:2016ioh,Lebiedowicz:2016ryp,Lebiedowicz:2018eui,Lebiedowicz:2016zka,Lebiedowicz:2018sdt,Lebiedowicz:2019jru,Lebiedowicz:2019boz}.
We want to couple two spin 2 pomeron fields $\Pom_{\kappa \lambda}$
to the $f_{1}$ vector field $U_{\alpha}$
which is, in equations, conveniently represented by an
antisymmetric second-rank tensor field
$\p_{\alpha} U_{\beta} - \p_{\beta} U_{\alpha}$.
The $l$ values of the couplings should be reflected
by $l$ derivatives.
Using these heuristic principles it is not difficult
to write down $\Pom \Pom f_{1}$ couplings which
fulfil all required properties.

In the following we shall first construct the $\Pom \Pom f_{1}$ coupling
corresponding to $(l,S) = (2,2)$.
For this we define the following rank 8 tensor function:
\begin{eqnarray}
\Gamma^{(8)}_{\kappa \lambda, \rho \sigma, \mu \nu, \alpha \beta} &=&
g_{\kappa \rho} g_{\mu \sigma} \varepsilon_{\lambda \nu \alpha \beta} +
g_{\lambda \rho} g_{\mu \sigma} \varepsilon_{\kappa \nu \alpha \beta} +
g_{\kappa \sigma} g_{\mu \rho} \varepsilon_{\lambda \nu \alpha \beta} +
g_{\lambda \sigma} g_{\mu \rho} \varepsilon_{\kappa \nu \alpha \beta} \nonumber \\
&&+
g_{\kappa \rho} g_{\mu \lambda} \varepsilon_{\sigma \nu \alpha \beta} +
g_{\sigma \kappa} g_{\mu \lambda} \varepsilon_{\rho \nu \alpha \beta} +
g_{\rho \lambda} g_{\mu \kappa} \varepsilon_{\sigma \nu \alpha \beta} +
g_{\sigma \lambda} g_{\mu \kappa} \varepsilon_{\rho \nu \alpha \beta}\nonumber \\
&&-
g_{\kappa \lambda} g_{\mu \rho} \varepsilon_{\sigma \nu \alpha \beta} -
g_{\kappa \lambda} g_{\mu \sigma} \varepsilon_{\rho \nu \alpha \beta} -
g_{\kappa \mu} g_{\rho \sigma} \varepsilon_{\lambda \nu \alpha \beta} -
g_{\lambda \mu} g_{\rho \sigma} \varepsilon_{\kappa \nu \alpha \beta}\nonumber \\
&&+ (\mu \leftrightarrow \nu)\,.
\label{A2}
\end{eqnarray}
For the Levi-Civita symbol we use the normalisation 
$\varepsilon_{0 1 2 3} = +1$.

It can be checked that $\Gamma^{(8)}$ satisfies the following relations:
\begin{eqnarray}
&&\Gamma^{(8)}_{\kappa \lambda, \rho \sigma, \mu \nu, \alpha \beta} =
\Gamma^{(8)}_{\lambda \kappa, \rho \sigma, \mu \nu, \alpha \beta} =
\Gamma^{(8)}_{\kappa \lambda, \sigma \rho, \mu \nu, \alpha \beta} \nonumber \\
&&\qquad \qquad \qquad =
\Gamma^{(8)}_{\kappa \lambda, \rho \sigma, \nu \mu, \alpha \beta} =
\Gamma^{(8)}_{\rho \sigma, \kappa \lambda, \mu \nu, \alpha \beta} =
-\Gamma^{(8)}_{\kappa \lambda, \rho \sigma, \mu \nu, \beta \alpha}\,;
\label{A4}\\ 
&&\Gamma^{(8)}_{\kappa \lambda, \rho \sigma, \mu \nu, \alpha \beta} 
\,g^{\kappa \lambda} = 0\,, \nonumber\\
&&\Gamma^{(8)}_{\kappa \lambda, \rho \sigma, \mu \nu, \alpha \beta} 
\,g^{\rho \sigma} = 0\,, \nonumber\\
&&\Gamma^{(8)}_{\kappa \lambda, \rho \sigma, \mu \nu, \alpha \beta} 
\,g^{\mu \nu} = 0\,.
\label{A5}
\end{eqnarray}

Now we define the $\Pom \Pom f_{1}$ coupling corresponding to $(l,S) = (2,2)$ as follows
\begin{eqnarray}
{\cal L}'_{\Pom \Pom f_{1}}(x) = \frac{g'_{\Pom \Pom f_{1}}}{32\,M_{0}^{2}}
\Big( \Pom_{\kappa \lambda}(x) 
\Big(\twosidep{\mu} \,\twosidep{\nu} \Big) 
\Pom_{\rho \sigma}(x) \Big)
\Big( \p_{\alpha} U_{\beta}(x) - \p_{\beta} U_{\alpha}(x) \Big)\,
\Gamma^{(8)\,\kappa \lambda, \rho \sigma, \mu \nu, \alpha \beta}\,.\qquad
\label{A6}
\end{eqnarray}
Here $\Pom_{\kappa \lambda}(x)$ is the effective field of the pomeron
and $U_{\alpha}(x)$ the field of the $f_{1}$ meson.
Furthermore we have introduced, for dimensional reasons,
in (\ref{A6}) a factor $M_{0}^{-2}$ with $M_{0} = 1$~GeV, 
and then $g'_{\Pom \Pom f_{1}}$ is a dimensionless coupling constant.
The asymmetric derivative has the form $\twosidep{\mu} = \rightsidep{\mu} - \leftsidep{\mu}$.
The $\Pom$ effective field satisfies the identities
\begin{eqnarray}
&&\Pom_{\kappa \lambda}(x) = \Pom_{\lambda \kappa}(x)\,, \nonumber \\
&&g^{\kappa \lambda}\, \Pom_{\kappa \lambda}(x) = 0\,.
\label{A7}
\end{eqnarray}
From (\ref{A6}) we get the ``bare'' $\Pom \Pom f_{1}$ vertex (\ref{vertex_pompomf1_A}).

Now we shall set up the $\Pom \Pom f_{1}$ coupling corresponding to 
$(l,S) = (4,4)$:
\begin{eqnarray}
{\cal L}''_{\Pom \Pom f_{1}}(x) &=& \frac{g''_{\Pom \Pom f_{1}}}{24 \cdot 32 \cdot M_{0}^{4}}
\Big( \Pom_{\kappa \lambda}(x)
\Big( \twosidep{\mu_{1}} \, \twosidep{\mu_{2}} \,
      \twosidep{\mu_{3}} \, \twosidep{\mu_{4}} \Big) 
\Pom_{\rho \sigma}(x) \Big)
\Big( \p_{\alpha} U_{\beta}(x) - \p_{\beta} U_{\alpha}(x) \Big) \nonumber \\
&& \times \Gamma^{(10)\,\kappa \lambda, \rho \sigma, \mu_{1} \mu_{2} \mu_{3} \mu_{4}, \alpha \beta}\,.
\label{A20}
\end{eqnarray}
Here we define the rank 10 tensor function
\begin{eqnarray}
\Gamma^{(10)}_{\kappa \lambda, \rho \sigma, \mu_{1} \mu_{2} \mu_{3} \mu_{4}, \alpha \beta} &=&
\left.
\Big\lbrace \Bigl[ 
\Bigl( g_{\kappa \mu_{1}} g_{\lambda \mu_{2}} 
-\frac{1}{4} g_{\kappa \lambda} g_{\mu_{1} \mu_{2}} \Bigr)
\Bigl( g_{\rho \mu_{3}} \varepsilon_{\sigma \mu_{4} \alpha \beta}
-\frac{1}{4} g_{\rho \sigma} \varepsilon_{\mu_{3} \mu_{4} \alpha \beta} \Bigr) 
\right. \nonumber \\ 
&& \left. + (\kappa \leftrightarrow \lambda) + (\rho \leftrightarrow \sigma) 
          + (\kappa \leftrightarrow \lambda, \rho \leftrightarrow \sigma) \Bigr] 
+ (\kappa, \lambda) \leftrightarrow (\rho, \sigma) 
\Big\rbrace 
\right. \nonumber \\ 
&&+\;{\rm all \;permutation \;of}\;\mu_{1}, \mu_{2}, \mu_{3}, \mu_{4} \,.
\label{A21}
\end{eqnarray}
$\Gamma^{(10)}$ (\ref{A21}) has the following properties:
\begin{eqnarray}
&&\Gamma^{(10)}_{\kappa \lambda, \rho \sigma, \mu_{1} \mu_{2} \mu_{3} \mu_{4}, \alpha \beta} =
\Gamma^{(10)}_{\lambda \kappa, \rho \sigma, \mu_{1} \mu_{2} \mu_{3} \mu_{4}, \alpha \beta} =
\Gamma^{(10)}_{\kappa \lambda, \sigma \rho, \mu_{1} \mu_{2} \mu_{3} \mu_{4}, \alpha \beta}
\nonumber \\ 
&&\qquad \qquad  \qquad \qquad \;=
\Gamma^{(10)}_{\rho \sigma, \kappa \lambda, \mu_{1} \mu_{2} \mu_{3} \mu_{4}, \alpha \beta} =
-\Gamma^{(10)}_{\kappa \lambda, \rho \sigma, \mu_{1} \mu_{2} \mu_{3} \mu_{4}, \beta \alpha}\,,
\label{A22}\\
&&\Gamma^{(10)}_{\kappa \lambda, \rho \sigma, \mu_{1} \mu_{2} \mu_{3} \mu_{4}, \alpha \beta} 
\;{\rm is \; totally \; symmetric \; in} \;\mu_{1}, \mu_{2}, \mu_{3}, \mu_{4}\,,
\label{A23}\\ 
&&\Gamma^{(10)}_{\kappa \lambda, \rho \sigma, \mu_{1} \mu_{2} \mu_{3} \mu_{4}, \alpha \beta} 
\,g^{\kappa \lambda} = 0\,, \nonumber\\
&&\Gamma^{(10)}_{\kappa \lambda, \rho \sigma, \mu_{1} \mu_{2} \mu_{3} \mu_{4}, \alpha \beta} 
\,g^{\rho \sigma} = 0\,.
\label{A24}
\end{eqnarray}

In (\ref{A20}) $g''_{\Pom \Pom f_{1}}$ is a dimensionless coupling constant.
From (\ref{A20})--(\ref{A24}) we get the ``bare'' $\Pom \Pom f_{1}$ vertex (\ref{vertex_pompomf1_B}).

\section{Different forms for the $\Pom \Pom f_1$ coupling as obtained in holographic QCD}
\label{sec:appendixB}
In (\ref{A6}) and (\ref{A20}) of Appendix~\ref{sec:appendixA} 
we have given a possible form for the $\Pom \Pom f_{1}$ couplings.
In the holographic framework another 
form is obtained.
In the Sakai-Sugimoto model \cite{Sakai:2004cn,Sakai:2005yt},
the coupling of singlet pseudoscalar and axial-vector mesons 
to two tensor glueballs is determined by
the gravitational CS action 
(describing axial-gravitational anomalies), 
as given in Eq.~(59) of \cite{Anderson:2014jia},
\begin{eqnarray}\label{SCS}
S_{{\rm CS}}&\supset & \frac{N_{c}}{1536\pi^{2}}\int d^{5}x\epsilon^{MNPQR}\text{Tr}(A_{M})R_{NPST}R_{QR}^{\quad TS}.
\end{eqnarray}
The (singlet component of the) axial-vector meson is contained in $\text{Tr}(A_{\mu})=A_{\mu}^{(0)}=U_\mu(x)\psi(Z)$, leading to
\begin{eqnarray}
S_{{\rm CS}}&\supset & \frac{N_{c}}{384\pi^{2}}\sqrt{\frac{N_{f}}{2}}\int d^{5}x\epsilon^{\mu\nu\rho\sigma}A_{\mu}^{(0)}R_{Z\nu ST}R_{\rho\sigma}^{\quad TS},
\end{eqnarray}
where $Z$ refers to the holographic direction.

Five-dimensional gravitons correspond to four-dimensional tensor glueballs, and their coupling to $f_1$
is obtained by expanding this term to second order in transverse-traceless metric perturbations
and integrating over radial wave functions.
Using the same notation as in Appendix~\ref{sec:appendixA} we derive the coupling Lagrangians,
\begin{eqnarray}
&&{\cal L}_{\rm CS}'(x) = \varkappa' \,U_{\alpha}(x)\,\varepsilon^{\alpha \beta \gamma \delta}\,
\Pom^{\mu}_{\;\;\beta}(x)\, \p_{\delta}\Pom_{\gamma \mu}(x)\,,
\label{A27} \\
&&{\cal L}_{\rm CS}''(x) = \varkappa'' \,U_{\alpha}(x)\,\varepsilon^{\alpha \beta \gamma \delta}\,
\Bigl( \p_{\nu}\Pom^{\mu}_{\;\;\beta}(x) \Bigr) 
\Bigl( \p_{\delta}\p_{\mu}\Pom^{\nu}_{\;\;\gamma}(x) - \p_{\delta}\p^{\nu}\Pom_{\gamma \mu}(x) \Bigr)\,,
\label{A28}
\end{eqnarray}
where
\begin{eqnarray}\label{kprimes}
 \varkappa' &=& -\frac{4.872\,\mathcal{N}\sqrt{N_{f}}}{\sqrt{N_{c}^{3}\lambda^{3}}} \,,\\
 \varkappa'' &=& \frac{27.434\,\mathcal{N}\sqrt{N_{f}}}{M_{\text{KK}}^{2}\sqrt{N_{c}^{3}\lambda^{3}}}
\end{eqnarray}
and $\mathcal{N}$ is a normalization constant that 
we leave undetermined because of the ambiguities \cite{Anderson:2014jia} in the reggeization of the tensor glueball into pomerons
(it would be unity for a purely flavour-singlet axial-vector meson when $\Pom_{\mu\nu}$ was replaced by the tensor glueball $T_{\mu\nu}$
normalized as in \cite{Brunner:2015oqa}).

The Sakai-Sugimoto model has two free parameters, a Kaluza-Klein mass scale $M_{\text{KK}}$ and the dimensionless 't Hooft
coupling $\lambda$ at this scale. Both, $\lambda$ and the normalization $\mathcal{N}$, drop out of the ratio
between the two $\Pom \Pom f_{1}$ couplings,
\begin{equation}\label{kapparatio_with_MKK}
 \frac{\varkappa''}{\varkappa'}=-\frac{5.631}{M_{\text{KK}}^{2}}\,.
\end{equation}
Usually \cite{Sakai:2004cn,Sakai:2005yt} $M_{\text{KK}}$ is fixed by matching the mass of the lowest vector meson to that of the physical $\rho$ meson,
leading to $M_{\text{KK}}=949\,\text{MeV}$.
However, this choice leads to a tensor glueball mass which is too low, $M_T \approx 1487\,\text{MeV}$.
The standard pomeron trajectory (\ref{pomtrajectory}) corresponds to a tensor glueball mass of $M_T\approx 1917.5\,\text{MeV}$,
whereas lattice gauge theory \cite{Morningstar:1999rf}
indicates a tensor glueball mass $M_T\simeq 2400\,\text{MeV}$ (or even higher \cite{Gregory:2012hu}).
The corresponding choices for $M_{\text{KK}}$
give ${\varkappa''}/{\varkappa'}=-\{6.25,3.76,2.44\}\,\text{GeV}^{-2}$, which motivates the range (\ref{kapparatiorange}) considered in the main text.


The ``bare'' vertices obtained from the coupling Lagrangians 
(\ref{A27}) and (\ref{A28}) read as follows:
\begin{eqnarray}
&&i\Gamma_{\kappa \lambda,\rho \sigma,\alpha}'^{\,{\rm CS}}(q_{1},q_{2})\mid_{\rm{bare}} =
\varkappa'\, \varepsilon_{\alpha \beta \gamma \delta}
\left( q_{1}^{\delta} \,g^{\kappa' \gamma} g^{\lambda' \rho'} g^{\sigma' \beta}
      +q_{2}^{\delta} \,g^{\kappa' \sigma'} g^{\lambda' \beta} g^{\rho' \gamma} \right)
\tilde{R}_{\kappa \lambda \kappa' \lambda'}\,\tilde{R}_{\rho \sigma \rho' \sigma'}\,,\quad
\label{A40}\\
&&i\Gamma_{\kappa \lambda,\rho \sigma,\alpha}''^{\,{\rm CS}}(q_{1},q_{2})\mid_{\rm{bare}} =
\varkappa''\, \varepsilon_{\alpha \lambda' \sigma' \delta} (q_{1} - q_{2})^{\delta}
\left[ q_{1 \rho'} q_{2 \kappa'} - (q_{1} \cdot q_{2}) g_{\kappa' \rho'} \right]
\tilde{R}_{\kappa \lambda}^{\;\;\;\;\kappa' \lambda'}\,\tilde{R}_{\rho \sigma}^{\;\;\;\;\rho' \sigma'}\,.\quad
\label{A66}
\end{eqnarray}
Here we define the tensor [unrelated to the Riemann tensor in (\ref{SCS})]
\begin{eqnarray}
\tilde{R}_{\mu \nu \kappa \lambda} = \frac{1}{2} g_{\mu \kappa} g_{\nu \lambda}
                           + \frac{1}{2} g_{\mu \lambda} g_{\nu \kappa}
                           - \frac{1}{4} g_{\mu \nu} g_{\kappa \lambda}\,.
\label{A40_aux}
\end{eqnarray}
In (\ref{A40}) and (\ref{A66}) we have taken out explicitly the traces in $(\kappa \lambda)$ and $(\rho \sigma)$.
The momenta and vector indices for these vertices
are oriented and distributed as in (\ref{vertex_pompomf1_A}) 
and (\ref{vertex_pompomf1_B}).

Now we consider the reaction (\ref{PomPom_to_f1}),
the fusion of two ``real pomerons'' (or two glueballs)
of mass $m$ giving an $f_{1}$ meson of mass squared $k^{2}$:
\begin{eqnarray}
&&\Pom(q_{1},\epsilon^{(1)}) + \Pom(q_{2},\epsilon^{(2)}) 
\to f_{1}(k,\epsilon)\,, \nonumber\\
&&q_{1} + q_{2} = k\,, \quad q_{1}^{2} = q_{2}^{2} = m^{2}\,.
\label{PomPom_to_f1_AppB}
\end{eqnarray}
Here $q_{1}$, $q_{2}$, and $\epsilon^{(1)}$, $\epsilon^{(2)}$
are the momenta and the polarisation tensors of the two
``real pomerons'', $k$ and $\epsilon$
are the momentum and the polarisation vector of the $f_{1}$.
We know from the results of Table~6 in Appendix~A of \cite{Lebiedowicz:2013ika}
that there are two independent amplitudes for the reaction
(\ref{PomPom_to_f1_AppB}).
Thus, for the reaction (\ref{PomPom_to_f1_AppB}) 
we expect to find an equivalence relation of the form
\begin{eqnarray}
{\cal L}'_{{\rm CS}} + {\cal L}''_{{\rm CS}} 
\wideestimates {\cal L}'_{\Pom \Pom f_{1}} 
             + {\cal L}''_{\Pom \Pom f_{1}}
\label{A72}
\end{eqnarray}
between the Lagrangians (\ref{A27}), (\ref{A28}), 
and (\ref{A6}), (\ref{A20}).
Of course, the respective coupling parameters must then satisfy
certain relations which we determined as
\begin{eqnarray}
g'_{\Pom \Pom f_{1}} &=& 
-\varkappa'\,\frac{M_{0}^{2}}{k^{2}}
-\varkappa''\,\frac{M_{0}^{2}(k^{2}-2 m^{2})}{2k^{2}} \,,
\nonumber\\
g''_{\Pom \Pom f_{1}} &=& 
\varkappa''\,\frac{2 M_{0}^{4}}{k^{2}} \,.
\label{A75}
\end{eqnarray}

The proof of (\ref{A72}) and (\ref{A75}) is given at the end of this Appendix. We note that the relation (\ref{A75}) involves $k^{2}$,
the invariant mass squared of the resonance $f_{1}$.
For a narrow resonance of mass $m_{f_{1}}$ we can set
$k^{2} = m_{f_{1}}^{2} = \rm{const}$.
Then (\ref{A75}) gives a linear relation of the couplings
$(\varkappa',\varkappa'')$ and 
$(g'_{\Pom \Pom f_{1}},g''_{\Pom \Pom f_{1}})$
with constant coefficients.
For a broad resonance $k^{2}$ varies.
Then we see from (\ref{A75}) that for 
constants $(\varkappa',\varkappa'')$ the couplings
$g'_{\Pom \Pom f_{1}}$ and $g''_{\Pom \Pom f_{1}}$ contain
additional form factors depending on $k^{2}$ and vice versa.

The strict equivalence relation (\ref{A72}) does
not hold any more for the scattering process (\ref{2to3_reaction})
where two pomerons with invariant masses
$t_{1} < 0$ and $t_{2} < 0$, and in general $t_{1} \neq t_{2}$,
collide to give an $f_{1}$ meson; see Fig.~\ref{fig:diagram}.
But for small $|t_{1}|$ and $|t_{2}|$ we can expect
the following approximate equivalence to hold:
\begin{eqnarray}
g'_{\Pom \Pom f_{1}} &\approx &
-\varkappa'\,\frac{M_{0}^{2}}{k^{2}}
-\varkappa''\,\frac{M_{0}^{2}(k^{2}-t_{1}-t_{2})}{2k^{2}} \,,
\nonumber \\
g''_{\Pom \Pom f_{1}} &\approx &
\varkappa''\,\frac{2 M_{0}^{4}}{k^{2}} \,.
\label{A78}
\end{eqnarray}
The reverse reads
\begin{eqnarray}
\varkappa' &\approx &
-g'_{\Pom \Pom f_{1}}\,\frac{k^{2}}{M_{0}^{2}}
-g''_{\Pom \Pom f_{1}}\,\frac{k^{2}(k^{2}-t_{1}-t_{2})}{4M_{0}^{4}} \,,
\nonumber\\
\varkappa'' &\approx &
g''_{\Pom \Pom f_{1}}\,\frac{k^{2}}{2 M_{0}^{4}} \,.
\label{A77}
\end{eqnarray}
Again, taking e.g., $g'_{\Pom \Pom f_{1}}$ and 
$g''_{\Pom \Pom f_{1}}$
as constants $\varkappa'$ and $\varkappa''$
will contain suitable form factors and vice versa.

We have made a numerical investigation of the above 
equivalence relations, (\ref{A78}) and (\ref{A77}),
for the case $g''_{\Pom \Pom f_{1}(1285)} = 0$ setting
\begin{eqnarray}
\varkappa' = -g'_{\Pom \Pom f_{1}(1285)} 
\frac{m^{2}_{f_{1}(1285)}}{M_{0}^{2}} \,.
\label{A41}
\end{eqnarray}
In Fig.~\ref{fig:2D_map} we show, in a two-dimensional plot, the ratio
\begin{eqnarray}
R(p_{t,1}, p_{t,2}) = \frac{d^{2}\sigma_{\varkappa'} / dp_{t,1}dp_{t,2}}
                                 {d^{2}\sigma_{(2,2)} / dp_{t,1}dp_{t,2}}
\label{ratio_pt1pt2}
\end{eqnarray}
for the $pp \to pp f_{1}(1285)$ reaction at 
$\sqrt{s} = 13$~TeV and $|{\rm y_{M}}| < 2.5$.
The ratio 1 occurs at $p_{t,1} = p_{t,2}$.
In the limited range of transverse momenta of the outgoing protons, 
$p_{t,1} \lesssim 0.6$~GeV and $p_{t,2} \lesssim 0.6$~GeV,
both approaches give similar contributions.
The deviations from the ratio 1 are here less than about 15\,\%.
The same remains true for larger $p_{t,1}$, $p_{t,2}$,
provided $\abs{p_{t,1} - p_{t,2}} \lesssim 0.4$~GeV. 
But clear differences
can be seen if one $p_{t}$ is large  and the other one is small.
We note that by adjusting the $t_{1,2}$ dependent form factors we could,
presumably, obtain the ratio $R(p_{t,1}, p_{t,2})$
in (\ref{ratio_pt1pt2}) even closer to 1
for a larger range of $p_{t,1}$ and $p_{t,2}$.
\begin{figure}[!ht]
\includegraphics[width=0.5\textwidth]{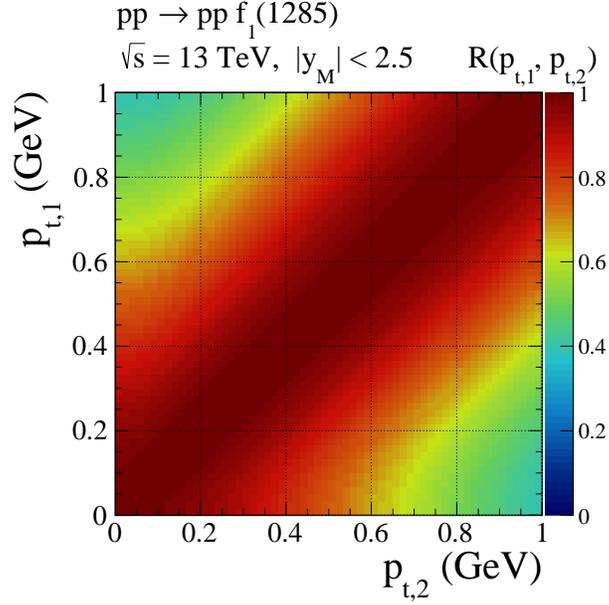}
\caption{\label{fig:2D_map}
\small
The ratio $R(p_{t,1}, p_{t,2})$ (\ref{ratio_pt1pt2})
for the $pp \to pp f_{1}(1285)$ reaction.
The calculation was done for $\sqrt{s} = 13$~TeV 
and with the cut on $|{\rm y_{M}}| < 2.5$.
No absorption effects were included here.}
\end{figure}

At the end of this Appendix we give the proof of (\ref{A72})
and (\ref{A75}). For this we study the reaction (\ref{PomPom_to_f1_AppB})
in the rest system of $f_{1}(k,\epsilon)$
choosing the direction of $\bqa$ as the $z$ axis.
We have then
\begin{eqnarray}
&&k =
\left( \begin{array}{c}
\sqrt{k^{2}} \\
0\\
\end{array} \right)\,, \quad
q_{1} =
\left( \begin{array}{c}
\frac{1}{2}\sqrt{k^{2}} \\
|\bqa|\,\bec\\
\end{array} \right)\,,  \quad
q_{2} =
\left( \begin{array}{c}
\frac{1}{2}\sqrt{k^{2}} \\
-|\bqa|\,\bec\\
\end{array} \right)\,, \nonumber \\
&&k^{2} = (q_{1} + q_{2})^{2} = 4 (m^{2} + |\bqa|^{2})\,,  \nonumber \\
&&q_{1}^{0} = q_{2}^{0} = \sqrt{m^{2} + |\bqa|^{2}} = 
\frac{1}{2} \sqrt{k^{2}}\,.
\label{B20}
\end{eqnarray} 
We shall evaluate the ${\cal T}$-matrix elements 
for (\ref{PomPom_to_f1_AppB}) in the basis
\begin{eqnarray}
\bra{f_{1}(k,\epsilon^{(M)})}{\cal T}
\ket{\Pom(q_{1},\epsilon_{1}^{(M_{1})}),\Pom(q_{2},\epsilon_{2}^{(M_{2})})}
\label{B21}
\end{eqnarray}
where $\epsilon^{(M)}$ and $\epsilon_{1}^{(M_{1})}$, $\epsilon_{2}^{(M_{2})}$
are polarisation vectors and tensors, respectively, corresponding
to definite eigenvalues of the angular momentum operator $J_{z}$.
In detail we choose for the $f_{1}$
\begin{eqnarray}
\bepsilon^{\,(M = \pm 1)} &=& \mp \frac{1}{\sqrt{2}} (\bea \pm i \beb)\,, \nonumber\\
\bepsilon^{\,(M = 0)} &=& \bec\,.
\label{B22}
\end{eqnarray}
Here the $J_{z}$ eigenvalues are $M$.
For the pomeron (1) we define the four-vectors
\begin{eqnarray}
(\chi_{1}^{\pm \, \mu}) &=& \mp \frac{1}{\sqrt{2}}
\left( \begin{array}{c}
0 \\
\bea \pm i \beb \\
\end{array} \right)\,, \nonumber \\
(\chi_{1}^{0 \, \mu}) &=& \frac{1}{m}
\left( \begin{array}{c}
|\bqa| \\
\bec\, q_{1}^{0}\\
\end{array} \right)
\label{B23}
\end{eqnarray} 
and the polarisation tensors $\epsilon_{1}^{(M_{1}) \,\mu\nu}$ ($M_{1} = -2, ..., 2$)
with eigenvalues $M_{1}$ of $J_{z}$ as
\begin{eqnarray}
\epsilon_{1}^{(2)\,\mu\nu} &=& \chi_{1}^{+ \, \mu} \chi_{1}^{+ \, \nu}\,, \nonumber\\
\epsilon_{1}^{(1)\,\mu\nu} &=& \frac{1}{\sqrt{2}} \left(\chi_{1}^{+ \, \mu} \chi_{1}^{0 \, \nu}
                                                       +\chi_{1}^{0 \, \mu} \chi_{1}^{+ \, \nu} \right)\,, \nonumber\\
\epsilon_{1}^{(0)\,\mu\nu} &=& \frac{1}{\sqrt{6}} \,\chi_{1}^{+ \, \mu} \chi_{1}^{- \, \nu}
                              +\sqrt{\frac{2}{3}} \,\chi_{1}^{0 \, \mu} \chi_{1}^{0 \, \nu}
                              +\frac{1}{\sqrt{6}} \,\chi_{1}^{- \, \mu} \chi_{1}^{+ \, \nu}\,, \nonumber\\
\epsilon_{1}^{(-1)\,\mu\nu} &=& \frac{1}{\sqrt{2}} \left(\chi_{1}^{0 \, \mu} \chi_{1}^{- \, \nu}
                                                        +\chi_{1}^{- \, \mu} \chi_{1}^{0 \, \nu} \right)\,, \nonumber\\
\epsilon_{1}^{(-2)\,\mu\nu} &=& \chi_{1}^{- \, \mu} \chi_{1}^{- \, \nu}\,.
\label{B24}
\end{eqnarray}
For the pomeron (2) we define the four-vectors
\begin{eqnarray}
(\chi_{2}^{\pm \, \mu}) &=& \mp \frac{1}{\sqrt{2}}
\left( \begin{array}{c}
0 \\
\bea \mp i \beb \\
\end{array} \right)\,, \nonumber \\
(\chi_{2}^{0 \, \mu}) &=& \frac{1}{m}
\left( \begin{array}{c}
|\bqa| \\
-\bec\, q_{1}^{0}\\
\end{array} \right)
\label{B25}
\end{eqnarray} 
and the polarisation tensors $\epsilon_{2}^{(M_{2}) \,\mu\nu}$
as in (\ref{B24}) but with $\chi_{1}$ everywhere replaced by $\chi_{2}$.
The $\epsilon_{2}^{(M_{2}) \,\mu\nu}$ are then the polarisation tensors
to the eigenvalues $M_{2}$ of ($-J_{z}$) where
$M_{2} \in \{-2, ..., 2 \}$.

Now the stage is set for the evaluation of the ${\cal T}$-matrix elements
(\ref{B21}) using either the couplings (\ref{A6}) plus (\ref{A20})
or (\ref{A27}) plus (\ref{A28}).
From angular momentum conservation only the elements with
\begin{eqnarray}
M = M_{1} - M_{2}
\label{B26}
\end{eqnarray} 
can be different from zero.
The calculations are straightforward but a bit lengthy.
We shall only give the results.
For this we define two ``reduced'' amplitudes
$\bra{M} \hat{T}^{(2,2)} \ket{M_{1},M_{2}}$ and
$\bra{M} \hat{T}^{(4,4)} \ket{M_{1},M_{2}}$; 
see Table~\ref{tab:appendixB}.
\begin{table}[]
\centering
\caption{The matrix elements 
$\bra{M} \hat{T}^{(2,2)} \ket{M_{1},M_{2}}$ and
$\bra{M} \hat{T}^{(4,4)} \ket{M_{1},M_{2}}$.
Matrix elements not listed are zero.}
\label{tab:appendixB}
\begin{tabular}{|c|c|c|c|c|}
\hline
$M$ & $M_{1}$ & $M_{2}$ & $\bra{M} \hat{T}^{(2,2)} \ket{M_{1},M_{2}}$ & $\bra{M} \hat{T}^{(4,4)} \ket{M_{1},M_{2}}$ \\
\hline
 1 & 2 & 1 & -1                                                          & 0\\
 1 & 1 & 0 & $ \frac{1}{\sqrt{6} \,m^{2}} (m^{2} + 4 |\bqa|^{2})$ & 1\\
 1 & 0 &-1 & $ \frac{1}{\sqrt{6} \,m^{2}} (m^{2} + 4 |\bqa|^{2})$ & 1\\
 1 &-1 &-2 & -1                                                          & 0\\
-1 & 1 & 2 &  1                                                          & 0\\
-1 & 0 & 1 & $-\frac{1}{\sqrt{6} \,m^{2}} (m^{2} + 4 |\bqa|^{2})$ &-1\\
-1 &-1 & 0 & $-\frac{1}{\sqrt{6} \,m^{2}} (m^{2} + 4 |\bqa|^{2})$ &-1\\
-1 &-2 &-1 &  1                                                          & 0\\
\hline
\end{tabular}
\end{table}

From the Lagrangians (\ref{A6}) plus (\ref{A20}), respectively
the vertices (\ref{vertex_pompomf1_A}) plus (\ref{vertex_pompomf1_B}),
we obtain for the matrix elements (\ref{B21})
\begin{eqnarray}
\bra{f_{1}(k,\epsilon^{(M)})}{\cal T}
\ket{\Pom(q_{1},\epsilon_{1}^{(M_{1})}),\Pom(q_{2},\epsilon_{2}^{(M_{2})})}
&=& g'_{\Pom \Pom f_{1}}\, \frac{k^{2} \,\sqrt{2} \,|\bqa|^{2}}{M_{0}^{2} \,m}\,
\bra{M} \hat{T}^{(2,2)} \ket{M_{1},M_{2}} \nonumber \\
&+& g''_{\Pom \Pom f_{1}}\, \frac{(k^{2})^{2} \,|\bqa|^{4}}{\sqrt{3} \,M_{0}^{4} \,m^{3}}\,
\bra{M} \hat{T}^{(4,4)} \ket{M_{1},M_{2}} \,.\qquad \quad
\label{B27}
\end{eqnarray}
Note that the $(l,S) = (2,2)$ coupling gives an amplitude
proportional to $|\bqa|^{2}$,
the $(l,S) = (4,4)$ term an amplitude
proportional to $|\bqa|^{4}$,
as it should be for these values of 
the orbital angular momentum $l$.

Now we consider the Lagrangians (\ref{A27}) plus (\ref{A28})
giving the vertices (\ref{A40}) and (\ref{A66}), respectively.
Here we get for the matrix elements (\ref{B21})
\begin{eqnarray}
&&\bra{f_{1}(k,\epsilon^{(M)})}{\cal T}
\ket{\Pom(q_{1},\epsilon_{1}^{(M_{1})}),\Pom(q_{2},\epsilon_{2}^{(M_{2})})} \nonumber \\
&&\qquad \qquad \qquad  = -\left(\varkappa' + \varkappa'' \frac{k^{2}-2m^{2}}{2} \right)\, 
\frac{\sqrt{2} \,|\bqa|^{2}}{m}\,
\bra{M} \hat{T}^{(2,2)} \ket{M_{1},M_{2}}\nonumber \\
&&\qquad \qquad \qquad \quad \, + \varkappa''\, 
\frac{2 k^{2} \,|\bqa|^{4}}{\sqrt{3} \,m^{3}}\,
\bra{M} \hat{T}^{(4,4)} \ket{M_{1},M_{2}} \,.
\label{B28}
\end{eqnarray}
Equating the expressions (\ref{B27}) and (\ref{B28})
we arrive at the relations (\ref{A75}) 
which are, therefore, proved.

\section{The $f_1$ mixing angle and relations between the $\Pom \Pom f_{1}(1285)$ and $\Pom \Pom f_{1}(1420)$ coupling constants}
\label{sec:appendixC}

The different magnitude of the 
coupling constants for the $\Pom \Pom f_1(1285)$ 
and $\Pom \Pom f_1(1420)$ interactions
can be expected to be related
to the internal structure of the mesons.

A commonly used model\footnote{This assumes that $f_1(1420)$ 
is a genuine $q\bar{q}$ resonance, which has been contested, 
however, in \cite{Debastiani:2016xgg,Liang:2020jtw}.
Alternatively, the less well established meson $f_1(1510)$
might appear in place of the $f_1(1420)$.}
is given by
\begin{eqnarray}
 f_1(1285)&=&\cos\phi_f \frac{u\bar u+d\bar d}{\sqrt2} - \sin\phi_f\; s\bar s\,, \nonumber\\
 f_1(1420)&=&\sin\phi_f \frac{u\bar u+d\bar d}{\sqrt2} + \cos\phi_f\; s\bar s\,,
\label{mixing_f1_states}
\end{eqnarray}
with a mixing angle $\phi_f$ parametrising the deviation from ``ideal'' mixing ($\phi_f = 0^\circ$), where
the heavier $f_1$ meson would be purely $s\bar s$.

Ideal mixing is often
assumed as a first approximation to account for the fact that $f_1(1420)$ decays dominantly
into $K\bar K\pi$ \cite{Tanabashi:2018oca}.
Radiative processes, however, indicate a deviation from ideal mixing of about $\phi_f\simeq +20^\circ$ \cite{Leutgeb:2019gbz}
which is consistent with the LHCb result \cite{Aaij:2013rja} of $\phi_f=\pm(24\pm3)^\circ$
and with other results pointing to a range of $+(20 \cdots 30)^\circ$~\cite{Dudek:2011tt,Cheng:2011pb}.

In the chirally symmetric Sakai-Sugimoto model the $\Pom \Pom f_1$ couplings
come exclusively from the axial-gravitational anomaly which involves
only the flavour-singlet combination
$(u \bar{u} + d \bar{d} + s \bar{s})/\sqrt{3}$. 
The assumption that this also holds true in real QCD
would give
\begin{equation}
\frac{g'_{\Pom \Pom f_{1}(1420)}}{g'_{\Pom \Pom f_{1}(1285)}}=
\frac{\sqrt{2} \sin \phi_f +\cos \phi_f }{\sqrt{2} \cos \phi_f -\sin \phi_f}\,.
\label{general_mixing_ratio}
\end{equation} 
and likewise for the couplings $g''$, $\varkappa'$, 
and $\varkappa''$.
Ideal mixing thus corresponds to
\begin{equation}
\frac{g'_{\Pom \Pom f_{1}(1420)}}{g'_{\Pom \Pom f_{1}(1285)}}\Big|_{\phi_f=0^\circ}=
\frac{1}{\sqrt{2}}\,,
\label{ideal_mixing_ratio}
\end{equation} 
while $\phi_f\simeq +20^\circ$ gives ratios larger than unity,
\begin{equation}
\frac{g'_{\Pom \Pom f_{1}(1420)}}{g'_{\Pom \Pom f_{1}(1285)}}\Big|_{\phi_f\simeq +20^\circ} \simeq
1.44\,.
\label{radiative_mixing_ratio}
\end{equation} 

But due to mass effects, 
the relation (\ref{general_mixing_ratio}) is expected to be
only qualitatively correct. 
And, indeed, we have seen in (\ref{f1_vs_f1p}) that
the WA102 data violate (\ref{ideal_mixing_ratio})
or (\ref{radiative_mixing_ratio})
by a factor of about 1.5 or 3, respectively.
If we blame this violation on the subleading reggeon exchanges
we get an indication that the true $\Pom \Pom f_{1}$ couplings
could differ from those given in 
(\ref{parameters_1285_a}) and (\ref{parameters_1420_a})
by a factor of this magnitude.

However, the assumption that the pomeron couples only
to the above flavour singlet $q\bar{q}$ combination
is questionable since it is based on the assumption
of exact flavour SU(3) symmetry.
In fact, the SU(3) flavour symmetry is known to be violated, also for diffractive processes.
For instance, the pomeron coupling to pions is different (larger)
than that for kaons (see, e.g., \cite{Donnachie:2002en}).
The same is true for the coupling of the pomeron
to $\rho^{0}$, $\omega$, and $\phi$ vector mesons;
see \cite{Lebiedowicz:2014bea,Lebiedowicz:2019boz}.

\section{Discussion of subleading exchanges}
\label{sec:appendixD}

In the main text we have assumed that the pomeron-pomeron fusion 
is the dominant reaction mechanism at the top WA102 energy 
$\sqrt{s} = 29.1$~GeV \cite{Barberis:1998by}. 
In fact the WA102 Collaboration measured $f_1(1285)$ and $f_{1}(1420)$ 
also at the significantly lower energy $\sqrt{s} = 12.7$~GeV.

For a complete theoretical discussion of all results
of the WA102 experiment we should consider also the lower energy and
include subleading reggeon-exchange contributions to $f_{1}$ CEP.
We list here the possible fusion reactions leading to 
an $f_{1}$ meson and involving such reggeons:
\begin{eqnarray}
&& \Pom f_{2 \Reg} + f_{2 \Reg} \Pom \to f_{1} \,, 
\label{C1}\\
&& f_{2 \Reg} f_{2 \Reg} \to f_{1} \,, 
\label{C2}\\
&& a_{2 \Reg} a_{2 \Reg} \to f_{1} \,, 
\label{C3}\\
&& \omega_{\Reg} \omega_{\Reg} \to f_{1} \,, 
\label{C4}\\
&& \rho_{\Reg} \rho_{\Reg} \to f_{1} \,, 
\label{C5}\\
&& \phi_{\Reg} \phi_{\Reg} \to f_{1} \,.
\label{C6}
\end{eqnarray}
Let us now discuss the effective couplings for these processes,
taking as a model the results of Appendix A;
see (\ref{A2})--(\ref{A20}).
Following \cite{Ewerz:2013kda} the $f_{2 \Reg}$ and $a_{2 \Reg}$
reggeons will be treated as effective second rank symmetric
traceless tensors, the $\omega_{\Reg}$, $\rho_{\Reg}$, 
and $\phi_{\Reg}$ as effective vectors.

Our coupling Lagrangians for (\ref{C2}) and (\ref{C3})
are then as in (\ref{A6}) and (\ref{A20})
but with the replacements
\begin{equation}
g'_{\Pom \Pom f_{1}} \to g'_{f_{2 \Reg} f_{2 \Reg} f_{1}} \,, \quad
g''_{\Pom \Pom f_{1}} \to g''_{f_{2 \Reg} f_{2 \Reg} f_{1}} \,,
\label{C7}
\end{equation}
and
\begin{equation}
g'_{\Pom \Pom f_{1}} \to g'_{a_{2 \Reg} a_{2 \Reg} f_{1}} \,, \quad
g''_{\Pom \Pom f_{1}} \to g''_{a_{2 \Reg} a_{2 \Reg} f_{1}} \,,
\label{C8}
\end{equation}
respectively.
All these couplings must be real.
For the process (\ref{C1}) there are more coupling possibilities
than the analogs of (\ref{A6}) and (\ref{A20}),
since $\Pom$ and $f_{2 \Reg}$ are distinct.
Indeed, using the methods of Appendix~A of \cite{Lebiedowicz:2013ika},
we find here six independent couplings.

For the process (\ref{C4}) we can rely on the general analysis of
two real vector particles giving an $f_{1}$ with $J^{P} = 1^{+}$ in
Appendix~B of \cite{Lebiedowicz:2013ika}.
From Table~8 there we find that there is only one possible coupling,
$(l,S) = (2,2)$, for this on-shell process.
A convenient coupling Lagrangian is easily written down
\begin{eqnarray}
{\cal L}'_{\omega_{\Reg} \omega_{\Reg} f_{1}}(x) = 
\frac{1}{M_{0}^{4}}
g_{\omega_{\Reg} \omega_{\Reg} f_{1}}
\Big( \omega_{\Reg\,\kappa \lambda}(x)
\twosidep{\mu} \twosidep{\nu} 
\omega_{\Reg\,\rho \sigma}(x) \Big)
\Big( \p_{\alpha} U_{\beta}(x) - \p_{\beta} U_{\alpha}(x) \Big)
g^{\kappa \rho} g^{\mu \sigma} \varepsilon^{\lambda \nu \alpha \beta},
\nonumber \\
\label{C9}
\end{eqnarray}
where
\begin{equation}
\omega_{\Reg\,\kappa \lambda}(x) = 
\p_{\kappa} \omega_{\Reg\,\lambda}(x) - 
\p_{\lambda} \omega_{\Reg\,\kappa}(x)
\label{C10}
\end{equation}
and $g_{\omega_{\Reg} \omega_{\Reg} f_{1}}$ is
a dimensionless coupling constant.
Similar coupling ans{\"a}tze apply to the processes 
(\ref{C5}) and (\ref{C6}).

The vertex following from (\ref{C9}) reads as follows:
\newline
\hspace*{0.65cm}\includegraphics[width=125pt]{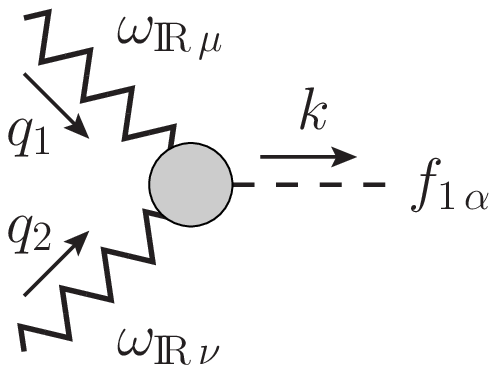}
\begin{eqnarray}
i\Gamma_{\mu \nu \alpha}^{(\omega_{\Reg} \omega_{\Reg} f_{1})}(q_{1},q_{2})\mid_{\rm{bare}} &=&
\frac{2 g_{\omega_{\Reg} \omega_{\Reg} f_{1}}}{M_{0}^{4}}
\Bigl[ (q_{1}-q_{2})^{\rho}
(q_{1}-q_{2})^{\sigma}
\varepsilon_{\lambda \sigma \alpha \beta}\,
k^{\beta} \nonumber\\
&& \times 
( q_{1 \kappa} \,\delta^{\lambda}_{\;\;\mu} - q_{1}^{\lambda}\, g_{\kappa \mu})
       ( q_{2}^{\kappa} \,g_{\rho \nu} - q_{2 \rho} \,\delta^{\kappa}_{\;\;\nu} )
+ (q_{1} \leftrightarrow q_{2}, \mu \leftrightarrow \nu) \Bigr].\qquad \;\;
\label{C101}
\end{eqnarray}
This vertex function satisfies the relations
\begin{eqnarray}
&&\Gamma_{\mu \nu \alpha}^{(\omega_{\Reg} \omega_{\Reg} f_{1})}(q_{1},q_{2}) =
\Gamma_{\nu \mu \alpha}^{(\omega_{\Reg} \omega_{\Reg} f_{1})}(q_{2},q_{1})\,, 
\nonumber \\
&&\Gamma_{\mu \nu \alpha}^{(\omega_{\Reg} \omega_{\Reg} f_{1})}(q_{1},q_{2})\,
q_{1}^{\mu} = 0\,, \nonumber \\
&&\Gamma_{\mu \nu \alpha}^{(\omega_{\Reg} \omega_{\Reg} f_{1})}(q_{1},q_{2})\,
q_{2}^{\nu} = 0\,, \nonumber \\
&&\Gamma_{\mu \nu \alpha}^{(\omega_{\Reg} \omega_{\Reg} f_{1})}(q_{1},q_{2})\,
(q_{1}+q_{2})^{\alpha} = 0\,.
\label{C102}
\end{eqnarray}
We shall use in the following the coupling (\ref{C9}) and the vertex
function (\ref{C101}) for $\omega$ reggeons as well as $\omega$ mesons.

As for the case of the $\Pom \Pom f_{1}$ coupling we find
it useful to consider the analog of the reaction (\ref{PomPom_to_f1_AppB}) here, 
the fusion of two real $\omega$ mesons giving an $f_{1}$ state
\begin{eqnarray}
&&\omega(q_{1},\epsilon_{1}^{(M_{1})}) + \omega(q_{2},\epsilon_{2}^{(M_{2})}) \to f_{1}(k,\epsilon^{(M)})\,. 
\label{C103}
\end{eqnarray}
For our purpose we consider fictitious $\omega$ mesons
of arbitrary mass $m \geqslant 0$ and a fictitious $f_{1}$
of mass $\sqrt{k^{2}} \geqslant 2m$.
We shall work again in the rest system of the $f_{1}$ and
choose the kinematics as in (\ref{B20}).
The polarisation vectors $\epsilon^{(M)}$
($M = \pm 1, 0$) for the $f_{1}$ are taken as in (\ref{B22}).
The polarisation vectors for the $\omega$ mesons 
are taken as follows
\begin{eqnarray}
\epsilon_{1}^{(M_{1})} = \chi_{1}^{(M_{1})}\,,\quad
\epsilon_{2}^{(M_{2})} = \chi_{2}^{(M_{2})}
\label{C104}
\end{eqnarray} 
with
$M_{1}, M_{2} \in \{ \pm 1, 0\}$ and
$\chi_{1}^{(M_{1})}$ and $\chi_{2}^{(M_{2})}$
as in (\ref{B23}) and (\ref{B25}), respectively.

After a straightforward calculation we find
\begin{eqnarray}
&&\bra{f_{1}(k,\epsilon^{(M)})}{\cal T}
\ket{\omega(q_{1},\epsilon_{1}^{(M_{1})}),\omega(q_{2},\epsilon_{2}^{(M_{2})})} = 
-\frac{4 g_{\omega_{\Reg}\omega_{\Reg}f_{1}}}{M_{0}^{4}}\,
k^{2} \,m \,|\bqa|^{2} \nonumber \\
&&\qquad \times
\left\lbrace \delta_{M,1} \left[ \delta_{M_{1},0} \, \delta_{M_{2},-1} 
                               + \delta_{M_{1},1} \, \delta_{M_{2},0} \right]
           - \delta_{M,-1} \left[ \delta_{M_{1},0}\, \delta_{M_{2},1} 
                                + \delta_{M_{1},-1}\,\delta_{M_{2},0} \right]
\right\rbrace \,. \quad
\label{C105}
\end{eqnarray}

Note that the amplitude (\ref{C105}) is proportional to $|\bqa|^{2}$
as it should be since it is derived from the $(l,S) = (2,2)$ coupling
(\ref{C9}). Furthermore, the amplitude (\ref{C105}) vanishes for $m = 0$
as it must be due to the Landau-Yang theorem \cite{Landau:1948kw,Yang:1950rg}.
Indeed, we can consider the production of an $f_{1}$ meson
by two virtual photons of mass squared $q^{2} \geqslant 0$.
For this we use the standard vector-meson-dominance (VMD) ansatz
for the coupling of the photons to the $\omega$ mesons
[see, e.g., (3.23) of \cite{Ewerz:2013kda}],
which then fuse to give the $f_{1}$.
In this case we get for the amplitude the same expression as in (\ref{C105})
with $m$ replaced by $\sqrt{q^{2}}$ 
and multiplied with the appropriate VMD factor
times a vertex form factor $F(q^{2}, q^{2}, k^{2})$
%
\begin{equation}
\left( e \frac{m_{\omega}^{2}}{\gamma_{\omega}} 
\Delta_{T}^{(\omega)}(q^{2})
\right)^{2}
F(q^{2}, q^{2}, k^{2})\,,
\label{C106}
\end{equation}
where $\Delta_{T}^{(\omega)}(q^{2})$ is the transverse part
of the $\omega$ meson propagator
[cf.~(3.2)--(3.4) of \cite{Ewerz:2013kda}].
All gauge invariance relations for these amplitudes are satisfied
due to (\ref{C102}) and the amplitudes vanish for $q^{2} \to 0$
in accord with the Landau-Yang theorem.

A different ansatz for the $\omega_{\Reg} \omega_{\Reg} f_{1}$ coupling
is obtained in the holographic approach~\cite{Domokos:2009cq}:
\begin{eqnarray}
&&{\cal L'}^{\,{\rm CS}}_{\omega_{\Reg} \omega_{\Reg} f_{1}}(x) = 
\varkappa_{\omega}\,
\varepsilon^{\alpha \beta \gamma \delta}\,
U_{\alpha}(x) \,\omega_{\Reg \beta}(x) \,\p_{\gamma} \, \omega_{\Reg \delta}(x)\,,
\label{C107}\\
&&i\Gamma_{\mu \nu \alpha}^{{\rm CS}\,(\omega_{\Reg} \omega_{\Reg} f_{1})}(q_{1},q_{2})\mid_{\rm{bare}} =
\varkappa_{\omega}\,
\varepsilon_{\alpha \mu \nu \rho}\,
(q_{1}-q_{2})^{\rho}\,
\label{C108}
\end{eqnarray}
with $\varkappa_{\omega}$ a dimensionless parameter.
For the vertex function (\ref{C108}) we find the relations
\begin{eqnarray}
&&\Gamma_{\mu \nu \alpha}^{{\rm CS}\,(\omega_{\Reg} \omega_{\Reg} f_{1})}(q_{1},q_{2}) =
\Gamma_{\nu \mu \alpha}^{{\rm CS}\,(\omega_{\Reg} \omega_{\Reg} f_{1})}(q_{2},q_{1})\,; 
\label{C109}\\
&&\Gamma_{\mu \nu \alpha}^{{\rm CS}\,(\omega_{\Reg} \omega_{\Reg} f_{1})}(q_{1},q_{2})\,
q_{1}^{\mu} = i \varkappa_{\omega}\,
\varepsilon_{\alpha \mu \nu \rho}\,
q_{1}^{\mu} q_{2}^{\rho} \neq 0\,,
\nonumber \\
&&\Gamma_{\mu \nu \alpha}^{{\rm CS}\,(\omega_{\Reg} \omega_{\Reg} f_{1})}(q_{1},q_{2})\,
q_{2}^{\nu} = -i \varkappa_{\omega}\,
\varepsilon_{\alpha \mu \nu \rho}\,
q_{1}^{\rho} q_{2}^{\nu} \neq 0\,;
\label{C110}\\
&&\Gamma_{\mu \nu \alpha}^{{\rm CS}\,(\omega_{\Reg} \omega_{\Reg} f_{1})}(q_{1},q_{2})\,
(q_{1}+q_{2})^{\alpha} = -i \varkappa_{\omega}\,
\varepsilon_{\alpha \mu \nu \rho}\,
(q_{1}+q_{2})^{\alpha} (q_{1}-q_{2})^{\rho} \neq 0\,.
\label{C111}
\end{eqnarray}

For the process (\ref{C103}) we find here
\begin{eqnarray}
&&\bra{f_{1}(k,\epsilon^{(M)})}{\cal T}
\ket{\omega(q_{1},\epsilon_{1}^{(M_{1})}),\omega(q_{2},\epsilon_{2}^{(M_{2})})} = 
\varkappa_{\omega} \frac{2 |\bqa|^{2}}{m} \nonumber \\
&&\qquad \times
\left\lbrace \delta_{M,1} \left[ \delta_{M_{1},0} \, \delta_{M_{2},-1} 
                               + \delta_{M_{1},1} \, \delta_{M_{2},0} \right]
           - \delta_{M,-1} \left[ \delta_{M_{1},0}\, \delta_{M_{2},1} 
                                + \delta_{M_{1},-1}\,\delta_{M_{2},0} \right]
\right\rbrace \,. \quad
\label{C112}
\end{eqnarray}
With constant $\varkappa_\omega$, these amplitudes diverge for $m \to 0$.
Here we cannot use the usual VMD relations
to relate these amplitudes to the ones for the fusion
of two virtual or real photons giving an $f_{1}$ meson.
Because of (\ref{C110}) the corresponding amplitudes for
$\gamma^{*} \gamma^{*} \to f_{1}$ would not satisfy
the necessary gauge invariance relations.

Vector-meson dominance is, in fact, realized in holographic QCD (for an extensive discussion
in the Sakai-Sugimoto model see \cite{Sakai:2005yt}).
The coupling to virtual or real photons involves bulk-to-boundary propagators
which correspond to sums over an infinite tower of massive vector mesons.
In place of the constant $\varkappa_\omega$ one obtains an asymmetric
transition form factor that does satisfy the Landau-Yang theorem and
which has been studied in \cite{Leutgeb:2019gbz}, where
good agreement with available data from the L3 experiment \cite{Achard:2001uu,Achard:2007hm} 
on single-virtual $\gamma \gamma^{*} \to f_{1}$ has been found.

Clearly the inclusion of all these subleading exchanges
(\ref{C1})--(\ref{C6}) would introduce many new coupling parameters
and form factors and would make a meaningful analysis of the WA102 data
practically impossible.
However, for the analysis of data from the COMPASS experiment,
which operates in the same energy range as previously
the WA102 experiment, it could be very worthwhile to study all
the above subleading exchanges in detail. 
In addition one also has to keep in mind 
that there should be a smooth transition from reggeon
to particle exchanges when going to very low energies.
Clearly, all these topics deserve careful analyses,
but they go beyond  the scope of the present paper.

Here we shall only discuss some rough estimates of subleading contributions
at the WA102 energy of $\sqrt{s} = 29.1$~GeV.

At the relatively low energies of the WA102 experiment 
the subleading reggeon exchanges are not excluded \textit{a priori}. 
Among those, the $\omega \omega \to f_1$
and $\rho^0 \rho^0 \to f_1$ exchanges are the most probable ones.
We know how the $\omega$ and $\rho^0$ couple to nucleons.
However, the coupling of $\omega \omega \to f_1$ and
$\rho^0 \rho^0 \to f_1$ is less known. 
Future experiments at HADES and
PANDA will provide new information there. 
The $\rho^0 \rho^0 \to f_1$ coupling constant 
can be obtained from the decays:
$f_1 \to \rho^0 \gamma$ and/or $f_1 \to \pi^+ \pi^- \pi^+ \pi^-$.
This issue will be discussed elsewhere.
The uncertainties related to form factors preclude, however,
strict predictions.
Fortunately, the following (almost model independent) observation 
explains the situation. 
It seems rather obvious that the reggeized-vector-meson-exchange or reggeon-reggeon-exchange contributions cannot
exceed the experimental data of the WA102 Collaboration 
\cite{Barberis:1998by}.
According to our estimates we find,
using subleading exchanges only,
that they allow a description of the $\sqrt{s} = 12.7$~GeV data
(see Table~\ref{tab:table1}) but then one misses the data for 
$\sqrt{s} = 29.1$~GeV by a factor of at least 5.
This is due to the energy dependence of the subleading contributions.
This means that the dominant contribution at $\sqrt{s} = 29.1$~GeV
is most probably related to the double-pomeron-exchange contribution 
considered in our paper.

To make this statement quantitative we proceed as follows.
Let ${\cal M}$ be the amplitude for the $\Pom \Pom \to f_{1}$ fusion
as calculated in the present paper with which we
could fit the WA102 data for $\sqrt{s} = 29.1$~GeV.
We assume now that the true $\Pom \Pom \to f_{1}$ fusion amplitude
is \mbox{$x {\cal M}$ ($x > 0$)} and that the reggeon amplitude is similar
in structure to the pomeron amplitude and given by $y {\cal M}$.
We must have then, precluding a complete sign change of the amplitudes,
\begin{equation}
x + y = 1 \,.
\label{D23}
\end{equation}
From the above estimates of the reggeon contributions alone we get
\begin{equation}
|y|^{2} \leqslant 0.20 \,.
\label{D24}
\end{equation}
For maximal constructive interference of pomeron and reggeon contributions
we get
\begin{equation}
\begin{split}
&y \leqslant \sqrt{0.20} = 0.45 \,,\\
&x = 1 - y \geqslant 0.55\,.
\end{split}
\label{D25}
\end{equation}
For destructive interference we would get $x > 1$.
The result (\ref{D25}) is the basis for the estimate that
the true $\Pom \Pom f_{1}$ couplings may be up to a factor of 2
smaller than the ones obtained in our present paper
from the comparison of the WA102 data at $\sqrt{s} = 29.1$~GeV
to our theory neglecting the reggeons.

\section{The $\phi_{pp}$ distributions for CEP of $f_1$- and $\eta$-type mesons at $\phi_{pp} = 0$ and $\pi$}
\label{sec:appendixE}

Here we discuss general properties of the $\phi_{pp}$ distributions
for CEP of $f_{1}$ mesons (\ref{2to3_reaction}) 
and for the analogous reaction with $\eta$-type mesons
\begin{eqnarray}
p(p_{a},\lambda_{a}) + p(p_{b},\lambda_{b}) \to
p(p_{1},\lambda_{1}) + \eta(k) + p(p_{2},\lambda_{2}) \,.
\label{E1}
\end{eqnarray}
Recall that $\phi_{pp}$ is the azimuthal angle between the transverse
momenta of the two outgoing protons in the overall
c.m. system (Fig.~\ref{fig:phi_pp}).
For the following arguments we work in this c.m. system.
\begin{figure}[!ht]
\includegraphics[width=3.5cm]{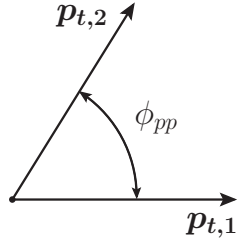}
\caption{\label{fig:phi_pp}
\small
Definition of the angle $\phi_{pp}$ ($0 \leqslant \phi_{pp} \leqslant \pi$).}
\end{figure}

\begin{figure}[!ht]
\includegraphics[width=10.0cm]{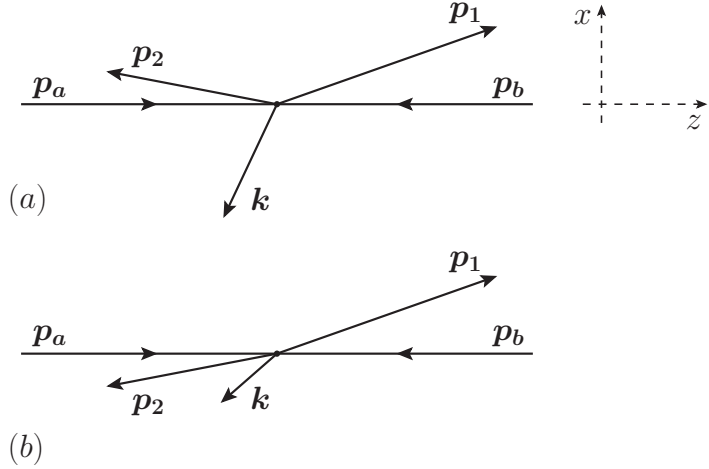}
\caption{\label{fig:CEP_system}
\small
Sketch of CEP reactions with (a) $\phi_{pp} = 0$ 
and (b) $\phi_{pp} = \pi$
and definition of the $x$ and $z$ axes.}
\end{figure}
For $\phi_{pp} = 0$ and $\pi$ the reaction (\ref{E1}) 
is planar (Fig.~\ref{fig:CEP_system}).
We choose the reaction plane as the $x z$ plane
of our coordinate system.
Note that this plane is a symmetry plane for our reaction
and we shall exploit this in the following.

In (\ref{2to3_reaction}) and (\ref{E1}) we have written our reactions
in terms of protons with definite helicities
$\lambda_{a}, \lambda_{b}, \lambda_{1}, \lambda_{2} \in \{ 1/2, -1/2 \}$.
Here we shall use protons with spin
$\tilde{\lambda} = \pm 1/2$ in the $y$ direction,
orthogonal to the reaction plane.
We have
\begin{eqnarray}
&&\ket{p(\bp,\tilde{\lambda})}_y = \frac{1}{\sqrt{2}}
\left( \ket{p(\bp, 1/2)} + i 2 \tilde{\lambda} \ket{p(\bp,-1/2)} \right)\,,
\nonumber\\
&&\tilde{\lambda} = \pm 1/2\,, \quad \bp = \bpa, \bpb, \bpaa, \bpbb\,.
\label{E2}
\end{eqnarray}
Now we consider a reflection $S$ on the $xz$ plane.
$S$ can be written as a parity transformation,
$P$, times a rotation $R_{2}(\pi)$ by $\pi$
around the $y$ axis
\begin{eqnarray}
S = R_{2}(\pi)\, P\,.
\label{E3}
\end{eqnarray}
For the proton states (\ref{E2}) this gives
\begin{eqnarray}
U(S) \ket{p(\bp,\tilde{\lambda})}_y = 
e^{i \pi \tilde{\lambda}} \ket{p(\bp,\tilde{\lambda})}_y \,.
\label{E4}
\end{eqnarray}

\begin{flushleft}
\textit{Assumption:}
\end{flushleft}
\vspace{-0.3cm}
Now we assume that at high energies there is $s$-channel
helicity conservation of the protons in (\ref{E1})
and, more strongly, helicity independence.
That is, we assume
\begin{eqnarray}
\bra{p(p_{1},\lambda_{1}), p(p_{2},\lambda_{2}), \eta(k)} {\cal T}
\ket{p(p_{a},\lambda_{a}), p(p_{b},\lambda_{b})} 
\propto \delta_{\lambda_{1} \lambda_{a}}\,
      \delta_{\lambda_{2} \lambda_{b}} \,.
\label{E5}
\end{eqnarray}

In our calculations for CEP reactions of $f_{1}$
we always used this high-energy approximation for the protons.
Transforming to the states (\ref{E2}) we also get there
\begin{eqnarray}
\bra{p(p_{1},\tilde{\lambda}_{1}), p(p_{2},\tilde{\lambda}_{2}), \eta(k)} {\cal T}
\ket{p(p_{a},\tilde{\lambda}_{a}), p(p_{b},\tilde{\lambda}_{b})}_{y} 
\propto \delta_{\tilde{\lambda}_{1} \tilde{\lambda}_{a}}\,
        \delta_{\tilde{\lambda}_{2} \tilde{\lambda}_{b}} \,.
\label{E6}
\end{eqnarray}

The next step is to apply to the ${\cal T}$-matrix element
(\ref{E6}) a reflection transformation $S$ (\ref{E3}).
With (\ref{E4}) and (\ref{E6}) we get for the pseudoscalar $\eta$
\begin{eqnarray}
&&\bra{p(p_{1},\tilde{\lambda}_{1}), p(p_{2},\tilde{\lambda}_{2}), \eta(k)} {\cal T}
\ket{p(p_{a},\tilde{\lambda}_{a}), p(p_{b},\tilde{\lambda}_{b})}_{y} \nonumber \\
&&= (-1) \exp \left[ i \pi \left(\tilde{\lambda}_{a}-\tilde{\lambda}_{1}+\tilde{\lambda}_{b}-\tilde{\lambda}_{2} \right) \right]
\bra{p(p_{1},\tilde{\lambda}_{1}), p(p_{2},\tilde{\lambda}_{2}), \eta(k)} {\cal T}
\ket{p(p_{a},\tilde{\lambda}_{a}), p(p_{b},\tilde{\lambda}_{b})}_{y}\nonumber \\
&&= 0 \,.
\label{E7}
\end{eqnarray}
This proves that under the above \textit{assumption} the distribution in
$\phi_{pp}$ must vanish for $\phi_{pp} = 0$ and $\pi$
in CEP of $\eta$-type mesons.

The $\phi_{pp}$ distributions in CEP of the $\eta$
of mass 548~MeV and $\eta'(958)$ were studied
in the WA102 experiment \cite{Barberis:1998ax} and,
using our theoretical framework, in \cite{Lebiedowicz:2013ika};
see Fig.~14 there.
The experimental distributions vanish 
for $\phi_{pp} = 0$, but at $\phi_{pp} = \pi$ a small
residual different from zero is visible.
According to our results this must be due to
contributions violating our assumptions concerning the helicities.

Finally we return to $f_{1}$ production (\ref{2to3_reaction}).
For an $f_{1}$ meson with $J^{P} = 1^{+}$ we can use
the Wigner basis with the $f_{1}$ polarisation vectors
$\bex, \bey, \bez$.
Under the reflection $S$ we have the following transformation
properties
\begin{eqnarray}
&&U(S) \ket{f_{1}(\bk,\bex)} = - \ket{f_{1}(\bk,\bex)}\,, \nonumber\\
&&U(S) \ket{f_{1}(\bk,\bey)} =  \quad \ket{f_{1}(\bk,\bey)}\,, \nonumber\\
&&U(S) \ket{f_{1}(\bk,\bez)} = - \ket{f_{1}(\bk,\bez)}\,.
\label{E8}
\end{eqnarray}
Using now the same argumentation as for $\eta$-type mesons above
we conclude that for $\phi_{pp} = 0$ and $\pi$
the produced $f_{1}$ must have the polarisation
in the $y$ direction, that is, transverse to the reaction plane.

To summarize: in this appendix we have shown the following \textit{theorem}.
Assuming $s$-channel helicity conservation and helicity independence
for CEP of $\eta$- and $f_{1}$-type mesons
the $\phi_{pp}$ distributions must vanish 
for $\phi_{pp} = 0$ and $\pi$ for the $\eta$ case
and the $f_{1}$ must be polarised transversely to the reaction plane
for these $\phi_{pp}$ values.

\acknowledgments
One of us (O.N.) thanks Carlo Ewerz
for useful discussions; J.L. and A.R.
thank Florian Hechenberger for collaboration in the early
stages of this project.
We thank Claude Amsler for pointing out 
an alternative explanation for the $f_{1}(1420)$.
This work was partially supported by
the Polish National Science Centre under Grant
No. 2018/31/B/ST2/03537
and by the Center for Innovation and Transfer of Natural Sciences 
and Engineering Knowledge in Rzesz\'ow (Poland).
J.L. was supported by the Austrian Science Fund FWF,
doctoral program Particles \& Interactions, Project No. W1252-N27.

\bibliography{refs}

\end{document}